\documentclass[]{aa}
\usepackage{graphicx}
\usepackage{txfonts}
\usepackage{amssymb}
\usepackage[figuresright]{rotating}

\let\3=\ss
\begin{document}
  
  \title{Spatially resolved detection of crystallized water ice in a T\,Tauri
    object\thanks{Based on 
      observations collected at the European Organisation for Astronomical
      Research in the Southern Hemisphere, Chile (proposal 077.C-0794(A).)}}
  \author{A. A. Schegerer \inst{1,2}, S. Wolf\inst{3}} 
  
  \offprints{A. A. Schegerer, \email{schegerer@mpia-hd.mpg.de}}
   
  \institute{Helmholtz Zentrum M\"unchen, German Research Center for
    Environmental Health, Ingolst\"adter Landstra\3e 1, 85758 Neuherberg,
    Germany \and Max-Planck-Institut f\"ur Astronomie, K\"onigstuhl 17, 69117
    Heidelberg, Germany \and Universit\"at Kiel, Institut
    f\"ur Theoretische Physik und Astrophysik, Leibnizstra{\ss}e 15, 24098
    Kiel, Germany }
  
  \date{Received $<>$ ; accepted $<>$ }
   
  \abstract
  {}
  {We search for frozen water and its processing around young stellar objects
    (YSOs of class I/II). We try to detect potential, regional
    differences in water ice evolution within YSOs, which is relevant to
    understanding the chemical structure of the progenitors of 
    protoplanetary systems and the evolution of solid
    materials. 
    Water plays an important role as a reaction
    bed for rich chemistry and is an indispensable requirement for
    life as known on Earth.}
  {We present our analysis of NAOS-CONICA/VLT spectroscopy of water ice at 3~$\mu$m for
    the T\,Tauri star YLW\,16\,A in the $\rho$~Ophiuchi molecular
    cloud. We obtained spectra for different regions of
    the circumstellar environment. The
    observed absorption profiles are deconvolved with the mass 
    extinction profiles of amorphous and crystallized ice measured in
    laboratory. We take into account both absorption and
    scattering by ice grains.}
  {Water ice in YLW16A is detected with optical depths of between  
    $\tau=1.8$ and $\tau=2.5$. The profiles 
    that are measured can be fitted predominantly by the extinction profiles of
    small grains (0.1~$\mu$m -- 0.3~$\mu$m) with a small  
    contribution from large grains ($<$$10$\%). However, an unambiguous 
    trace of grain growth cannot be found. We detected 
    crystallized water ice spectra that have their origin in different regions
    of the circumstellar environment of the T\,Tauri star
    YLW\,16\,A. The crystallinity increases in the upper layers of the
    circumstellar 
    disk, while only amorphous grains exist in the bipolar
    envelope. As in studies of silicate grains in 
    T\,Tauri objects, the higher crystallinity in the upper layers of
    the outer
    disk regions implies that water ice crystallizes and remains
    crystallized close to the disk
    atmosphere where water ice is shielded against hard irradiation.
    }
  {}  
  \keywords{Infrared: stars -- Accretion disks -- Astrochemistry }
  
  \authorrunning{Schegerer \& Wolf}
  \titlerunning{Crystallized water ice in T\,Tauri objects}
  \maketitle
  
  \section{Introduction}\label{section:introduction}
  The importance of water (ice) to protoplanetary systems 
  is justified by the following properties:  
  liquid water is a proper solvent for many minerals and organic
  molecules and allows the formation of complex organic molecules. Water is
  the reaction bed for the 
  photochemical synthesis of both aromatic compounds (Bernstein et
  al.~\cite{bernstein}) and amino acids (Munoz Caro et
  al.~\cite{munoz}), even in its 
  solid state. Therefore, rich deposits of water (ice) in
  protoplanetary disks are assumed to be a requirement for the formation of 
  life as known on Earth. The importance of water
  ice is in addition emphasized by the assumption that ice planets, such as
  Neptune, 
  form outside the snowline where water condenses on dust grains. The 
  mass fraction of solid matter in
  protoplanetary disks abruptly increases at the
  snowline at least by one order of magnitude
  (Stevenson \& 
  Lunine~\cite{stevenson}).  
  
  The water ice band at $3.08\,\mathrm{\mu m}$, which is caused by a
  vibrational excitation of the OH bond in the water molecule, was
  primarily discovered in the envelopes of numerous deeply embedded
  protostars (Gillett \& Forrest~\cite{gillett}; Gillett et
  al.~\cite{gillettII}; Merrill et al.~\cite{merrill}). Frozen water that can
  be found in
  dust particles is the most frequent ice molecule in YSOs  
  (Ehrenfreund et al.~\cite{ehrenfreund})\footnote{Fourteen different ice
    compounds, such as CO, CO$_\mathrm{2}$, or CH$_\mathrm{4}$ ice are known. Ice
    molecules are weakly bounded just by van-der-Waals forces or by
    hydrogen bonds. A chemical compound that can be found in metals,
    minerals, and diamonds does not form.}. The huge water ice deposit results
  from its molecular 
  property: in contrast to nonpolar, volatile ice molecules,
  such as CO, CO$_\mathrm{2}$, N$_\mathrm{2}$, and O$_\mathrm{2}$ ice, 
  H$_\mathrm{2}$O ice has a high sublimation temperature of
  $\sim$$120\,\mathrm{K}$ depending on the ambient pressure
  (Davis~\cite{davis}). The polarity of the water molecule is
  responsible for its stable adsorption by condensation seeds such as silicate
  dust grains. Cosmic irradiation and chemical reactions could force the
  evaporation of water ice (L\'eger et al.~\cite{leger}; Hartquist \&
  Wolf~\cite{hartquist}). However, these reactions are impeded by
  regions with strong shielding and low temperatures 
  (Tielens \&
  Hagen~\cite{tielens}). This is why the water ice band is
  commonly observed in the spectra of deeply embedded protostars
  (Boogert et al.~\cite{boogert}). However, it is assumed that larger deposits
  of water ice exist in the optically
  thick regions of circumstellar disks (Fig.~1 in Chiang et
  al.~\cite{chiang}). This assumption may be supported by the
  detection of the water ice band at $46.5\,\mathrm{\mu m}$
  (Creech-Eakman et al.~\cite{creech-eakman}) and $3.08\,\mathrm{\mu
    m}$ (Leinert et al.~\cite{leinert}; Terada et al.~\cite{terada}) in
  the spectra of several T\,Tauri objects. 
  
  The water ice band at $3.08\,\mathrm{\mu m}$ in L band\footnote{The L band
    covers a wavelength interval between $2.8\,\mathrm{\mu m}$ and
    $4.2\,\mathrm{\mu m}$ in the near-infrared range.} is assumed
  to be an excellent indicator of
  the evolutionary status of circumstellar disks (van de
  Bult~\cite{bult}). In a protostellar molecular cloud, water ice has
  a primarily amorphous structure, as many previous studies have shown (e.g., Thi et
  al.~\cite{thi}; Pontoppidan et al.~\cite{pontoppidan}). As soon as the
  temperature exceeds a level of $\sim$$80\,\mathrm{K}$, water ice
  begins to crystallize (Hagen et al.~\cite{hagen}). This conversion can
  be detected by monitoring the profile of the optical depth, i.\/e., the logarithm of the
  absorption band: as the 
  crystallinity increases, the profile narrows and its maximum 
  shifts to longer wavelengths (e.g., Smith et
  al.~\cite{smith}). But the crystalline structure can also be destroyed, e.g., by
  hard irradiation (Kouchi \&
  Kuroda~\cite{kouchi}). In addition to crystallization, grain growth of ice
  particles can be inferred from a modification of the optical depth
  profile. An increase in the dust grain radius results in a  
  higher scattering efficiency, a broadening of the absorption band
  at long wavelengths, and a shift of the maximum to longer
  wavelengths. These changes are comparable to those related to the prominent 
  $10\,\mathrm{\mu m}$ silicate band (e.g., Bouwman et
  al.~\cite{bouwman}), but the origin of the spectral bands is
  different. While the emission band of silicate has its origin in the
  warm disk atmosphere, the absorption band of water ice originates in
  more embedded, cold regions closer to the disk midplane and far from the
  central star. 
  
  The water ice band interferes with absorption bands of additional
  materials that are dissolved in water ice or form a complex
  compound with the water molecule: ammonia hydrate
  H$_\mathrm{2}$O.NH$_\mathrm{3}$ (minimum at $3.47\,\mathrm{\mu m}$:
  Merrill et al.~\cite{merrill}; Mukai et al.~\cite{mukai}; Dartois \&
  d'Hendecourt~\cite{dartois}), ammonia NH$_\mathrm{3}$
  ($2.96\,\mathrm{\mu m}$: d'Hendecourt et al.~\cite{dhendecourt}; Dartois \&
  d'Hendecourt~\cite{dhendecourt}), methanol CH$_\mathrm{3}$OH
  ($3.54\,\mathrm{\mu m}$: Dartois et al.~\cite{dartois}),
  hydrocarbons with the vibrational excitations of the CH,
  CH$_\mathrm{2}$, and CH$_\mathrm{3}$ bond ($3.38\,\mathrm{\mu m}$,
  $3.42\,\mathrm{\mu m}$, and $3.48\,\mathrm{\mu m}$: Duley \&
  Williams~\cite{duley}; Chiar et al.~\cite{chiar}), and other chemical
  compounds (Ehrenfreund et al.~\cite{ehrenfreund}). The
  absorption bands of these molecules are narrow and interfere predominantly
  with the  
  water ice band at longer wavelengths around $3.6\,\mathrm{\mu m}$. They can
  therefore be distinguished from the water ice band. To study the chemical
  evaluation of ices in solar-mass systems, a survey in the infrared
  wavelength range where $41$ low-luminosity YSOs were included was 
  conducted by using the infrared spectrograph on the Spitzer
  Space Telescope (Boogert et al.~\cite{boogertII}).   

  The scientific scope of this study is to look for spatial variations in the
  water ice band profile at $3.08\,\mathrm{\mu m}$ within a T\,Tauri star 
  to detect possible, regional differences in water ice evolution. In 
  Sect.~\ref{section:observation} and \ref{section:datareduction}, we
  scrutinize in detail our spectroscopic observations and data
  reduction, respectively. The derived water ice profiles are presented in 
  Sect.~\ref{section:results}. The ice profiles that could be derived are
  analyzed in Sect.~\ref{section:modelresults} by considering both the 
  absorption profiles of ice grains of different size and 
  crystallization degree that are measured in laboratory. This
  well-established approach of deconvolution is described 
  in Sect.~\ref{section:approach}. The question of whether evolved, i.\/e., large
  and/or crystyllized ice grains may finally be found is discussed in 
  Sect.~\ref{section:graingrowth} and \ref{section:crystalline?}. We
  conclude this study with a summary in Sect.~\ref{section:summary}.
  
  \section{Observations}\label{section:observation}  
  \begin{table*}[!t]
    \centering
    \begin{minipage}{1.0\textwidth}
      \caption{Overview of our observations with
        NAOS-CONICA. Coordinates (Cutri et al.~\cite{cutri}),
        L band magnitude, spatial FWHM, date, observing
        time ({\@}$UT${\@}), as well as the mean airmass ({\@}$AM${\@}) during
        the 
        observations are listed. The object YLW\,16\,A was observed twice
        using two different slit orientations: parallel (p) and
        orthogonal (o) to the rotational axis of the YSO. The quantity $T_\mathrm{tot}$ is
        the total exposure time. Directly before or after the
        observation, the standard star HR\,6070 was observed. }
      \label{table:observation}
      \begin{center}
        \begin{tabular}{lrrrrrccr} 
          \hline\hline
          object & $\alpha(\mathrm{J2000})$ & $\delta(\mathrm{J2000})$ & L
          [mag] & $FWHM$ [''] & 
          date & $UT$ & $AM$ & $T_\mathrm{tot}$\\ \hline 
          HR\,6070 & $16\,18\,17.9$ & $-28\,36\,50$ & $4.807$ & -- & April
          $28^{th}, 2006$ & $7:00 - 7:02$ & $1.0$ & $2.0\,\mathrm{s}$\\ 
          \object{YLW\,16\,A} (p) & $16\,27\,28.0$ & $-24\,39\,34$ & $6.7$ & $0.5$ & April
          $28^{th}, 2006$ & $7:26 - 8:08$ & $1.0$ &  
          $60 \times 3 \times 4.0\,\mathrm{s}$\smallskip\\      
          YLW\,16\,A (o) & $16\,27\,28.0$ & $-24\,39\,34$ & $6.7$ & $0.5$ & August
          $10^{th}, 2006$ & $0:45 - 2:00$ & $1.1$ &  
          $26 \times 50 \times 2.8\,\mathrm{s}$ \\
          HR\,6070 & $16\,18\,17.9$ & $-28\,36\,50$ & $4.807$ & -- & August
          $10^{th}, 2006$ & $2:51 - 2:59$ & $1.3$ & 
          $2.0\,\mathrm{s}$\bigskip\\ 
        \end{tabular}
      \end{center}
    \end{minipage}
  \end{table*}

  Our observations with the near-infrared (NIR) spectrometer CONICA (Lenzen et
  al.~\cite{lenzen}) were performed at the $8.2\,\mathrm{m}$ YEPUN telescope
  of the 
  Very Large Telescope (VLT) in spring and summer $2006$. We used the
  L54\_1\_SL mode where a wavelength range of between 
  $2.60\,\mathrm{\mu m}$ and $4.20\,\mathrm{\mu m}$ is covered. Observations
  with CONICA are 
  supported by the adaptive optics NAOS (Rousset et al.~\cite{rousset}).
  A linear, spectral dispersion of $3.16\,\mathrm{nm/pixel}$ could be 
  reached. The spatial resolution that could theoretically be reached
  was $0.054${\arcsec}. Depending on the weather conditions during our
  observations, the pixel scale was $\sim$$0.12${\arcsec}. The slit of the 
  spectrometer had a width of $0.172${\arcsec} and a length of
  $28${\arcsec}. 

  We selected YSOs with quasi edge-on disks to be able to 
  observe optically thick regions that shield
  ice from hard irradiation. Images, polarimetric maps, and previous object
  models 
  were used as selection criteria. A
  visual extinction of $A_\mathrm{V}>3\,\mathrm{mag}$ is 
  sufficient for an effective shielding of water ice against irradiation
  (Murakawa et al.~\cite{murakawa}). 

  A lunar occultation observation of YLW\,16\,A at $2\,\mathrm{\mu m}$ (Simon et
  al.~\cite{simon}) showed that the object is a single extended source and
  strongly inclined. Two
  brightness peaks are separated by $\sim$$0.5${\arcsec} in a K band image
  obtained by NICMOS at the Hubble Space Telescope (HST; Allen et
  al.~\cite{allen}). These peaks are assigned to the optically thin bipolar
  envelope. 

  Table~\ref{table:observation} provides an overview of our observations
  of the target YLW\,16\,A and the corresponding standard star HR\,6070. The
  brightness of the standard star, an A\,0\,V star, is known from the
  standard star catalog of van 
  Bliek et al.~(\cite{bliek}). For the preparation of the observations, the L
  band brightness of the target   
  was estimated in this study by considering the H band magnitudes and the 
  (H-K) and (K-N) colors of additional YSOs of the same star-forming
  region (Allen et al.~\cite{allen}).
  
  Our target was observed using the nodding mode with a nodding angle
  of $10${\arcsec}. Thus, the target was successively
  observed in different detector areas where the sky background could be
  eliminated by subtracting subsequent exposures. 
  The chopping mode, which provides a more effective
  elimination of short-term variations in the sky, was not available in
  the observation mode used. 
  
  The target was observed twice, with two orthogonal 
  orientations of the slit. The observation sequence started with a slit
  orientation parallel to the rotational axis of the object (position:
  p-parallel). In a second 
  exposure, the slit was rotated by $90^{\circ}$ (position: o-orthogonal). The
  photometric center served as a rotation center. This procedure is
  assumed to allow the determination of the spatial 
  distribution of water ice in the different regions of the circumstellar
  environment.  

  In the appendix of this publication, we list additional observations of YSOs
  within this program. These YSOs also belong to the $\rho$ Ophiuchi region. 
   
  \section{Data reduction}\label{section:datareduction}
\begin{figure}[t]
   \hfill{}\includegraphics[scale=0.5]{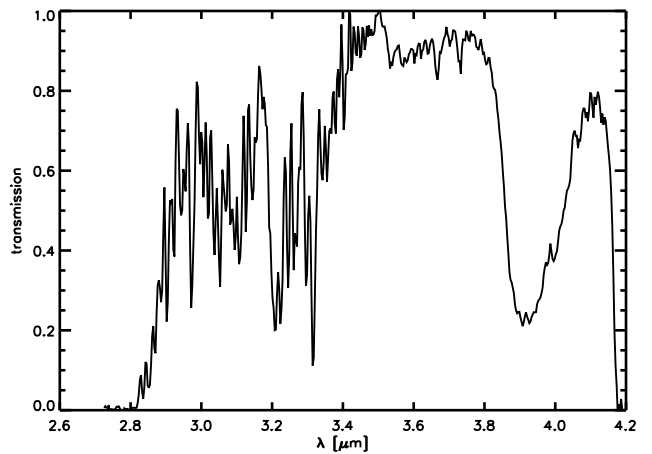}\hfill{}
   \caption{\label{figure:transmission} Transmission of atmosphere and
     instrument derived from the spectrum of the 
     standard star HR\,6070.}
 \end{figure}

  When reducing the spectroscopic data obtained with
  NAOS-CONICA, 
  particular problems have to be taken into account. The specific steps of the
  reduction are therefore considered in detail:  
  \renewcommand{\labelenumi}{\roman{enumi}.}
  \begin{enumerate}
  \item Bad pixels are localized and replaced by the median
    of the pixels within the surrounding $3 \times 3$ detector segment. 
  \item The background of detector, telescope, and sky are eliminated by
    the 
    subtraction of successive exposures for different nodding positions.
    Owing to unavailability of the chopping mode that can clear short-term variations
    in the sky, telluric features cannot be completely removed in the
    object spectra. The
    resulting images are flatfielded. 
  \item The spatial and spectral dimensions of a spectroscopic
    exposure are 
    assigned to the rows and columns of the image matrix,
    respectively, to a first
    approximation. A Gaussian function is fitted to each row of the
    array to determine the maximum of the spatial intensity
    distribution. The counting rates of the pixels within $4 \times FWHM$
    (full width at half maximum) of
    each Gaussian function are added. Alternatively,
    the Gaussian functions are divided into seven single intervals resulting in
    spatially 
    adjacent spectra. For a pixel scale of
    $0.12${\arcsec} and the source distance of $160\,\mathrm{pc}$, the single
    intervals have widths of about $20\,\mathrm{AU}$, $20\,\mathrm{AU}$,
    $30\,\mathrm{AU}$, and $40\,\mathrm{AU}$, respectively, symmetrically
    arranged of increasing
    width at increasing distance from the brightness maximum. The
    interval of the central region has a width of $20\,\mathrm{AU}$, the
    intervals of the outermost regions have widths of $40\,\mathrm{AU}$. 
  \item The wavelength calibration of the L band spectra acquired with
    NAOS-CONICA is difficult because observations of specific calibration lamps
    could not be performed in this observing mode. Intrinsic stellar emission lines
    such as the Pf$\gamma$-line at $3.74\,\mathrm{\mu m}$ (Wallace \&
    Hinkle~\cite{wallace}) do not clearly emerge from the background
    noise. Therefore, the measured spectra were cross-correlated with the
    telluric features of a sky spectrum obtained from 
    NSF/NOAO.\footnote{The sky spectrum is available at \\ {\tt 
        http://www.eso.org/sci/facilities/paranal/instruments/}.}     
    As the shapes of the asymmetrically curved telluric lines are not known,
    the accuracy of the wavelength calibration cannot be better than
    the accuracy provided by a pixel width. 
  \item To perform subsequent flux calibration, the standard star HR\,6070 was
    observed in the immediate 
    vicinity ($\sim$$1$\degr), directly before or after the observation of the 
    target. This ensured that instrumental and
    the atmospheric transmission could be determined, simultaneously. The 
    division by a black-body function
    $B_{\nu}(T_\mathrm{eff})$ with $T_\mathrm{eff}=9980\,\mathrm{K}$ for
    HR\,6070 (Allen~\cite{allenII}) helps to reduce the influence of the
    continuum of the standard star. Considering the template spectrum
    of an F\,0\,V star in L band (Rayner et al.~\cite{rayner}), the
    template spectrum of an A\,0\,V star up to $3.0\,\mathrm{\mu m}$
    (Pickles~\cite{pickles}) and the amplitude of the 
    absorption depth of the water ice feature, potential stellar 
    lines are assumed to be neglected, i.\/e., we do not correct for
    potential absorption lines in the spectrum of 
    HR\,6070. Figure~\ref{figure:transmission}
    presents the normalized transmission of sky and instrument that were
    obtained from the observation of the standard star HR\,6070. Spectral
    shifts of the standard star spectra on a sub-pixel scale allowed an improved
    elimination of the telluric lines in the spectrum of the target. To remove
    atmospheric residuals and 
    data points affected by poor atmospheric transmission, sections of the
    spectra with less than $35$\% transmission are not considered in the
    analysis. These vary from
    exposure to exposure depending on weather conditions. 
  \item By considering the photometric L' band spectra\footnote{The L'
      band covers the NIR range between 
      $3.4\,\mathrm{\mu m}$ and $4.2\,\mathrm{\mu m}$.} and the filter
    curves that were formerly used to perform the photometric measurements of the
    standard star (van Bliek et al.~\cite{bliek}), a photometric flux
    calibration of the observed spectra of the target is 
    possible. We note that the resulting scale
    factor of the photometric 
    calibration can vary by more than $15$\% during a night
    (Przygodda~\cite{przygodda}). 
 \end{enumerate}

 \section{Results}\label{section:results}
\begin{figure*}[!tb]
    \centering
    \resizebox{0.44\textwidth}{!}{\includegraphics{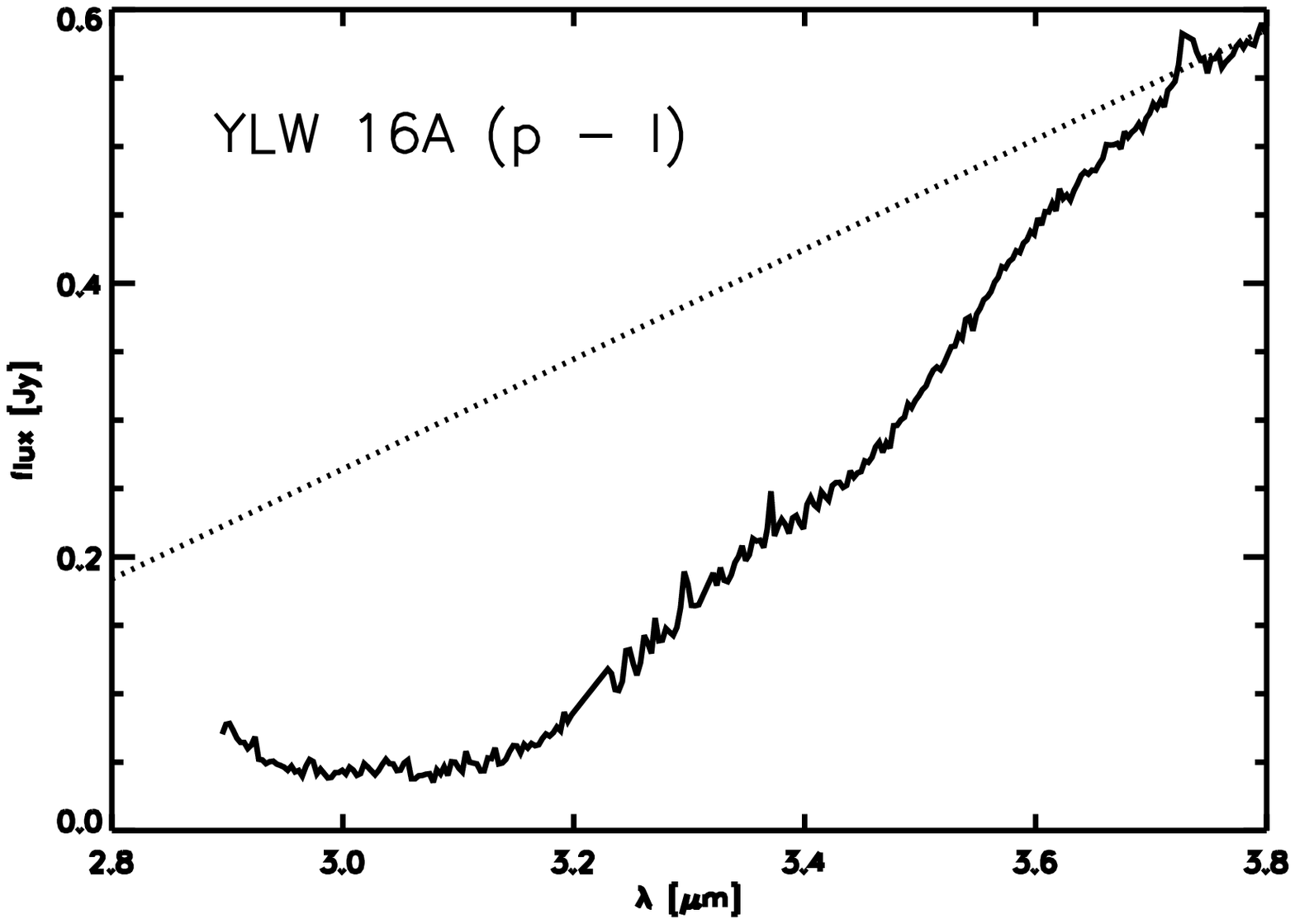}}
    \resizebox{0.44\textwidth}{!}{\includegraphics{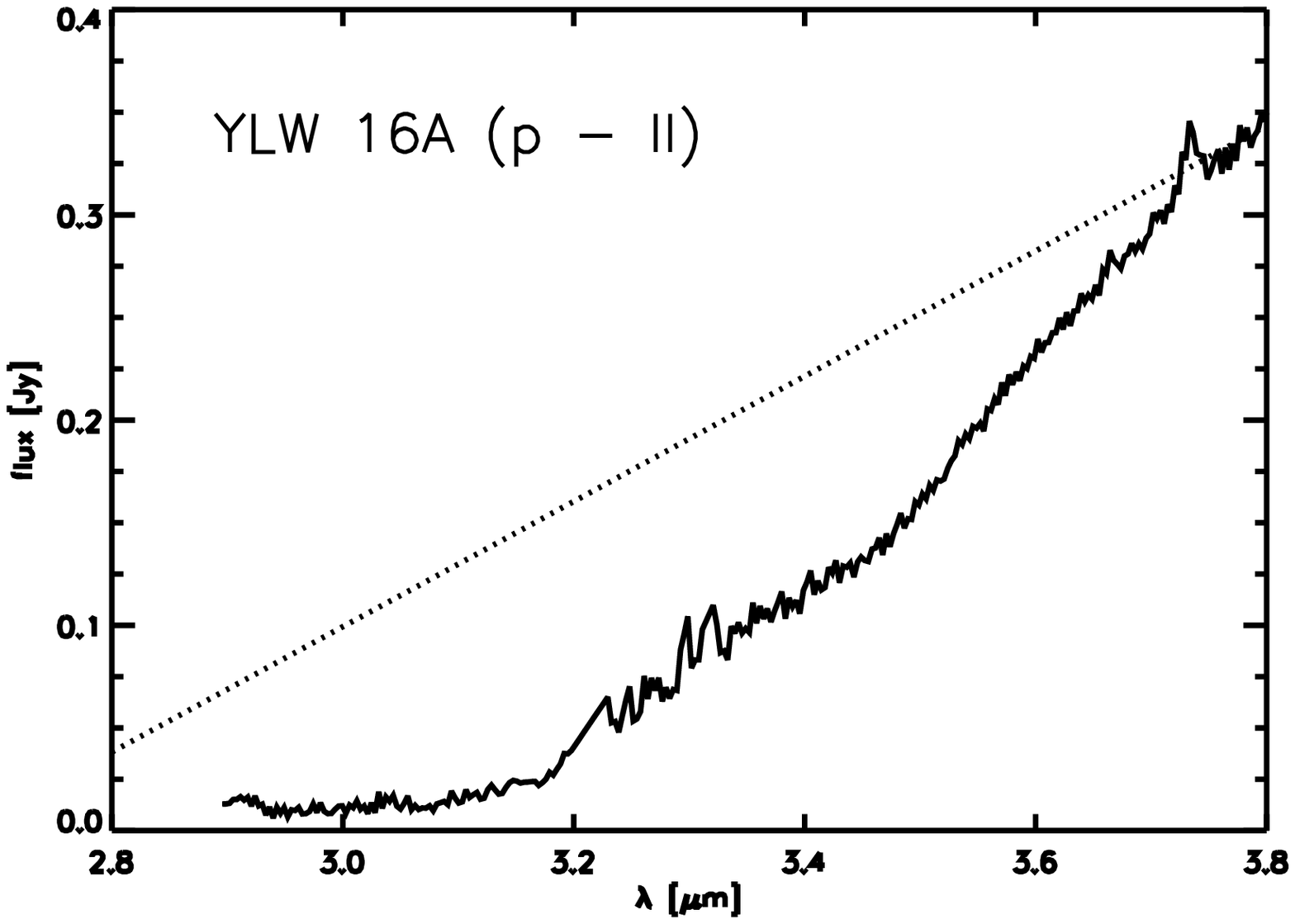}}\newline
    \resizebox{0.44\textwidth}{!}{\includegraphics{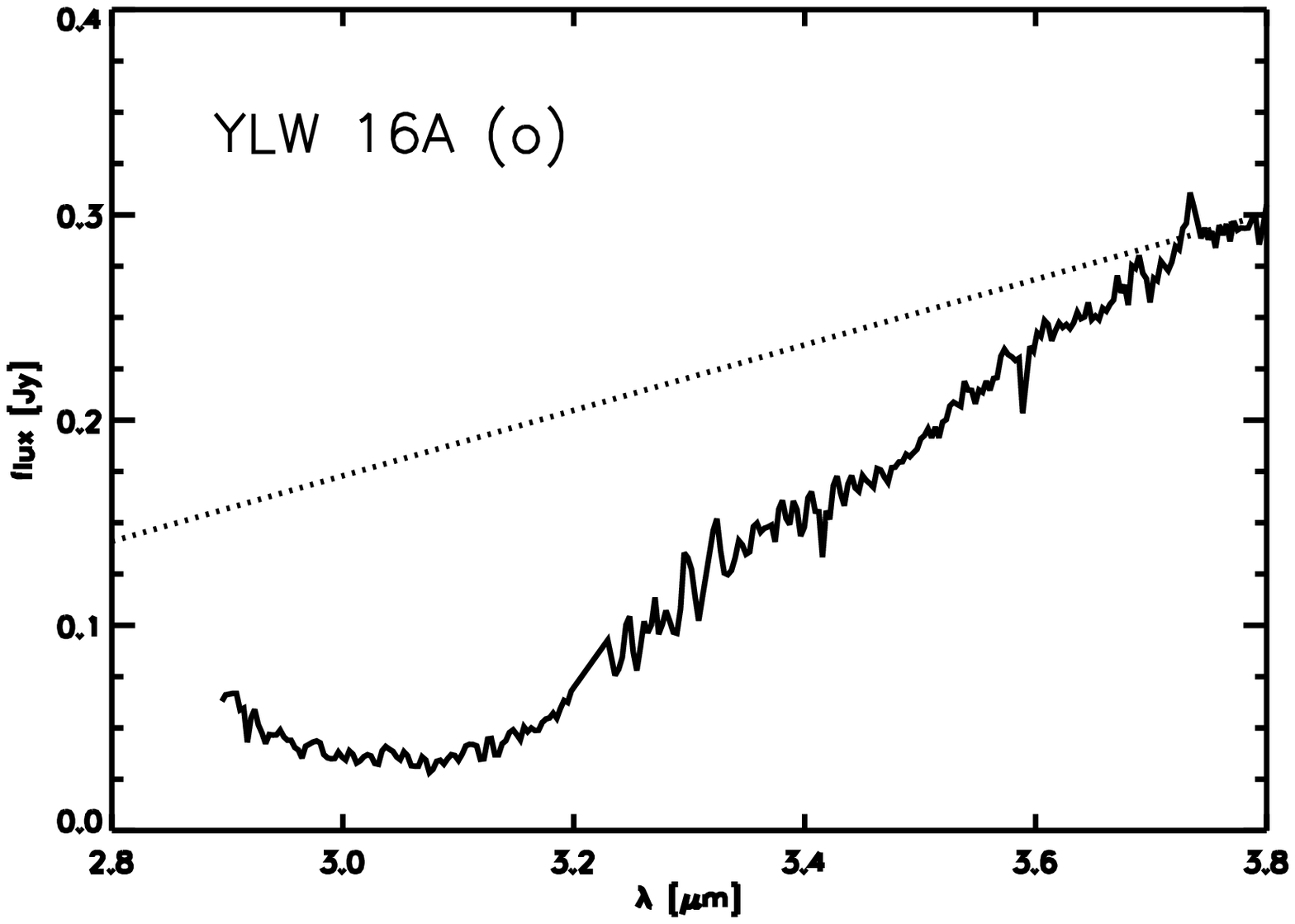}}
    \caption{Extracted and photometrically calibrated L band spectra of the
      source YLW\,16\,A for slit orientations parallel and orthogonal to
      its rotational axis. The counting rates of
      $4 \times FWHM$ of the (spatial) widths of the spectra 
      were determined. The dotted lines represent the ``best-fit'' spectral
      continua that we assume in our modeling approach.}
    \label{figure:result-spek} 
  \end{figure*}

\begin{figure*}[!tb]
    \centering
    \resizebox{0.44\textwidth}{!}{\includegraphics{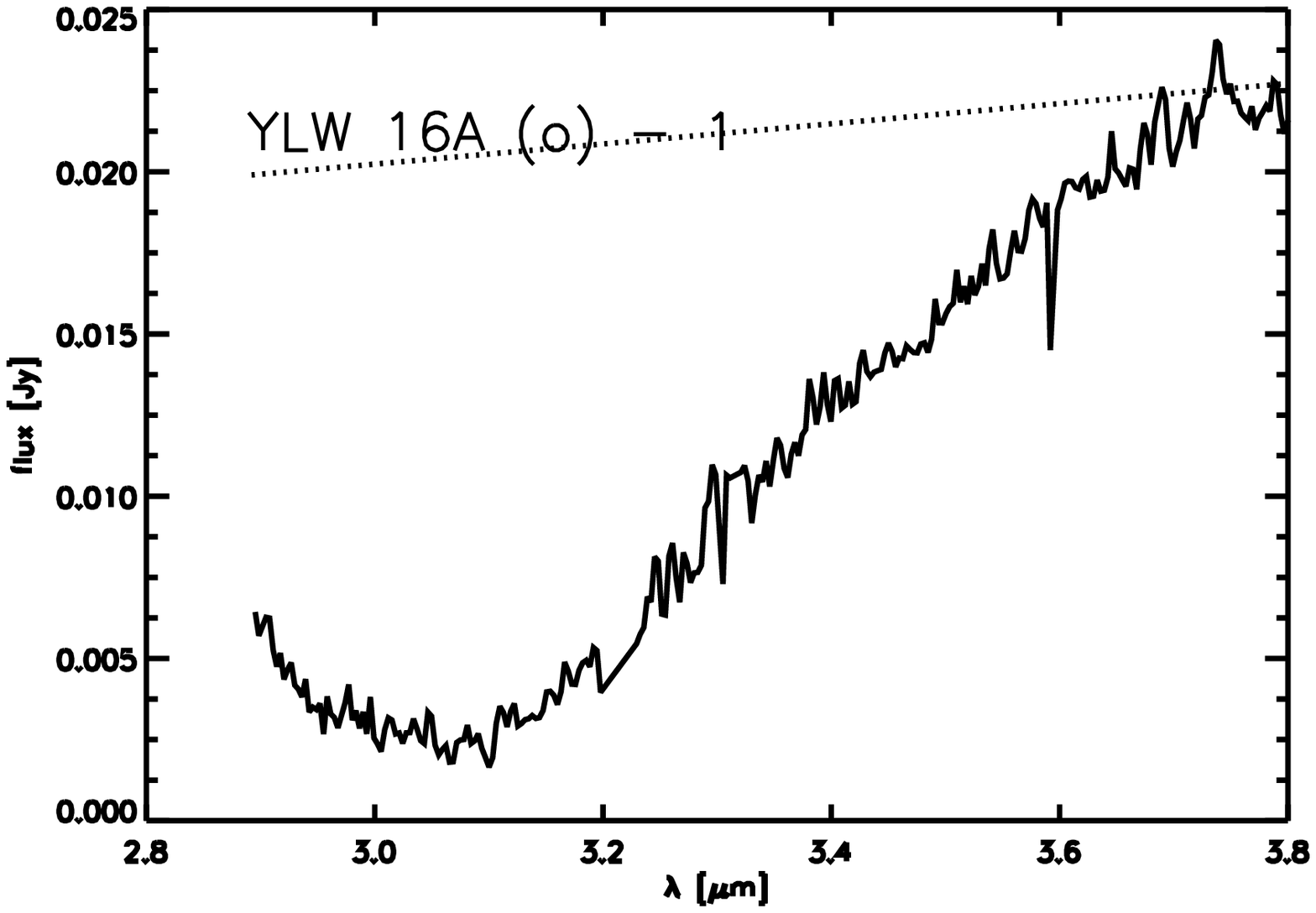}} 
    \resizebox{0.44\textwidth}{!}{\includegraphics{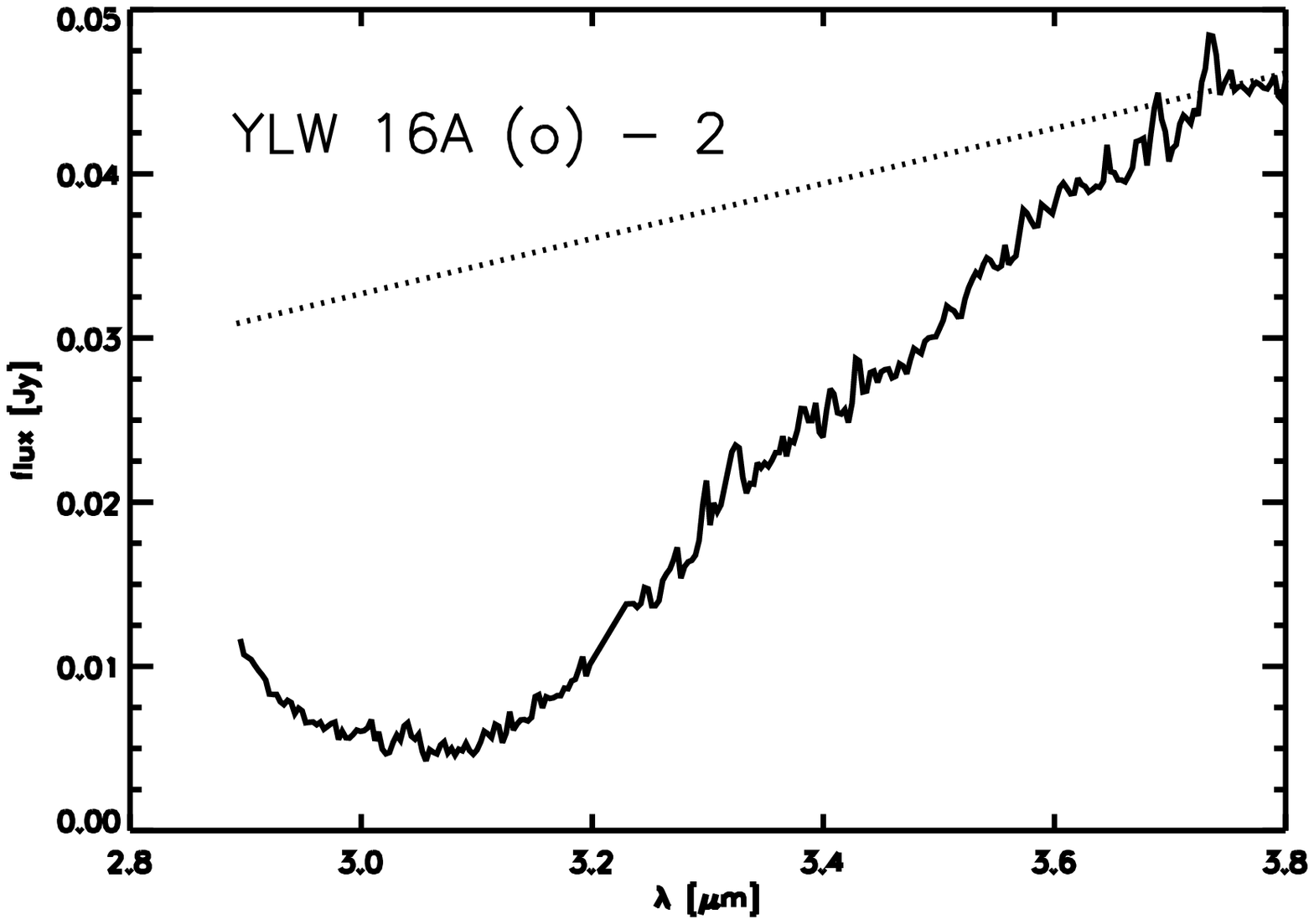}}\newline
    \resizebox{0.44\textwidth}{!}{\includegraphics{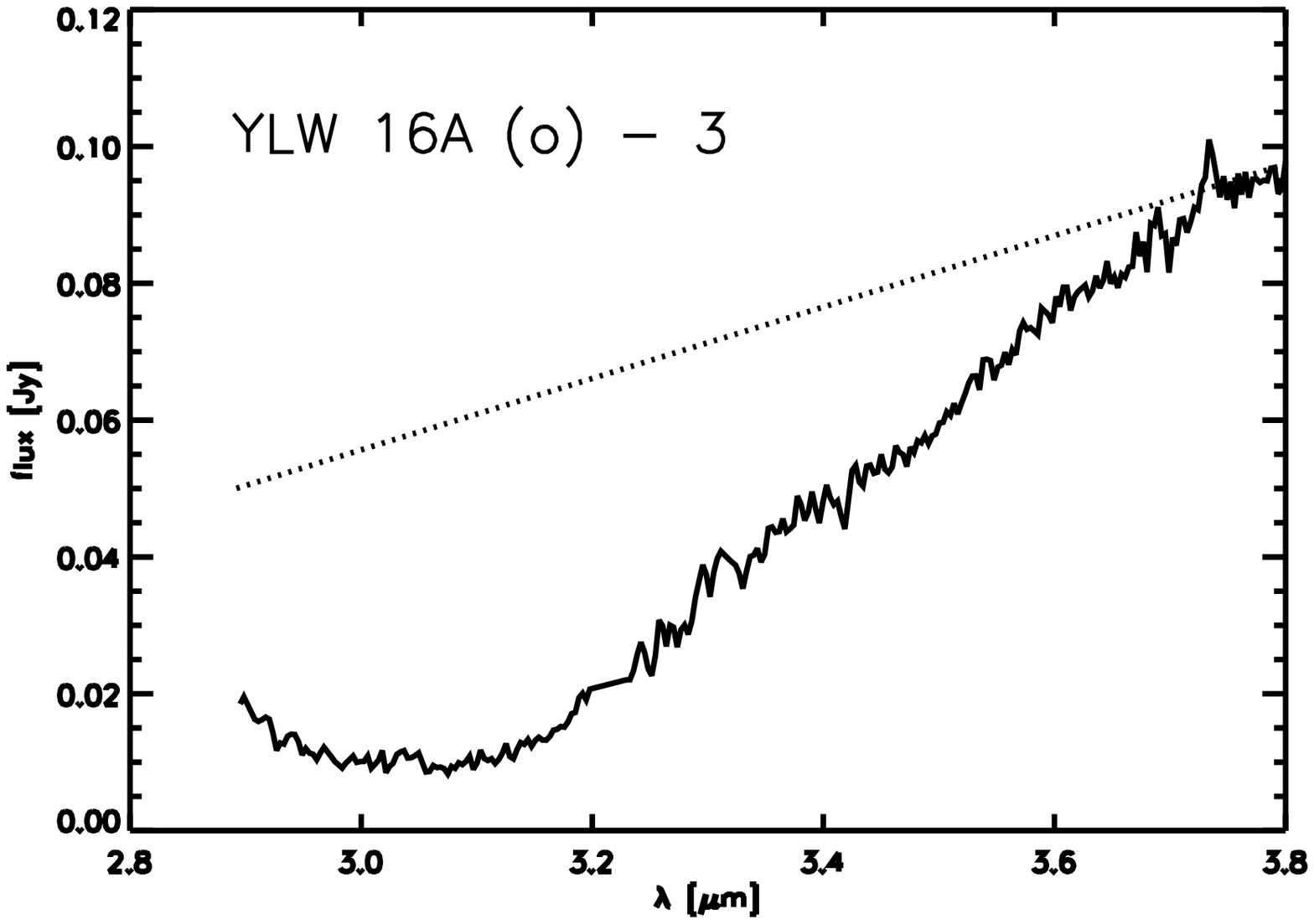}}
    \resizebox{0.44\textwidth}{!}{\includegraphics{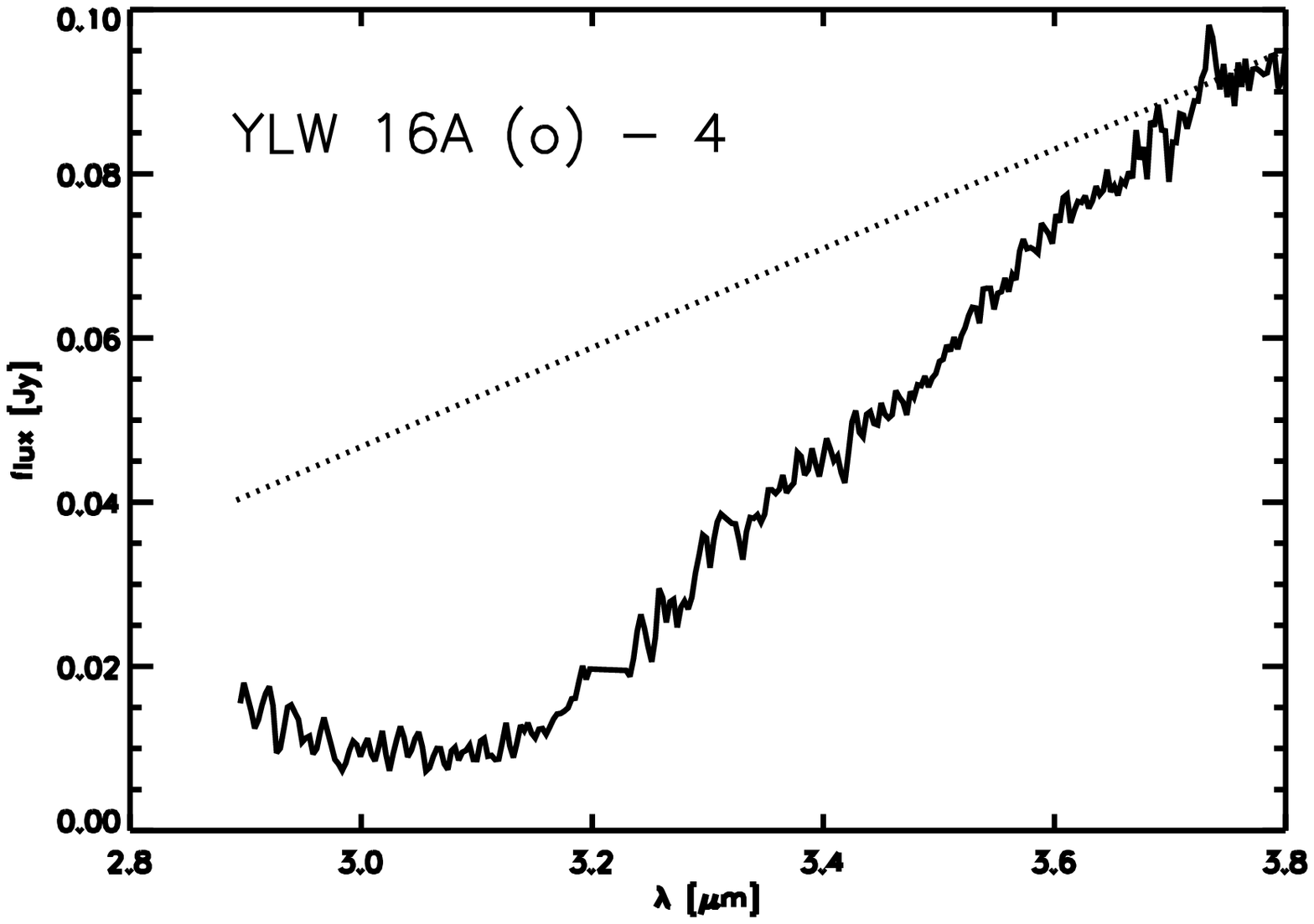}}\newline
    \resizebox{0.44\textwidth}{!}{\includegraphics{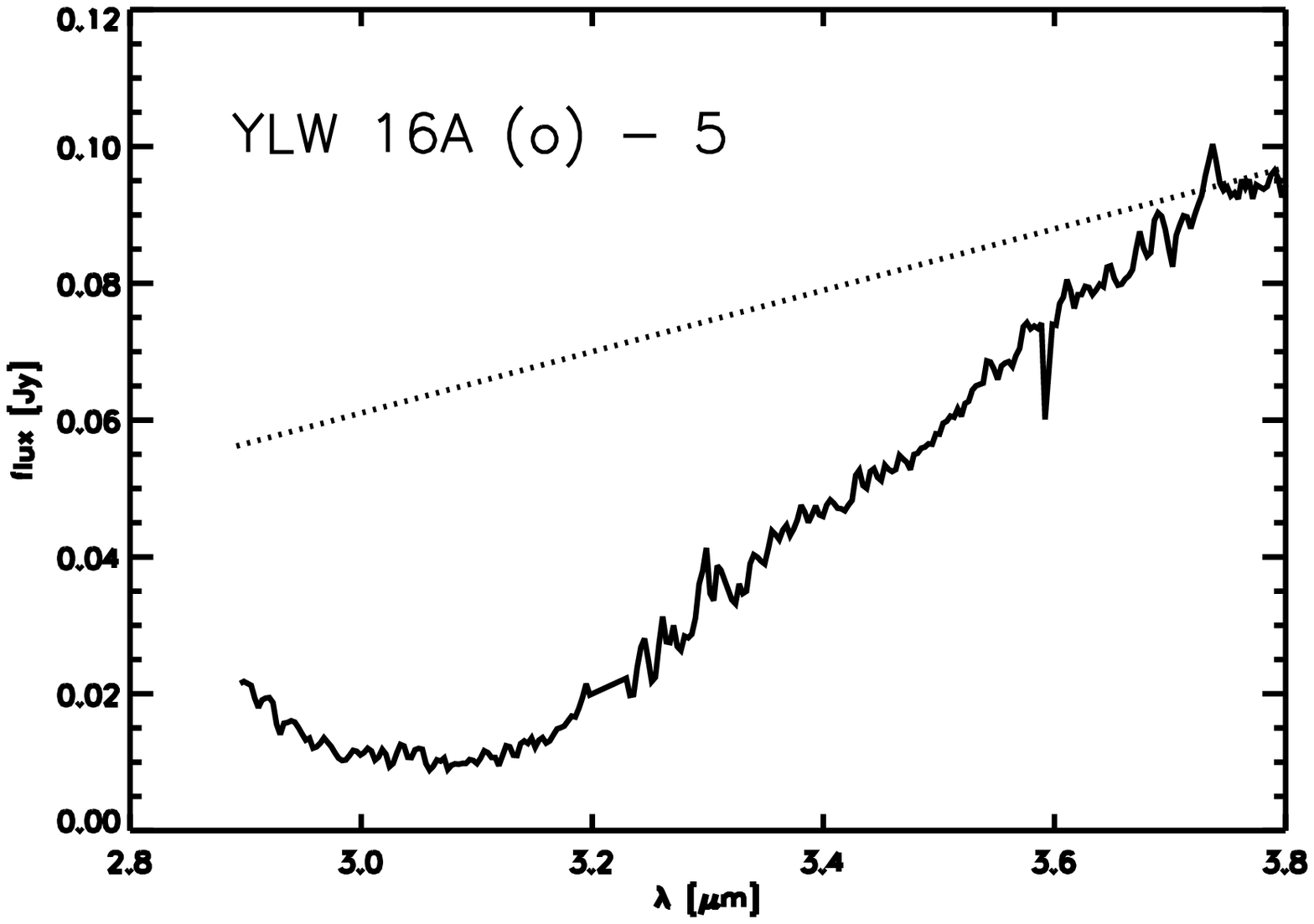}}
    \resizebox{0.44\textwidth}{!}{\includegraphics{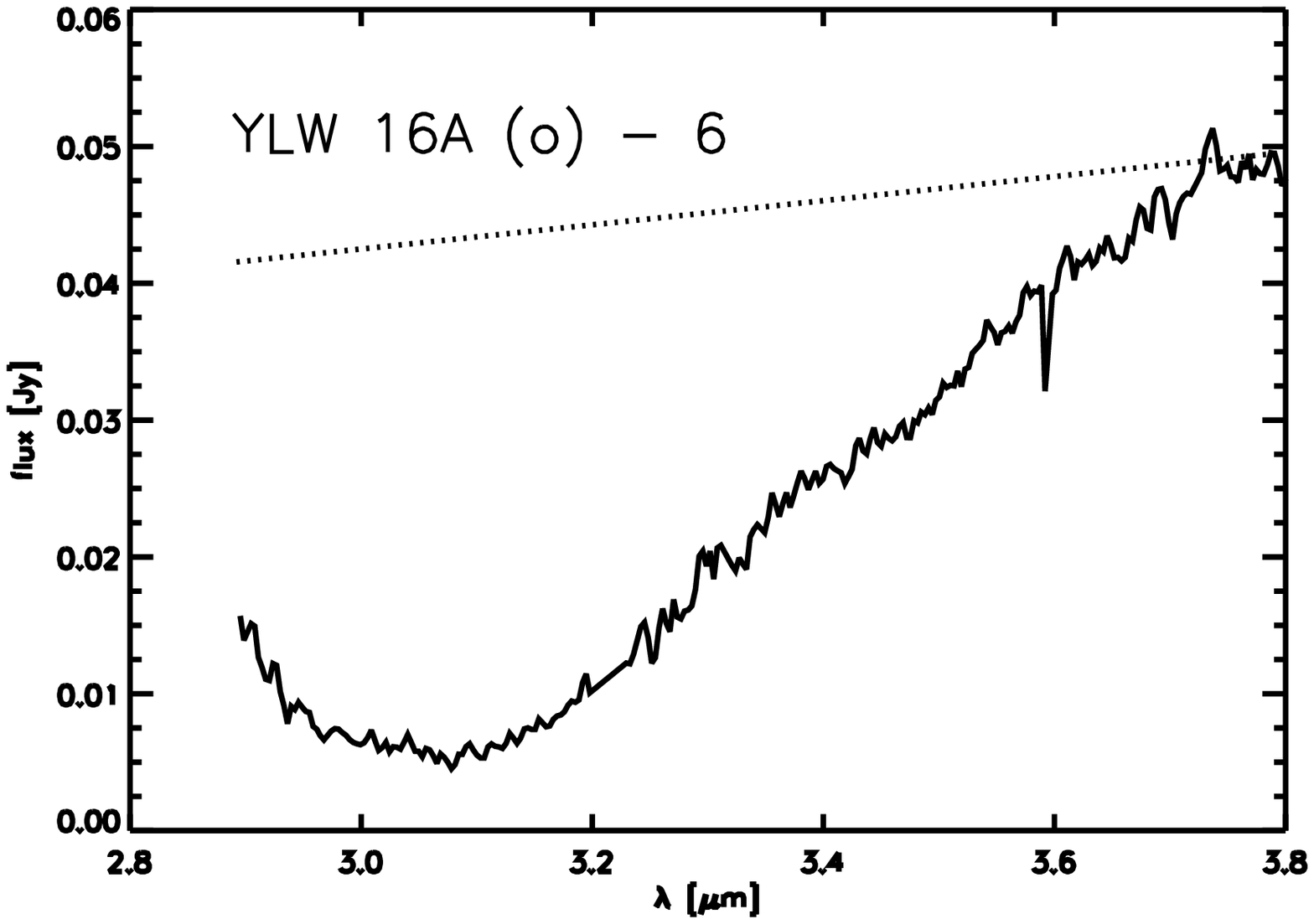}}\newline
    \resizebox{0.44\textwidth}{!}{\includegraphics{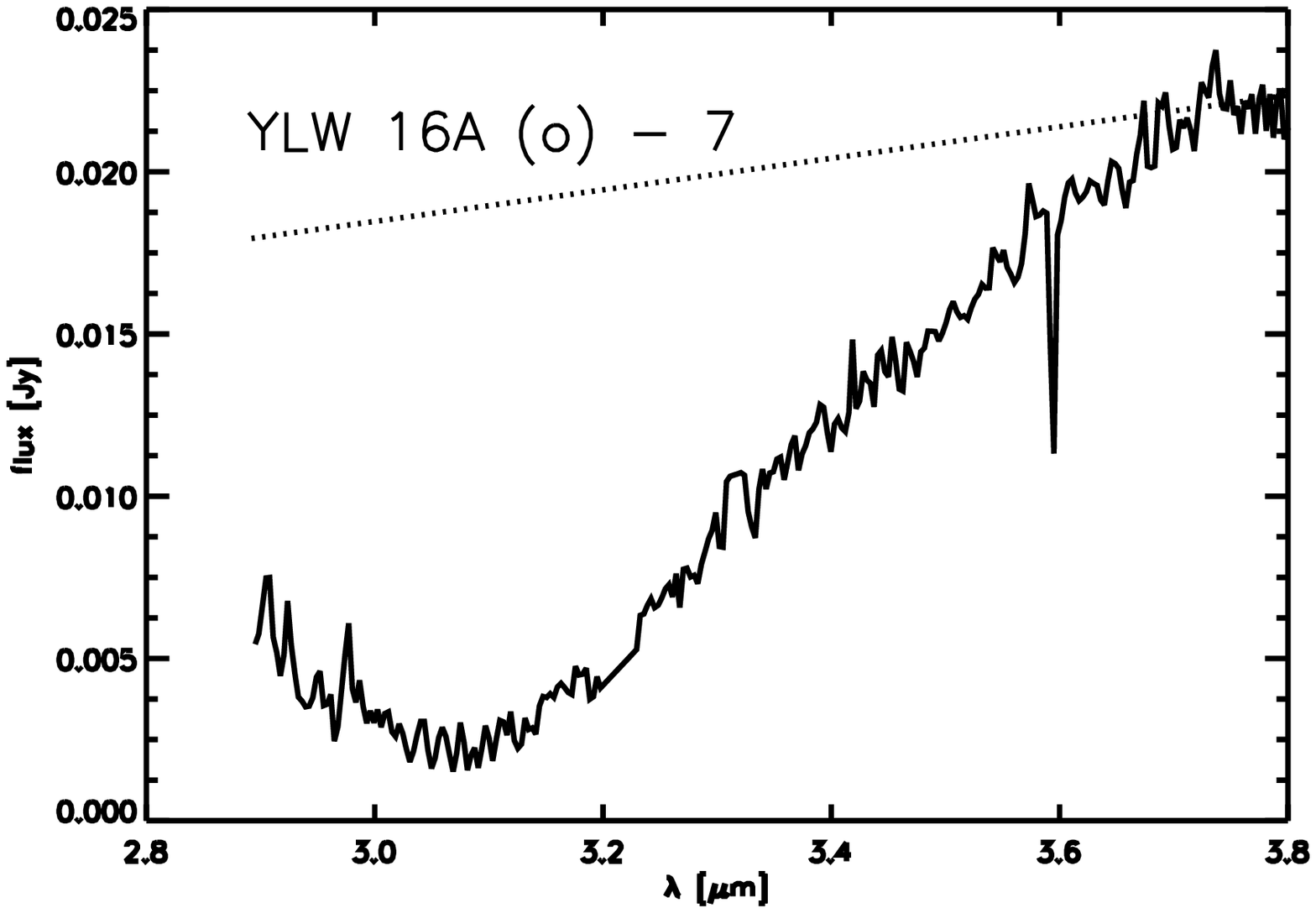}} 
    \caption{Extracted and photometrically calibrated L band spectra of YLW\,16\,A
      derived from seven spatially adjacent stripes cut from the
      acquired 
      spectra. The spectra $1$ to $7$ represent spectra derived from
      source regions of widths $40\,\mathrm{AU}$, $30\,\mathrm{AU}$,
      $20\,\mathrm{AU}$,  $20\,\mathrm{AU}$, $20\,\mathrm{AU}$,
      $30\,\mathrm{AU}$, and $40\,\mathrm{AU}$, respectively.}
    \label{figure:result-spekII} 
  \end{figure*}

  Figure~\ref{figure:result-spek} shows the resulting L band
  spectra of the target for which the counting rates of $4 \times FWHM$ of the
  spectrum were determined
  (Sect.~\ref{section:datareduction}). To provide spatial
  information, Fig.~\ref{figure:result-spekII} shows the spectra
  derived from seven
  spatially adjacent stripes of the total spectra measured in orthogonal
  orientations. Spectra 1 to 7 represent spectra derived from
  source regions with projected widths of 
  $40\,\mathrm{AU}$, $30\,\mathrm{AU}$, $20\,\mathrm{AU}$,  $20\,\mathrm{AU}$,
  $20\,\mathrm{AU}$, $30\,\mathrm{AU}$, and $40\,\mathrm{AU}$, respectively.

  The NICMOS/HST-images of the class-I-object YLW\,16\,A exhibit a bipolar
  envelope above and below its circumstellar disk (Allen et
  al.~\cite{allen}). Both bipolar components yield two separate spectra
  I and II in the (p)-orientation of the slit. 

  The photometric flux can be
  compared to photometric measurements of the Infrared Array Camera (IRAC)
  onboard of the Spitzer 
  satellite. We perform photometry on the pipeline mosaic available at the
  Spitzer 
  archive.\footnote{{\tt http://irsa.ipac.caltech.edu/Missions/spitzer.html}}
  The photometric
  calibration of IRAC is based on an 
  aperture whose radius is set to be $10$\,pixels. The
  flux amounts to $0.79\,\mathrm{Jy}$ at $3.6\,\mathrm{\mu m}$ after
  background subtraction. 
  In contrast to NAOS-CONICA, IRAC does 
  not allow us to spatially resolve the geometrical structure of
  YLW\,16\,A. Therefore, the
  different spatial resolution powers of IRAC and NAOS-CONICA
  explain the differences between the L band fluxes. 
  
  All spectra derived from our NAOS-CONICA observations display the
  broad water-ice band absorption feature with its minimum at $\sim$$3.1\,\mathrm{\mu m}$.
  As the atmospheric transmission is low at wavelengths of
  $<$$3.0\,\mathrm{\mu m}$ (Fig.~\ref{figure:transmission}), the water
  ice feature is truncated at this lower wavelength end of the spectra.  

  \section{Modeling of the water ice band}\label{section:modeling}
  \subsection{Modeling approach}\label{section:approach}
  By definition, the depth of the absorption band is the natural
  logarithm of the ratio of the  
  measured flux $F_\mathrm{obs}$ to the continuum flux $F_\mathrm{cont}$ above
  the absorption band. It depends on the path $L$ through the
  absorbing material, its (mean) extinction  
  $Q_\mathrm{ext}$, and the particle density $\rho$. Whittet~(\cite{whittet})
  derived the optical depth $\tau$ of spherical particles with a mean radius
  $a$ and a mean material 
  density $\eta$ to be 
  \begin{eqnarray} 
    \tau & = & \ln \left( \frac{F_\mathrm{obs}}{F_\mathrm{cont}} \right) = 
    \nonumber \\
    & = & L \rho \frac{Q_\mathrm{ext} \pi a^2 }{4/3 \pi a^3 \eta}=L \rho 
    \times \kappa_\mathrm{ext}=K_{0} \sum_{i} \left( K_{i} \times \kappa_{{\rm
          ext}; i} \right). 
    \label{eq:opticaldepth}
  \end{eqnarray}
  The quantity $\kappa_\mathrm{ext}$
  represents the (total) mass extinction. The extinction results from
  a linear  
  superposition of mass extinctions $\kappa_{\mathrm{ext};i}$ of
  material $i$ that are weighted with the coefficient $K_{i}$ (Martin~\cite{martin};
  \v{S}olc~\cite{solc}). The quantity $K_{0} = L \rho$ is a constant for each
  object.  

  To determine the ice components that effectively contribute to the
  $3\,\mathrm{\mu m}$-absorption band, extinction profiles of ice grains of
  different size and crystallinity are linearly combined to fit the
  ice profile. As a first approximation, scattering effects can
  only be neglected if the condition 
  \begin{eqnarray}
    \label{eq:scattering}
    \hfill{}
    \frac{2 \pi a}{\lambda} < \frac{1}{|m|}
    \hfill{}
  \end{eqnarray}
  is fulfilled (Dartois et al.~\cite{dartois}), 
  where the quantity $|m|$ is the complex refractive index. For $|m| \approx 1.35$
  (Querry et al.~\cite{querry})
  and $\lambda = 3.5\,\mathrm{\mu m}$, we obtain $a < 0.41\,\mathrm{\mu m}$. This
  result justifies the consideration of scattered radiation. Furthermore, the
  bipolar envelope of the object
  YLW\,16\,A appears in scattered light in the corresponding
  K band images obtained with NICMOS/HST. In contrast to the
  absorption profiles, the 
  extinction profiles are asymmetric, i.\/e., more bell-shaped at longer
  wavelengths (Dartois \& d'Hendecourt~\cite{dartois}). 

\begin{figure*}[tb]
    \resizebox{0.33\textwidth}{!}{\includegraphics{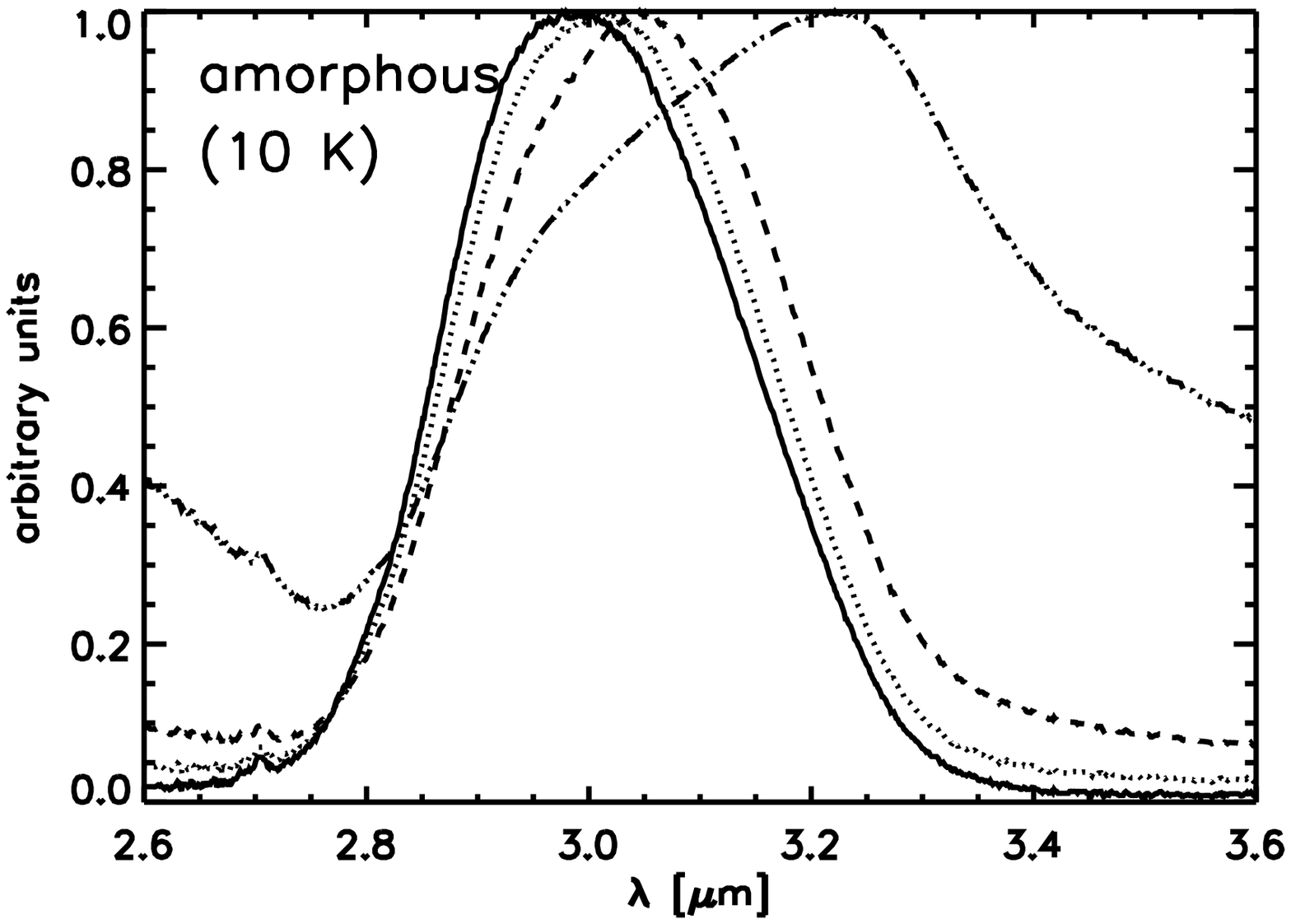}}
    \resizebox{0.33\textwidth}{!}{\includegraphics{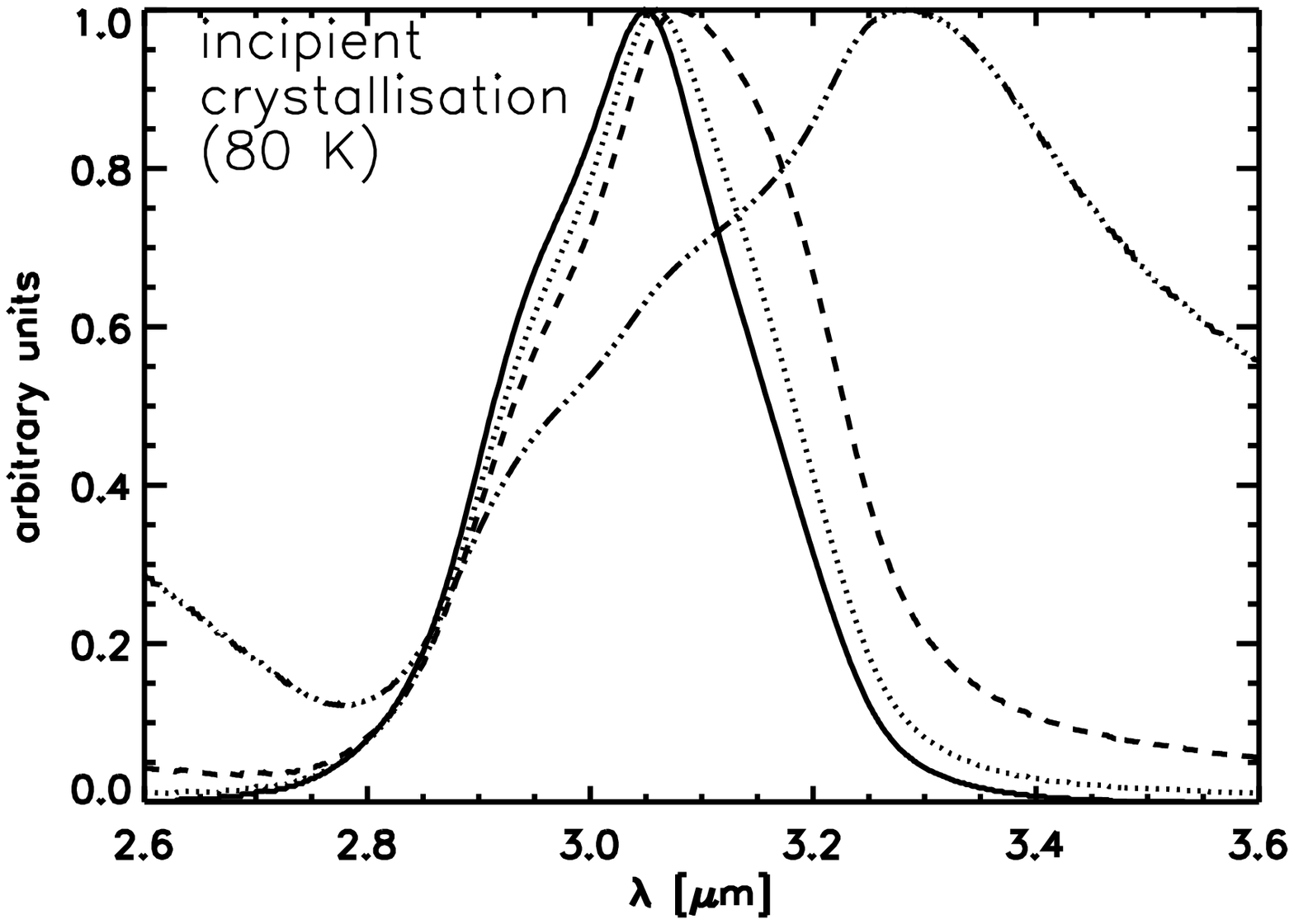}}
    \resizebox{0.33\textwidth}{!}{\includegraphics{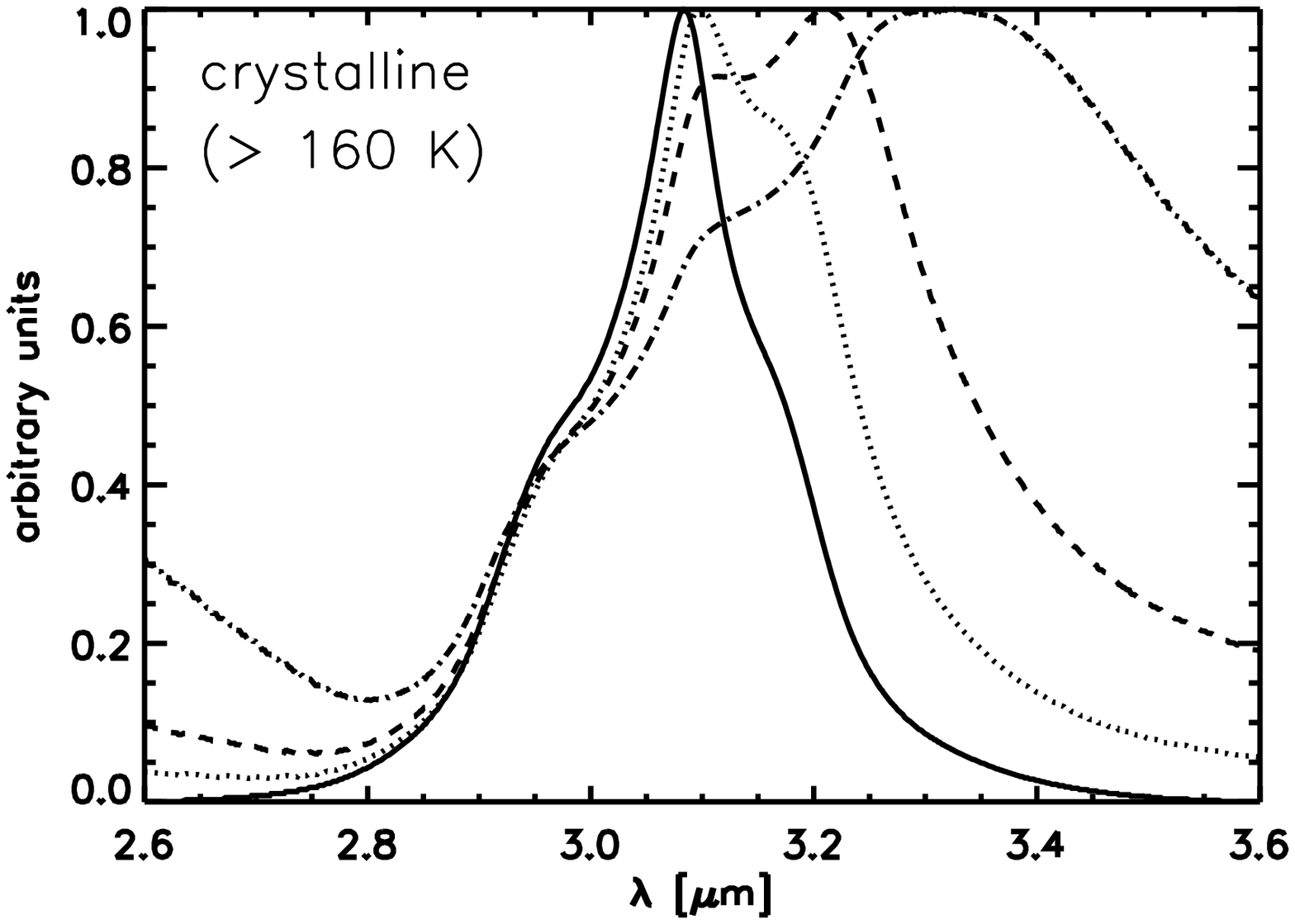}} 
    \caption{\label{figure:extinctionlines} 
      Normalized extinction profiles $\kappa_{\mathrm{ext};i}$ of amorphous ({\it
        left panel}) and crystallized ({\it right panel}) water
      ice for particle radii $a$ of $0.1\,\mathrm{\mu m}$ (solid curve),
      $0.5\,\mathrm{\mu m}$ (dotted line), $0.8\,\mathrm{\mu m}$
      (dashed curve), and $1.5\,\mathrm{\mu m}$ (dot-dashed curve). The profiles
      were normalized by their maximum amplitudes. The {\it
        middle panel} shows the profile of water ice with a temperature of
      $80\,\mathrm{K}$ where crystallization starts. A shift of the maximum
      towards longer wavelengths is caused by grain growth and
      increasing crystallization degree.}
  \end{figure*}

  For our fitting routine, we assume opacities of entirely amorphous
  ($T=10\,\mathrm{K}$) and crystallized ($T>140\,\mathrm{K}$) water ice. The
  ice grain radii $a$ are $0.1\,\mathrm{\mu m}$, $0.3\,\mathrm{\mu
    m}$, $0.5\,\mathrm{\mu m}$, $0.8\,\mathrm{\mu m}$, and
  $1.5\,\mathrm{\mu m}$. In this study, grains with $a \geqq 0.5\,\mathrm{\mu m}$ and
  $a < 0.5\,\mathrm{\mu m}$ are called large and small grains,
  respectively. A restriction to two grain radii (e.g.,
  $0.1\,\mathrm{\mu m}$ and $0.8\,\mathrm{\mu m}$) only increases the deviation
  between model and measurement but does not modify the resulting mass ratios
  $K_{i}$. Figure~\ref{figure:extinctionlines} shows the mass
  extinction $\kappa_\mathrm{ext}(\lambda)$ of 
  amorphous and crystallized water ice used in our fitting approach
  (Eq.~\ref{eq:opticaldepth}). Assuming spherical, compact ice 
  particles, the mass 
  extinction curves are derived using the program MIEX (Wolf \& 
  Voshchinnikov~\cite{wolfII}), which is based on Mie scattering. This program 
  calculates the extinction profiles by considering the   
  grain size and refractive indices (Schmitt et al.~\cite{schmitt};
  Dartois \& d'Hendecourt~\cite{dartois}). 

  An unambigous determination of the continuum flux $F_\mathrm{cont}$ above the
  absorption band is difficult. Owing to reduced atmospheric transmission
  (Fig.~\ref{figure:transmission}), parts of the spectra at
  short-wavelengths are missing. The determination of the continuum at 
  longer wavelengths ($>$$3.6\,\mathrm{\mu m}$) is also difficult because some
  telluric lines could not be removed entirely during data reduction, producing
  noisy spectra. In previous
  investigations of the water ice band (Thi et al.~\cite{thi};
  Dartois et al.~\cite{dartoisII}; Boogert et al.~\cite{boogertII}), a Planck
  function $B_{\nu}(T)$ or a spline function was 
  fitted to the continuum and -- if available -- to photometric data 
  points in adjacent spectral bands to determine the continuum. In
  addition, studies were performed where 
  the spectral energy distribution of the entire
  infrared wavelength range was fitted using disk models (Pontoppidan
  et al.~\cite{pontoppidanII}). Assuming a spatial resolution of
  $0.12${\arcsec} ($\sim$$20\,\mathrm{AU}$) for our NAOS-CONICA observations,
  which is smaller than the true spatial 
  extension of the object, only a fraction of the total flux of the
  disk is
  detected. Therefore, when comparing with IRAS observations (Sect.~\ref{section:results}),
  photometric measurements do not have to represent the absolute
  level of the L band spectra measured with NAOS-CONICA.  

  Apart from either the Planck function $B_{\nu}(T)$ or a spline function, we
  use a straight line to determine the 
  continuum. The straight line is rotated successively around
  different spectral points resulting in different optical depths
  (Eq.~\ref{eq:opticaldepth}). The spectral points about which the line is rotated 
  are the sampling points at $3.6\,\mathrm{\mu m}$ and  $3.8\,\mathrm{\mu m}$,
  respectively. The latter point is used because the
  water ice band and other  
  compounds (Sect.~\ref{section:introduction}) do not contribute to the
  spectrum around $3.8\,\mathrm{\mu m}$. The angular
  step used to represent the rotation 
  depends on the depth of the absorption profile. Smaller angular steps were  
  not found to improve the fit to the optical depth $\tau$. We found that the
  resulting optical depth  
  $\tau$ is strongly affected by the gradient 
  of the continuum's line. We finally look for the
  line at a certain rotational center whose corresponding linear combination
  of extinction profiles 
  $\kappa_{\mathrm{ext}; i}$ reproduce the optical depth $\tau$ the
  most successfully, i.\/e.,
  with the smallest reduced chi-square $\chi^2$ (Eq.~\ref{eq:chi}). We use the
  fitting routine presented 
  in Schegerer et al.~(\cite{schegerer}). Representative continuum lines are
  drawn in the spectra shown in Figures~\ref{figure:result-spek} and~\ref{figure:result-spekII}.   

  \begin{figure*}[!tb]
    \resizebox{0.48\textwidth}{!}{\includegraphics{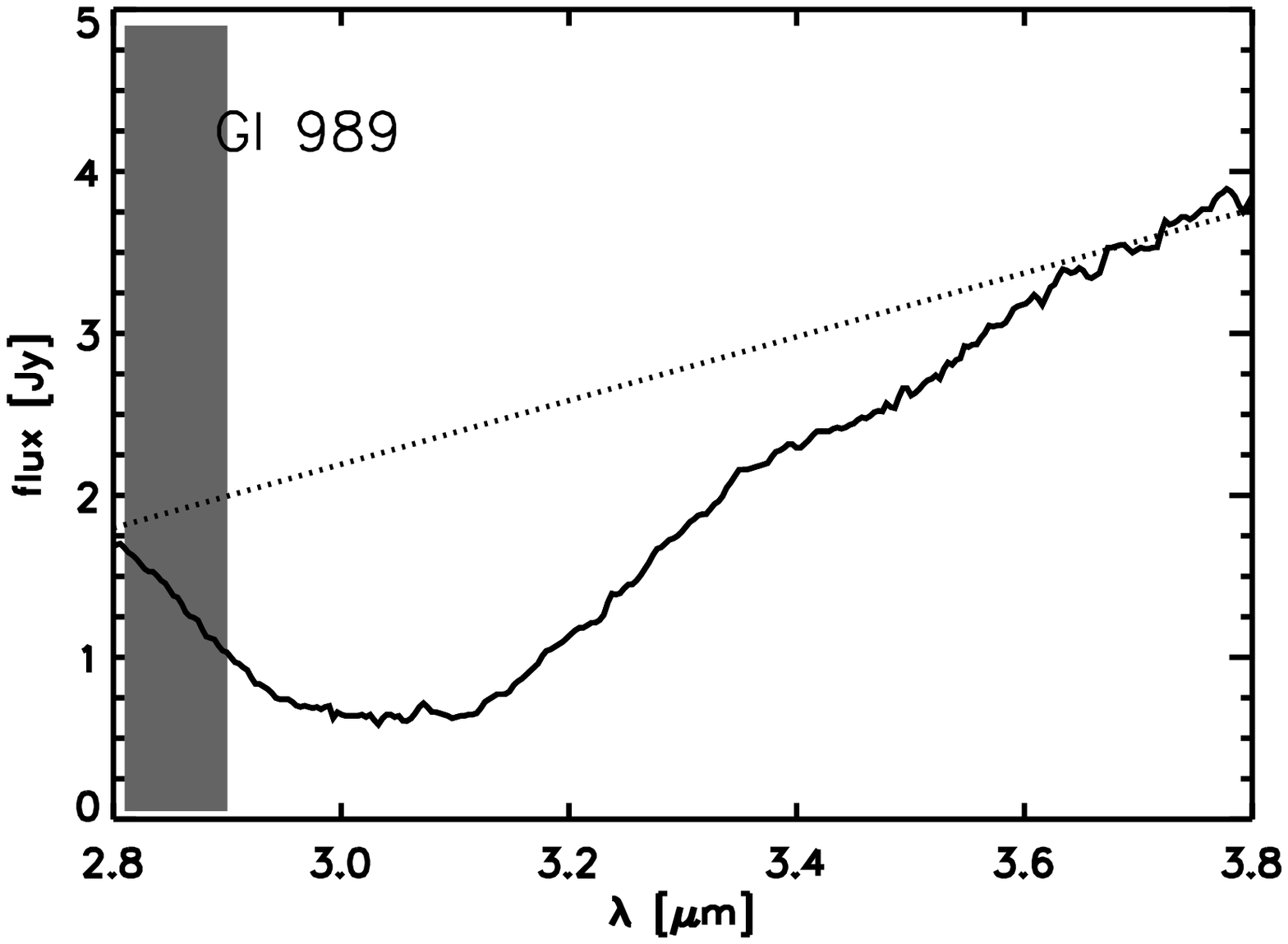}}
    \resizebox{0.48\textwidth}{!}{\includegraphics{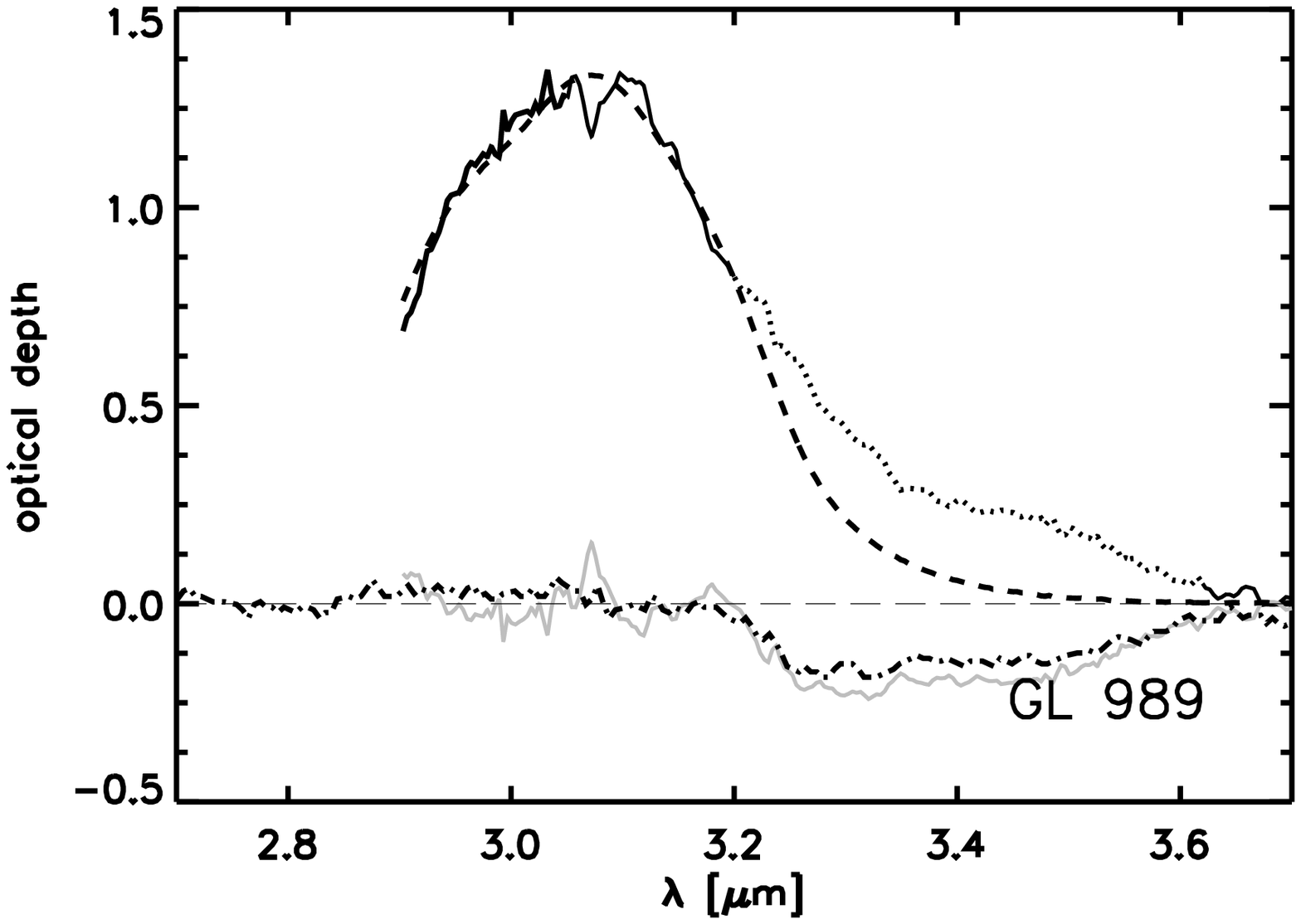}} 
    \caption{\label{figure:gl989} {\it Left panel:} L band spectrum of the object
      GL\,989 measured with the ISO satellite. Because of the lower
      transmission of the sky, the spectral interval at short wavelengths that
      is indicated by gray color is difficult to observe using
      NAOS-CONICA or any 
      other instrument on the ground because of the lower transmission of the sky. The
      dotted line represents the likely 
      linear continuum. \newline
      {\it Right:} Derived optical depth $\tau$ of the water ice band of
      GL\,989 truncated at $2.9\,\mathrm{\mu m}$. The absorption between 
      $3.2\,\mathrm{\mu m}$ and $3.6\,\mathrm{\mu m}$ that is not only caused by
      water ice, is not considered by the modeling
      (dotted curve). The linear combination of ice profiles
      $\kappa_{\mathrm{ext}; i}$ that represents the
      band with the smallest deviation from the measurement, i.\/e., smallest
      $\chi^2$, is plotted by 
      the dashed curve. The thin gray curve represents the deviaton of the model from
      measured data. For 
      comparison, the deviation for the modeling of
      Dartois 
      \& d'Hendecourt~(\cite{dartois}) is also plotted (dot-dashed curve). The
      long-dashed curve represents the zero level.}
  \end{figure*}
  The determinations of both the continuum and the ice components
  that effectively contribute to the absorption profile were tested by
  analyzing the spectrum of the YSO \object{GL\,989}. Its L band
  spectral distribution was previously
  investigated by Dartois \& d'Hendecourt~(\cite{dartois}). In
  contrast to our measurements, the source was formerly observed with the ISO satellite
  (Infrared Space Observatory) where the spectrum at short wavelengths could
  be retained. But when this spectrum is cut at short wavelengths up to a wavelength of
  $2.9\,\mathrm{\mu m}$, the resulting
  spectrum corresponds to the NAOS-CONICA spectra that we acquired
  during different weather conditions. 
  The results of the analysis of the
  $3\,\mathrm{\mu m}$ spectra of GL\,989 with and without the short-wavelength
  interval are finally compared. Potential differences 
  can then be considered
  as an approximation of the systematic error caused by the missing
  interval of the NAOS-CONICA spectra up to $2.9\,\mathrm{\mu m}$. We
  note that 
  Dartois \& d'Hendecourt~(\cite{dartois}) used a Planck function to 
  determine the underlying continuum of GL\,989 assuming a temperature
  of 
  $T=500\,\mathrm{K}$. Furthermore, they considered extinction profiles of ice
  grains with a continous size distribution. 

  Figure~\ref{figure:gl989} shows the results. Independent of
  the rotational center used, the fitting
  results obtained from the spectra that are truncated at
  $2.9\,\mathrm{\mu m}$ differ by $<$$15\%$ from the results derived from
  spectra for which the lower end of the L 
  band is considered. The mass contribution of crystallized ice can be
  neglected in all spectra. The mass contribution
  of amorphous, large grains depends strongly on the
  rotational center used.  Both the non-truncated and truncated spectra exhibit
  only marginal evidence of large water ice grains ($<$$1\%$) when a rotational
  center at $3.8\,\mathrm{\mu m}$ is used for the linear continuum. In contrast, the
  contribution of amorphous, large grains with a size of
  $0.5\,\mathrm{\mu m}$ is 
  $>$$85 \%$ when a rotational center at $3.6\,\mathrm{\mu m}$ is used.  
  The fitting results deviate by between $\chi^2=0.0015$ and $\chi^2=0.003$ from the
  measurement (Eq.~\ref{eq:chi}). The deviation from the measurement is larger
  for fits with the rotational center at $3.6\,\mathrm{\mu m}$.

  \subsection{Modeling results}\label{section:modelresults}
  Table~\ref{table:result-water} and 
  Figures~\ref{figure:result-water1} and~\ref{figure:result-water2} 
  summarize our modeling results. The quality of each fit is characterized
  by $\chi^{2}$ (e.g., Press et al.~\cite{press}) with 
  \begin{eqnarray} 
    \chi^2=\frac{1}{N_{\rm \lambda}-f} \sum^{\rm N_{\rm \lambda}}_{\rm i=1}
    ({\cal F}^{\rm model}_{\rm \nu}(\lambda_{\rm i})-{\cal F}^{\rm data}_{\rm \nu}
      (\lambda_{\rm i}))^2. 
    \label{eq:chi} 
  \end{eqnarray}
  The data with the largest $\chi^2$ have the lowest 
  signal-to-noise ratios. In Eq.~\ref{eq:chi}, $N_{\rm \lambda}$
  is the number of data points, $f$ the degree of freedom, and ${\cal F}^{\rm
    model}_{\rm \nu}(\lambda)$ and ${\cal F}^{\rm
    data}_{\rm \nu}(\lambda)$ are the modeled and measured fluxes,
  respectively.   

  The features of additional compounds (e.\/g., of ammonia
  hydrate) is superimposed on fractions 
  of the ice absorption band at a wavelength of $\sim$$3.4\,\mathrm{\mu
    m}$. The spectral region between $3.25\,\mathrm{\mu m}$ and
  $3.65\,\mathrm{\mu m}$ is not considered by our modeling approach
  (dotted curves).     
  
  In contrast to our results for 
  spectra with the parallel slit orientation of the slit, a larger amount of
  crystalline grains ($28\%$) contributes to the  
  spectrum of the orthogonal slit orientation ($3^{\mathrm{rd}}$ line
  in the Table~\ref{table:result-water}). We point out
  that this spectrum can be fitted by a larger amount of large,
  i.\/e., $0.5\,\mathrm{\mu m}$-sized amorphous grains, if crystallized grains
  are excluded from the fit. As the fit then worsens by $35\%$, this finding
  can be considered as evidence of crystalline dust grains, as the
  maximum of the extinction profile of larger 
  amorphous grains shifts to longer wavelengths similar to the shift of the
  extinction profile of crystalline grains.  
 
  In Fig.~\ref{figure:result-water2}, the spectra $1$ to $7$ correspond to 
  different regions within the source that could be spatially resolved by our
  observations with NAOS-CONICA for the orthogonal slit orientation
  (Sect.~\ref{section:datareduction}). Because 
  of the too noisy data, we do not consider the optical depth derived from the
  observation where the spectroscopic slit was orientated parallel to the
  rotational axis. As we obtain tighter fits to the derived optical
  depth, i.\/e., smaller $\chi^2$, we focus only on the results
  obtained from fits where the continuum's line is rotated around
  $3.8\,\mathrm{\mu m}$ but without forgetting the corresponding results
  obtained for a rotational center at $3.6\,\mathrm{\mu m}$. 

  The
  spectra $3$, $4$, and $5$ that were derived from the most central region of
  the source around the photocenter appear to contain the
  largest contribution of non-evolved, i.\/e., amorphous, small grains, while
  the spectra from the outer source regions exhibit a greater
  contribution from crystallized water ice. These crystalline, peripheral grains
  also contributes to the spectrum where the counting rates of $4 \times FWHM$
  were summarized ($3^{\mathrm{rd}}$ line in the Table~\ref{table:result-water}). 

  Considering Fig.~\ref{figure:result-spekII} and the flux
  scales used, it is conspicuous that different gradients are found
  for fitting the continua of spectra that originate in different
  source regions. However, if steeper gradients were used to fit continua of
  spectra originating in outer source regions, even a larger contribution
  of crystallized grains would be found. On the other hand, a shallower 
  gradient for the continua  
  derived in the more central source regions would favor a larger contribution
  of small amorphous grains. In any case, the fit is poorer when
  gradients other than those are used.  

  The fit results obtained for spectra in which a rotational center at
  $3.6\,\mathrm{\mu m}$ is used, have qualitatively the same increasing
  contribution of crystalline grains towards the outer regions
  ($m_\mathrm{c}=57\%$, $0.25\%$, $98\%$ at the regions 1, 3, and 7,
  respectively). However, as already mentioned in 
  Sect.~\ref{section:approach}, the contribution of amorphous, large (i.\/e.,
  $0.5\,\mathrm{\mu m}$-sized) grains has increased at the expense of
  amorphous, small grains, in particular in the more
  central regions 4, and 5, where on average $m_\mathrm{a;l}\approx 88\%$.

  The column density $N_\mathrm{A}$ of the absorbing ice material can be
  determined by an integrating of the extinction profiles of amorphous
  water ice that are fitted to the optical depth $\tau(\lambda)$. According
  to Whittet~(\cite{whittet}), we obtain 
  \begin{eqnarray} 
    N_\mathrm{A}(\mathrm{H}_{\mathrm{2}}\mathrm{O [ice]}) =
    \int^{3.8\,\mathrm{\mu m}}_{2.8\,\mathrm{\mu m}}  
    \frac{\tau(\lambda)}{A} \frac{d\lambda}{\lambda^2}. 
    \label{eq:column-density}
  \end{eqnarray}
  The integral extinction cross-section $A$ of amorphous water ice at a
  temperature of 
  $T=10\,\mathrm{K}$ is $A = 2.0 \times 10^{-18}\,\mathrm{m}$/molecule (Hagen \&
  Tielens~\cite{hagenII}). As the O-H stretching mode
  is very strong and dominates over other species (Tielens et
  al.~\cite{tielensII}), the $3\,\mathrm{\mu m}$ band provides
  an accurate water ice column density.    
  Teixeira \& Emerson~(\cite{teixeira})
  found the following empirical relation between the column density
  $N_\mathrm{A}(\mathrm{H}_{\mathrm{2}}\mathrm{O [ice]})$ and the visual
  extinction $A_\mathrm{V}$:  
  \begin{eqnarray} 
     N_\mathrm{A}(\mathrm{H}_{\mathrm{2}}\mathrm{O [ice]}) = (A_\mathrm{V} -
     2.1) \times 10^{21} \mathrm{m^{-2}}. 
    \label{eq:column-density-av}
  \end{eqnarray}
  When considering spectra obtained from different source regions, the
  quantities $N_\mathrm{A}$ and $A_\mathrm{V}$ are higher  
  by $18\%$ and $16\%$, respectively, in spectra obtained from
  outer than for those from the inner regions. 
 
  \begin{table*}[!t]
    \centering
    \caption{Relative mass contribution of amorphous-small
      ($m_\mathrm{a;s}$: $0.1\,\mathrm{\mu m}$ and $0.3\,\mathrm{\mu m}$),
      amorphous-large ($m_\mathrm{a;l}$: $0.5\,\mathrm{\mu m}$, $0.8\,\mathrm{\mu m}$,
      and $1.5\,\mathrm{\mu 
      m}$) and crystallized ($m_\mathrm{c}$: all particle radii) water
      ice grains derived from our modeling approach by using a
      rotational center at $3.8\,\mathrm{\mu m}$. The position
      angles ($PA$) of the spectroscopic slit, 
      measured from North to East, are additionally listed. The
      regions $1$ to $7$ represent spectra that originate in source regions with widths of
      $40\,\mathrm{AU}$, $30\,\mathrm{AU}$, $20\,\mathrm{AU}$,
      $20\,\mathrm{AU}$, $20\,\mathrm{AU}$, $30\,\mathrm{AU}$, and
      $40\,\mathrm{AU}$. The spectra $3-5$ originate closer to the
      photocenter than the spectra $1$, $2$, $6$, and$7$. In contrast,
      for the first three exposures listed, the 
      counting rates within $4 \times FWHM$ of the spatial dimension 
      are summarized. The quantities
      $N_\mathrm{A}$ and $A_\mathrm{V}$ represent the column density of water
      ice particles and the visual extinction derived along the line of
      sight (Eqs.~\ref{eq:column-density}, and~\ref{eq:column-density-av}). 
    }
    \label{table:result-water}
    \centering
    {\small \begin{tabular}{lccccccc} 
        \hline\hline
        exposure & $PA$ [$^{\circ}$] & $N_\mathrm{A}$ [$10^{22}$ m$^{-2}$] &
        $A_\mathrm{V}$ [mag] & 
        $m_\mathrm{a;s}$ [\%] & $m_\mathrm{a;l}$ [\%] & $m_\mathrm{c}$ [\%] &
        $\chi$ \\ \hline 
        YLW\,16\,A (p - I) & $178$ & $1.7$ & $19$ & $97$  & -- & $3$ &
        $0.0074$ \\ 
        YLW\,16\,A (p - II) & $178$ & $2.5$ & $27$ & $>$$99$ & $<$$1$ & $<$$1$ &
        $0.020$ \\ 
        YLW\,16\,A (o) & $88$ & $2.2$ & $24$ & $70$ & $1.7$ & $28$ & $0.078$
        \medskip\\
        YLW\,16\,A (o) - 1 & $88$ & $2.0$ & $22$ & $32$ & $1.1$ & $67$ & $0.011$ \\ 
        YLW\,16\,A (o) - 2 & $88$ & $1.9$ & $20$ & $31$ & $1.4$ & $68$ &
        $0.0046$ \\  
        YLW\,16\,A (o) - 3 & $88$ & $1.7$ & $19$ & $93$ & $1.0$ & $5.7$ &
        $0.0042$ \\  
        YLW\,16\,A (o) - 4 & $88$ & $1.8$ & $20$ & $78$ & $10$ & $12$ &
        $0.010$ \\  
        YLW\,16\,A (o) - 5 & $88$ & $1.7$ & $19$ & $87$ & $4.9$ & $7.8$ &
        $0.0079$ \\  
        YLW\,16\,A (o) - 6 & $88$ & $2.0$ & $22$ & $84$ & $7.6$ & $8.2$ & $0.0042$ \\ 
        YLW\,16\,A (o) - 7 & $88$ & $2.0$ & $22$ & $4.5$ & $3.2$ & $92$ &
        $0.0087$ \\  
        \hline
      \end{tabular}}  
  \end{table*}

  \begin{figure*}[!tb]
    \centering
    \resizebox{0.44\textwidth}{!}{\includegraphics{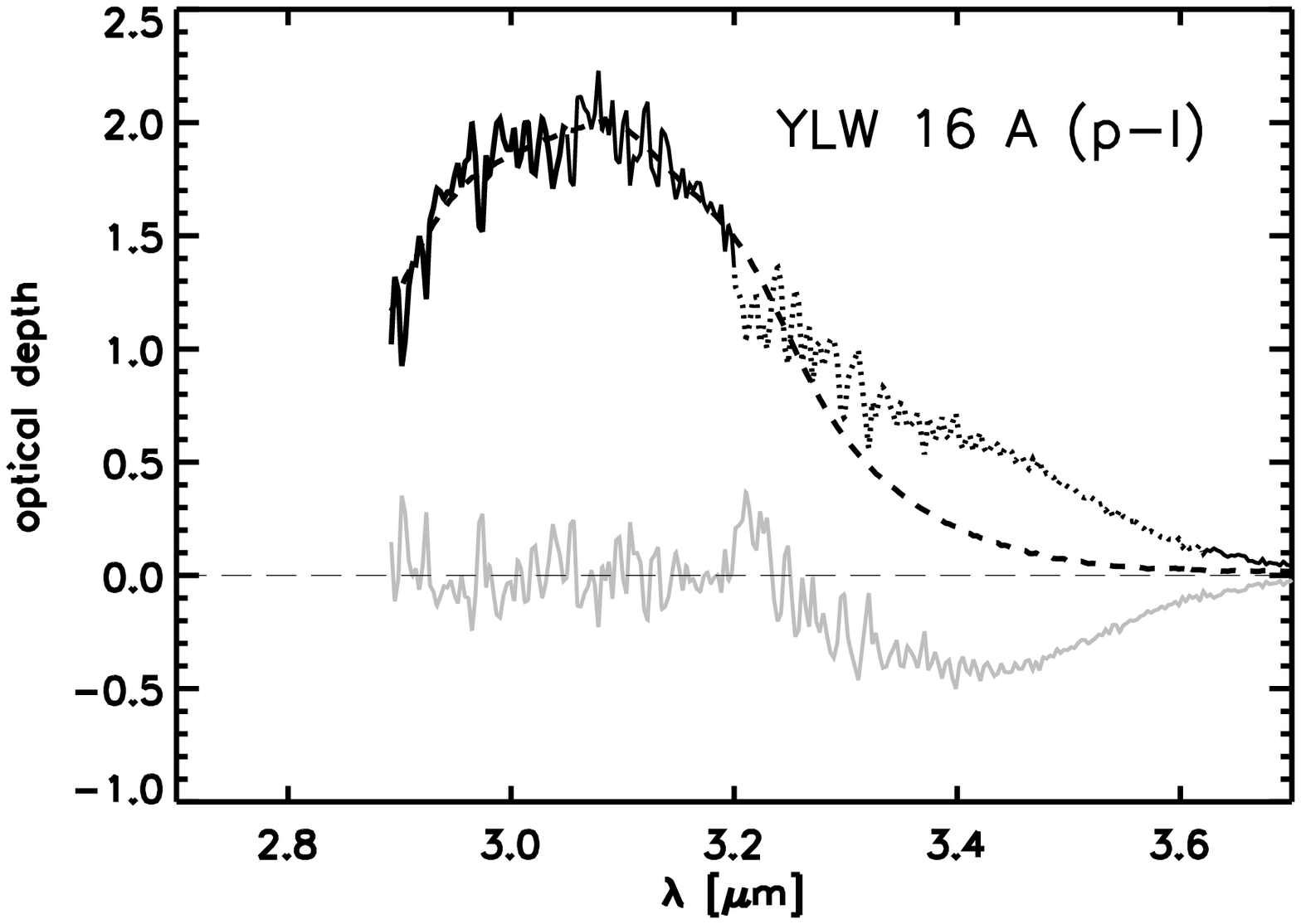}}
    \resizebox{0.44\textwidth}{!}{\includegraphics{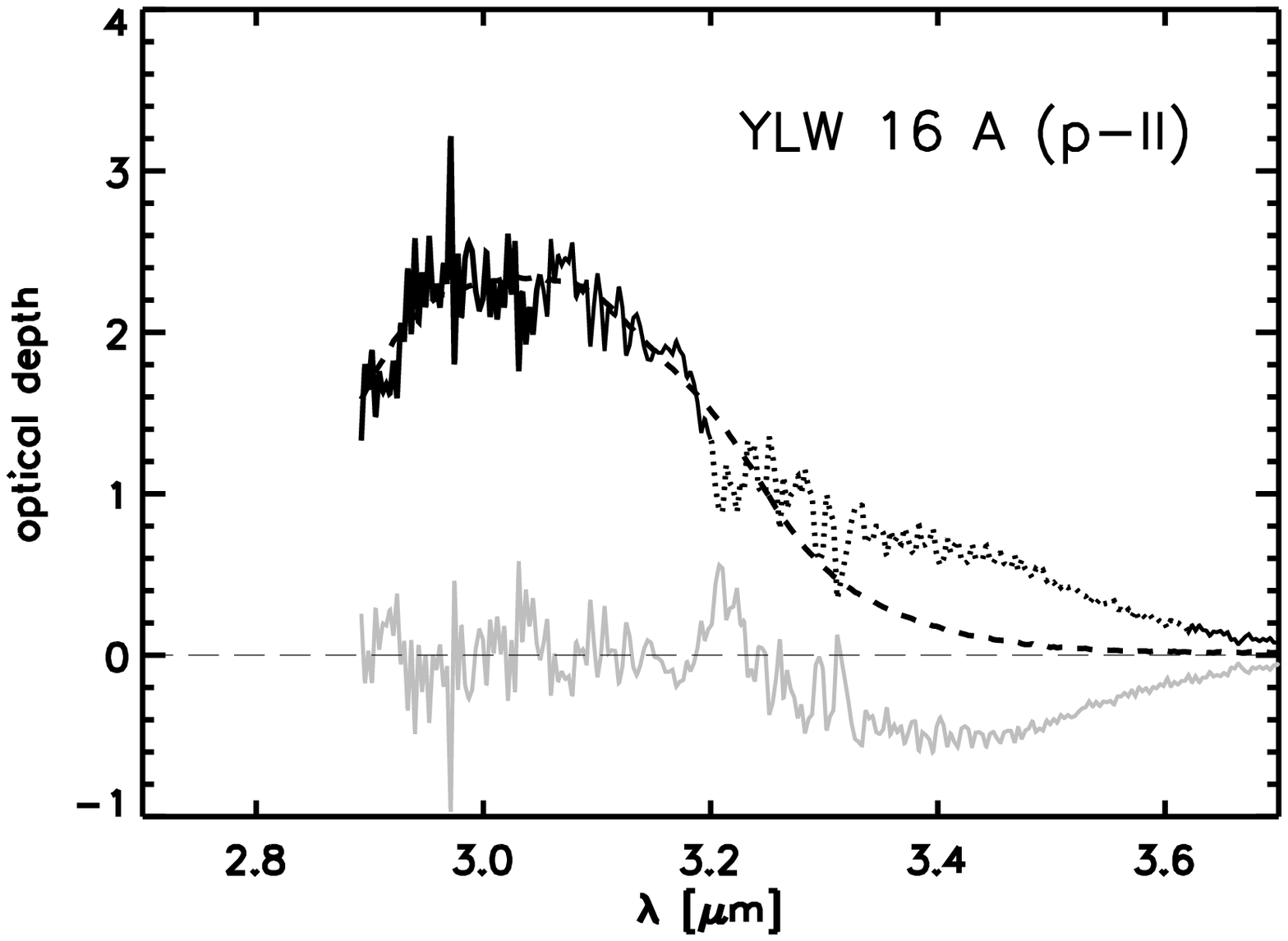}}\newline
    \resizebox{0.44\textwidth}{!}{\includegraphics{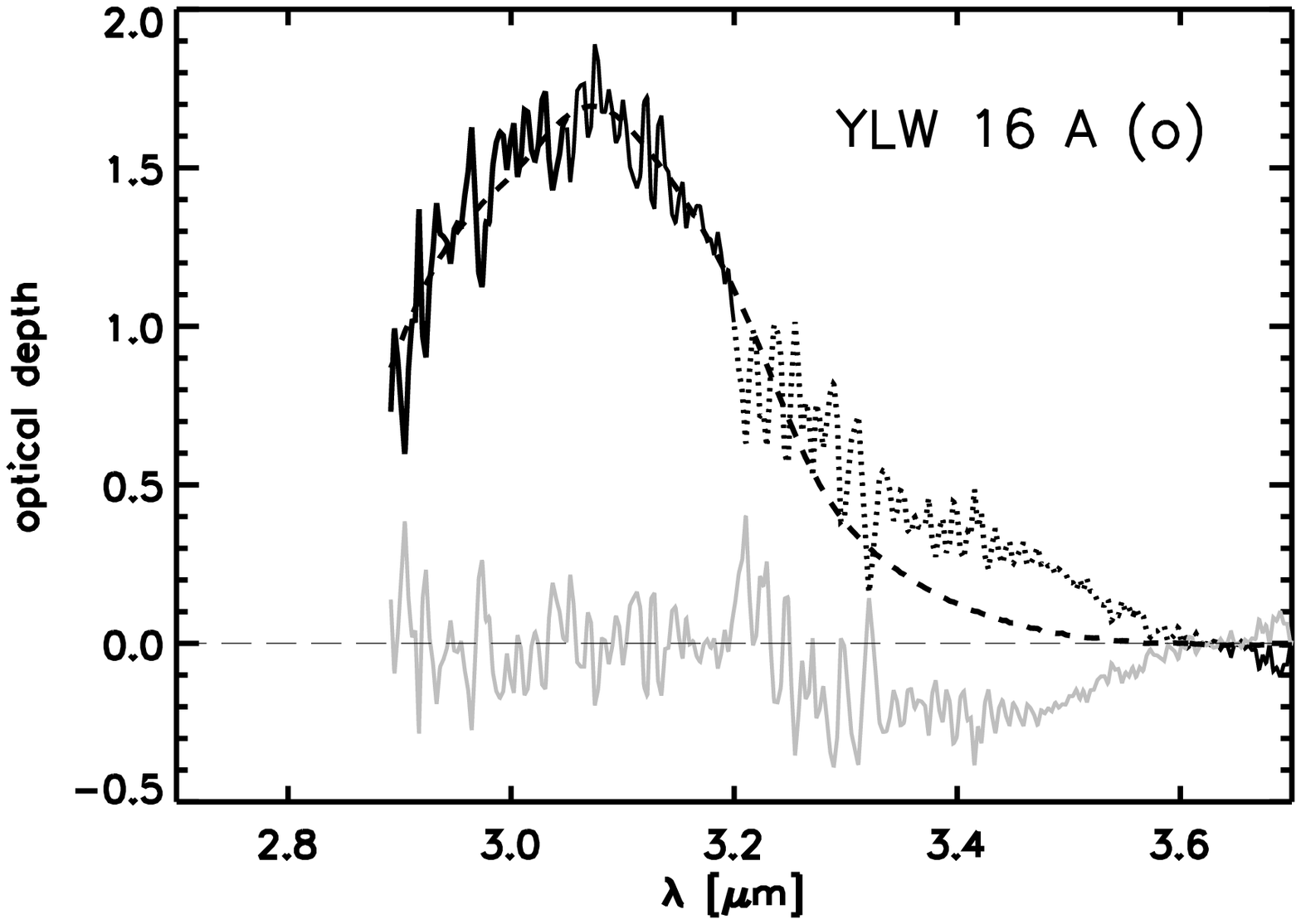}}
    \caption{Optical depth $\tau(\lambda)$ derived from the spectra of
      Fig.~\ref{figure:result-spek} considering Eq.~\ref{eq:opticaldepth}. Dotted
      curves represent the wavelength interval that is not considered in the 
      modeling approach. The water ice band in the interval between
      $3.2\,\mathrm{\mu m}$ and $3.6\,\mathrm{\mu m}$ is superimposed on the 
      absorption bands of ammonia hydrate and other compounds. The models are
      represented by dashed 
      curves (Table~\ref{table:result-water}). The
      gray curves around the zero level are deviations from the measured
      optical depth. 
      The source is not spatially
      resolved in the optical depths shown here.}
    \label{figure:result-water1}
  \end{figure*}
  \begin{figure*}[!tb]
    \centering    
    \resizebox{0.44\textwidth}{!}{\includegraphics{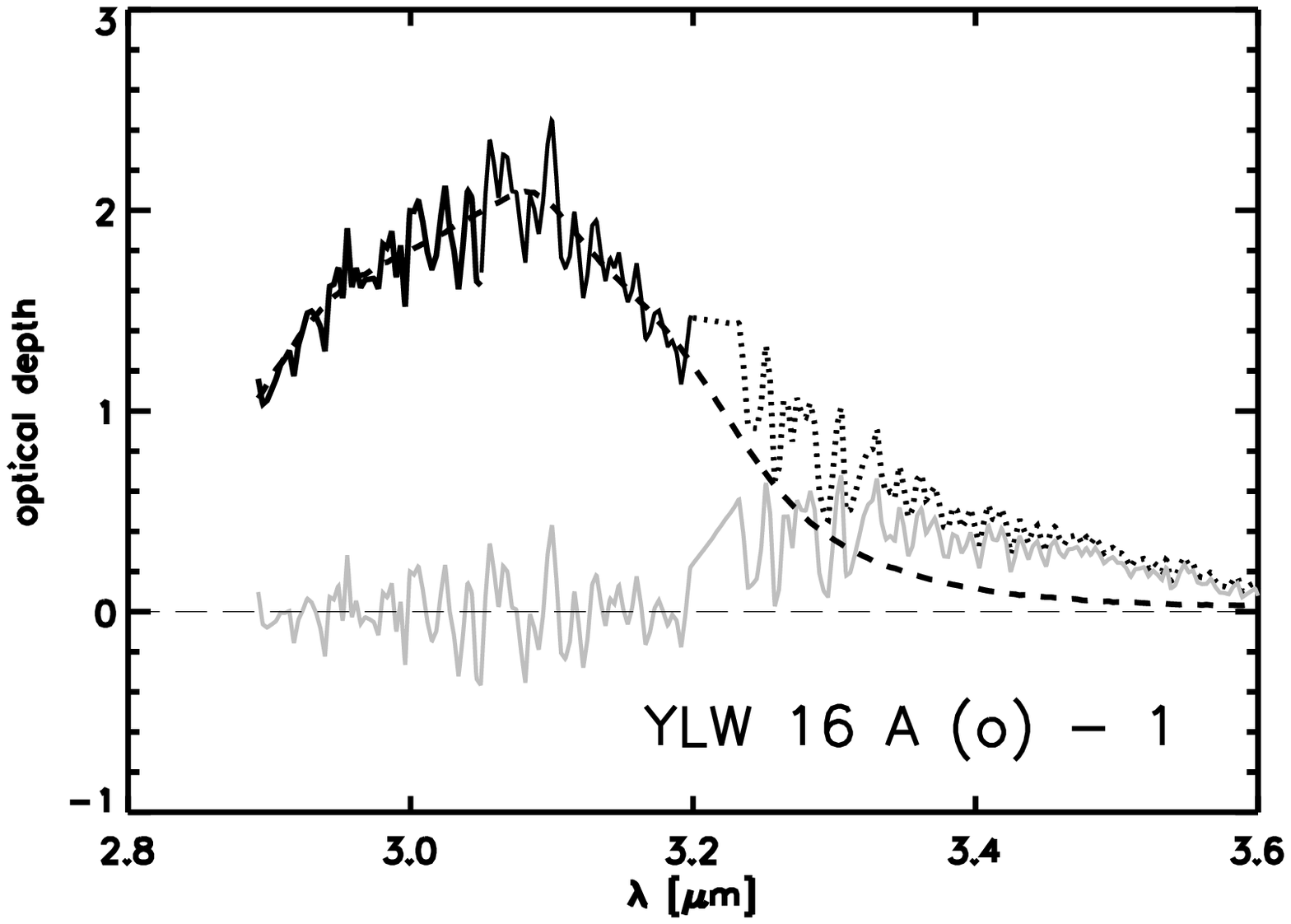}}
    \resizebox{0.44\textwidth}{!}{\includegraphics{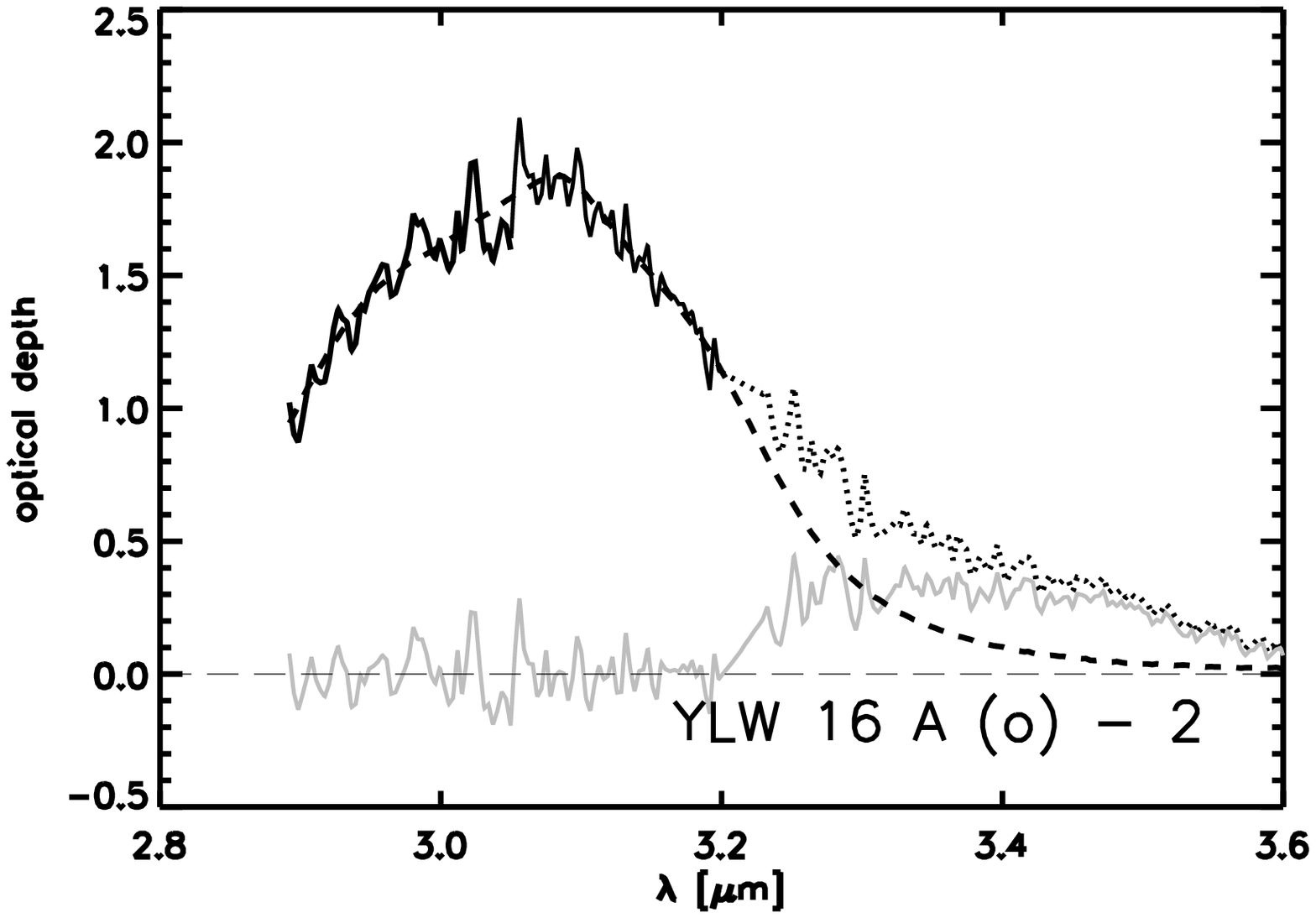}}\newline
    \resizebox{0.44\textwidth}{!}{\includegraphics{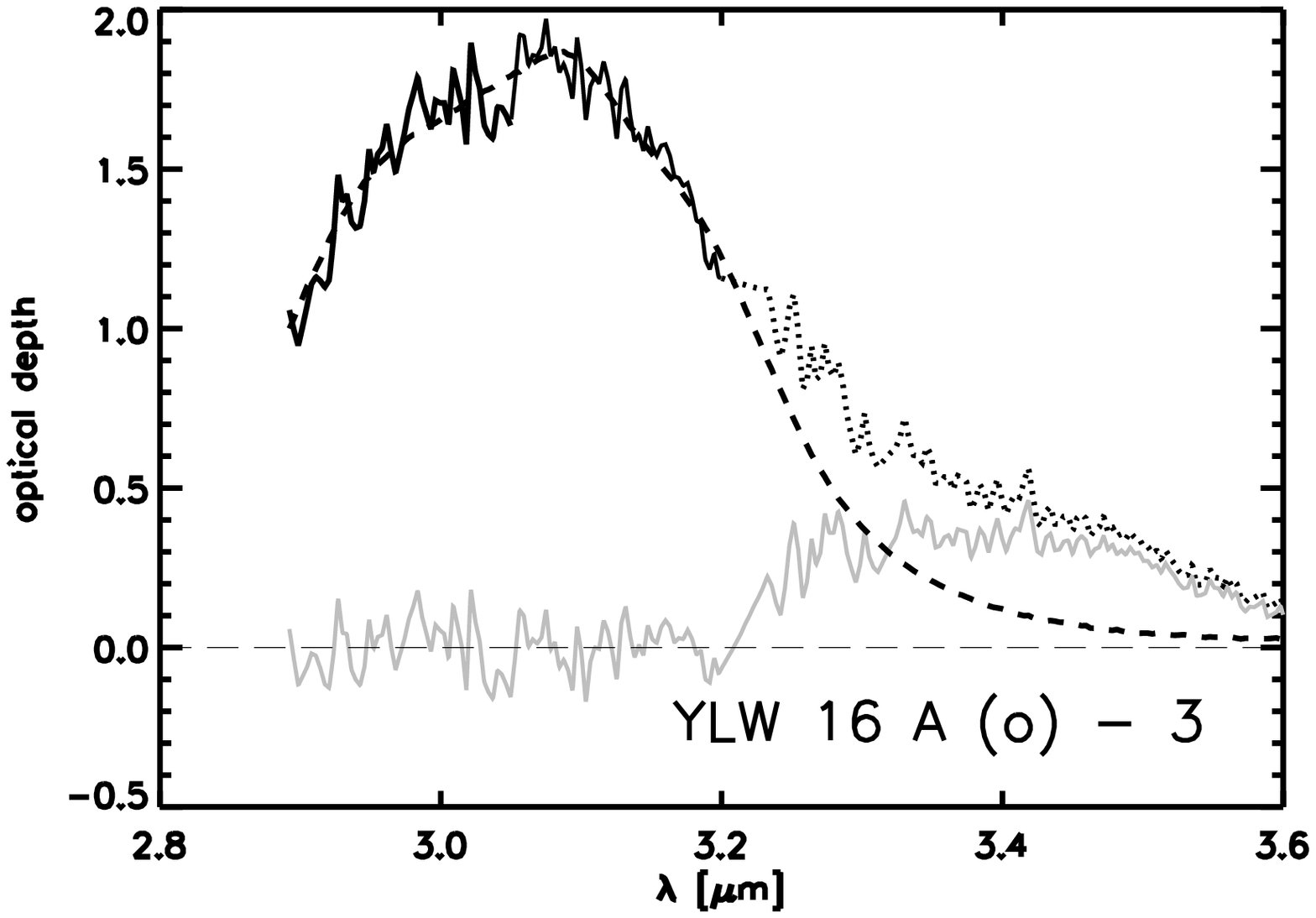}}
    \resizebox{0.44\textwidth}{!}{\includegraphics{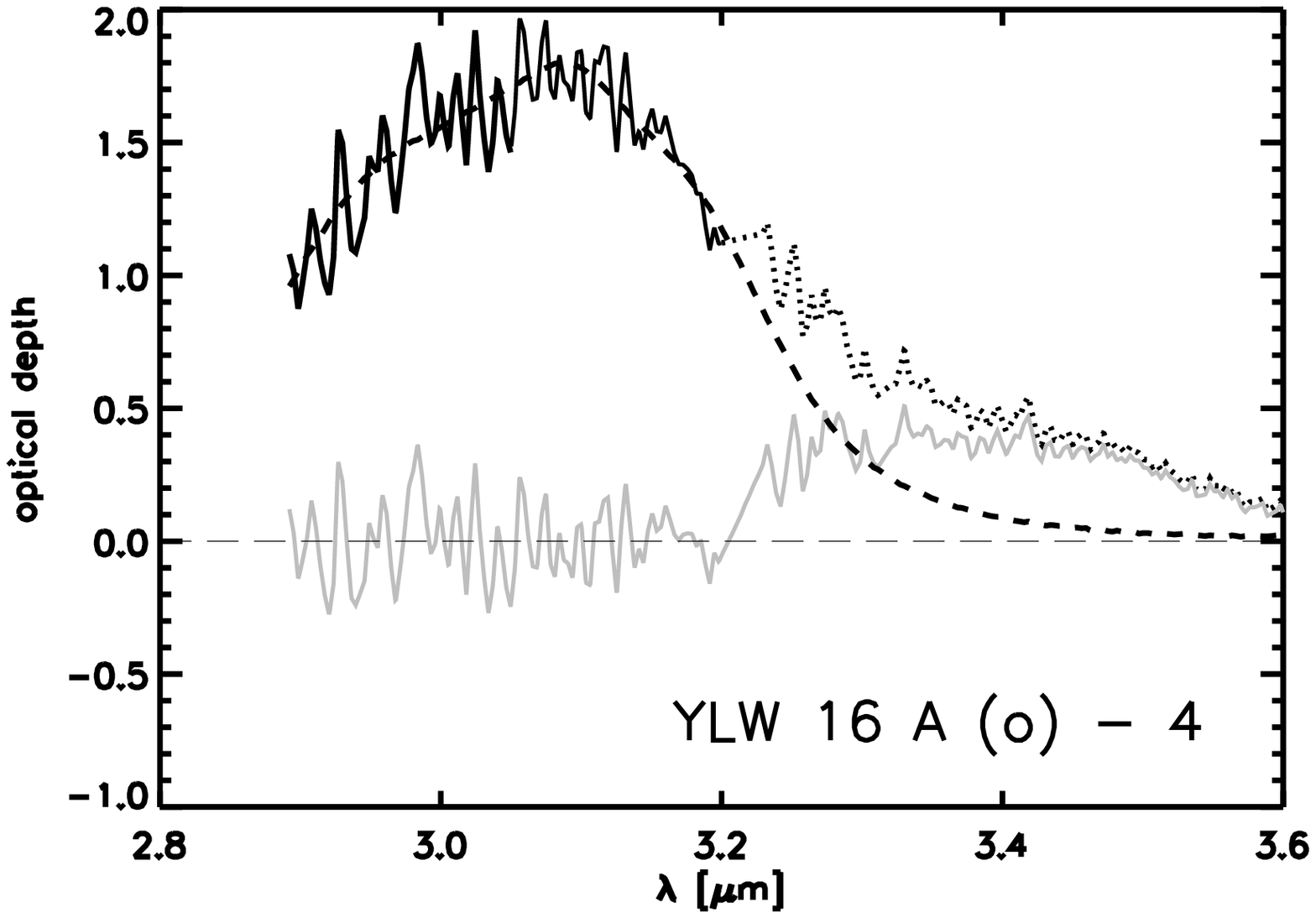}}\newline
    \resizebox{0.44\textwidth}{!}{\includegraphics{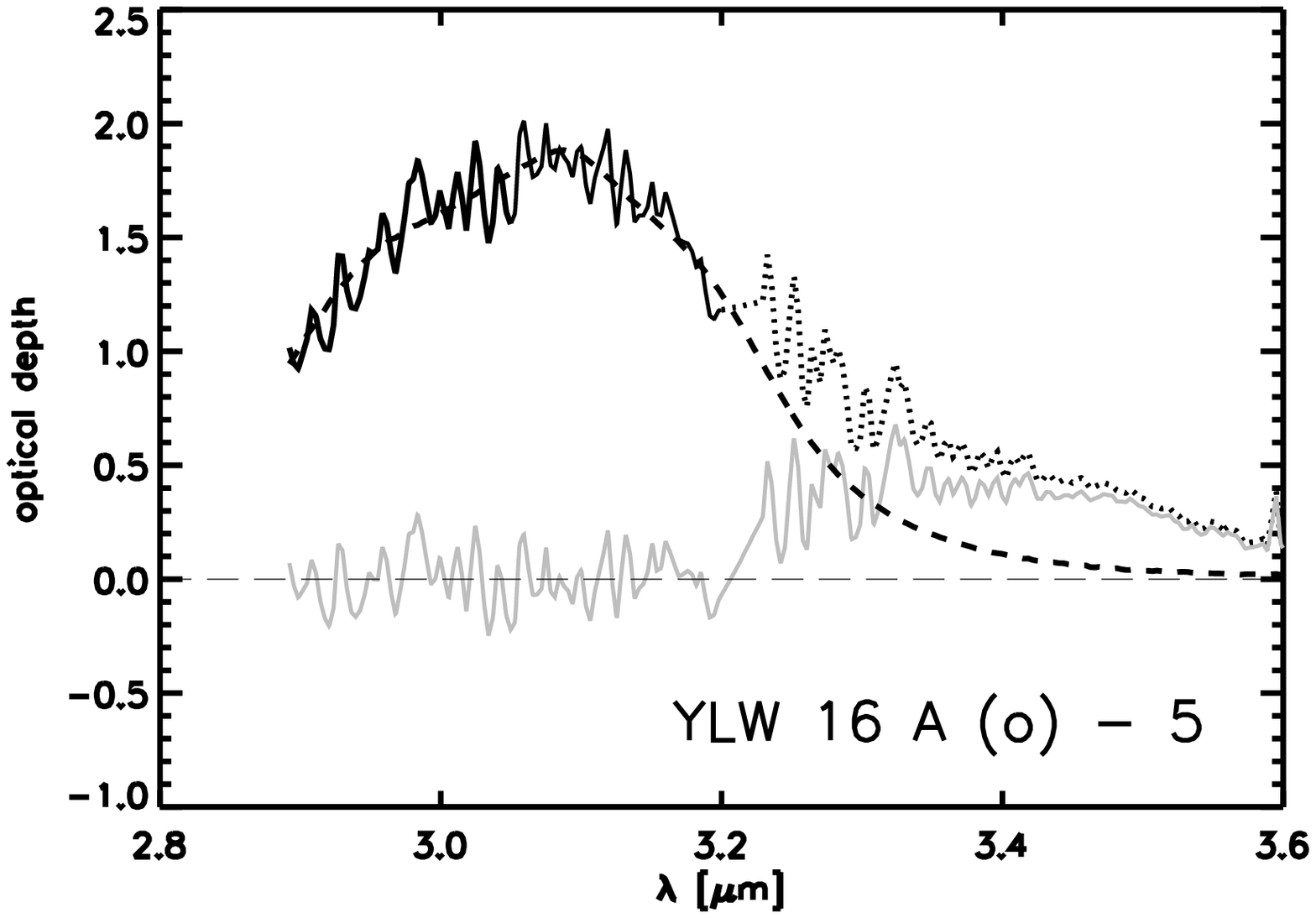}}
    \resizebox{0.44\textwidth}{!}{\includegraphics{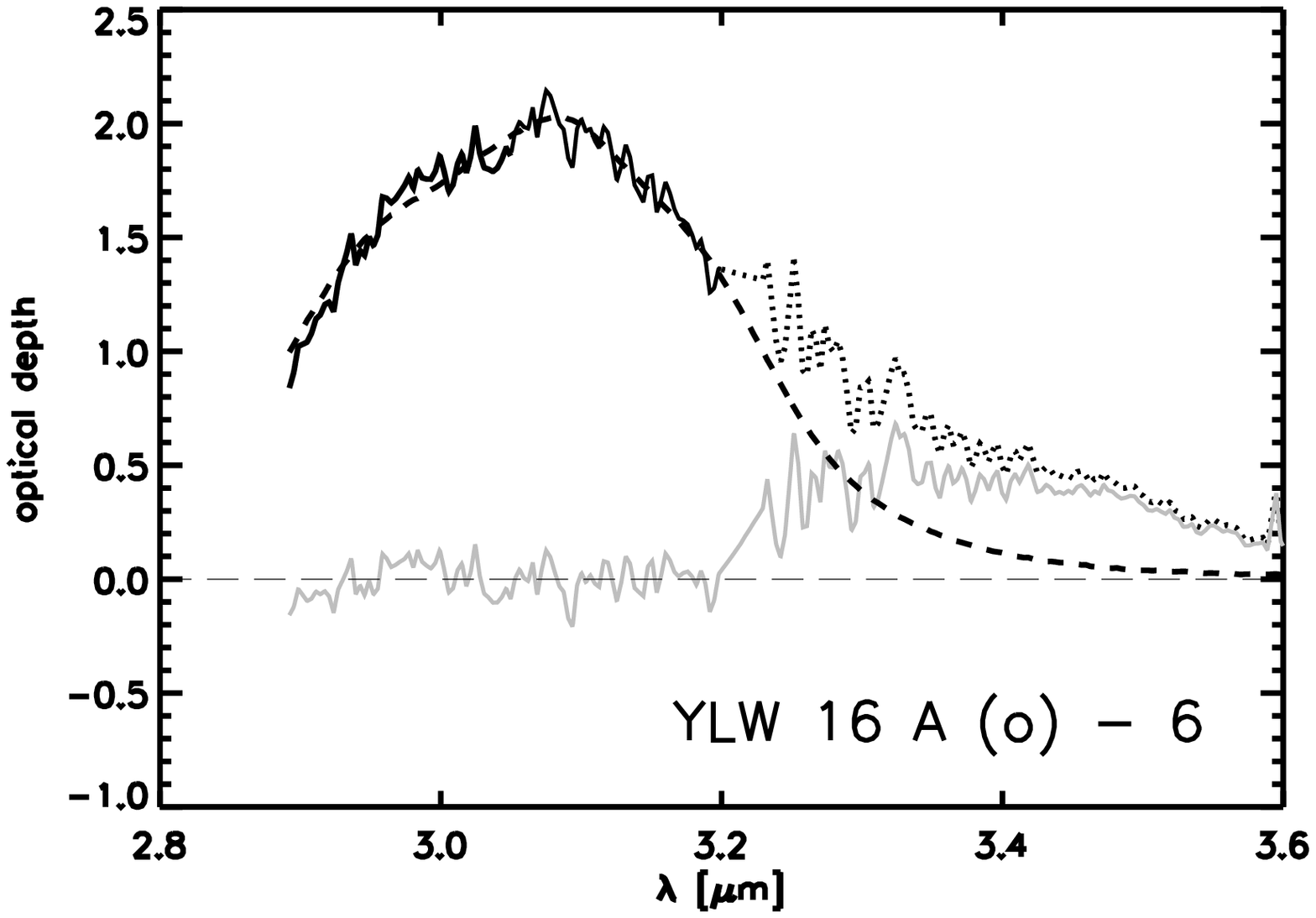}}\newline 
    \resizebox{0.44\textwidth}{!}{\includegraphics{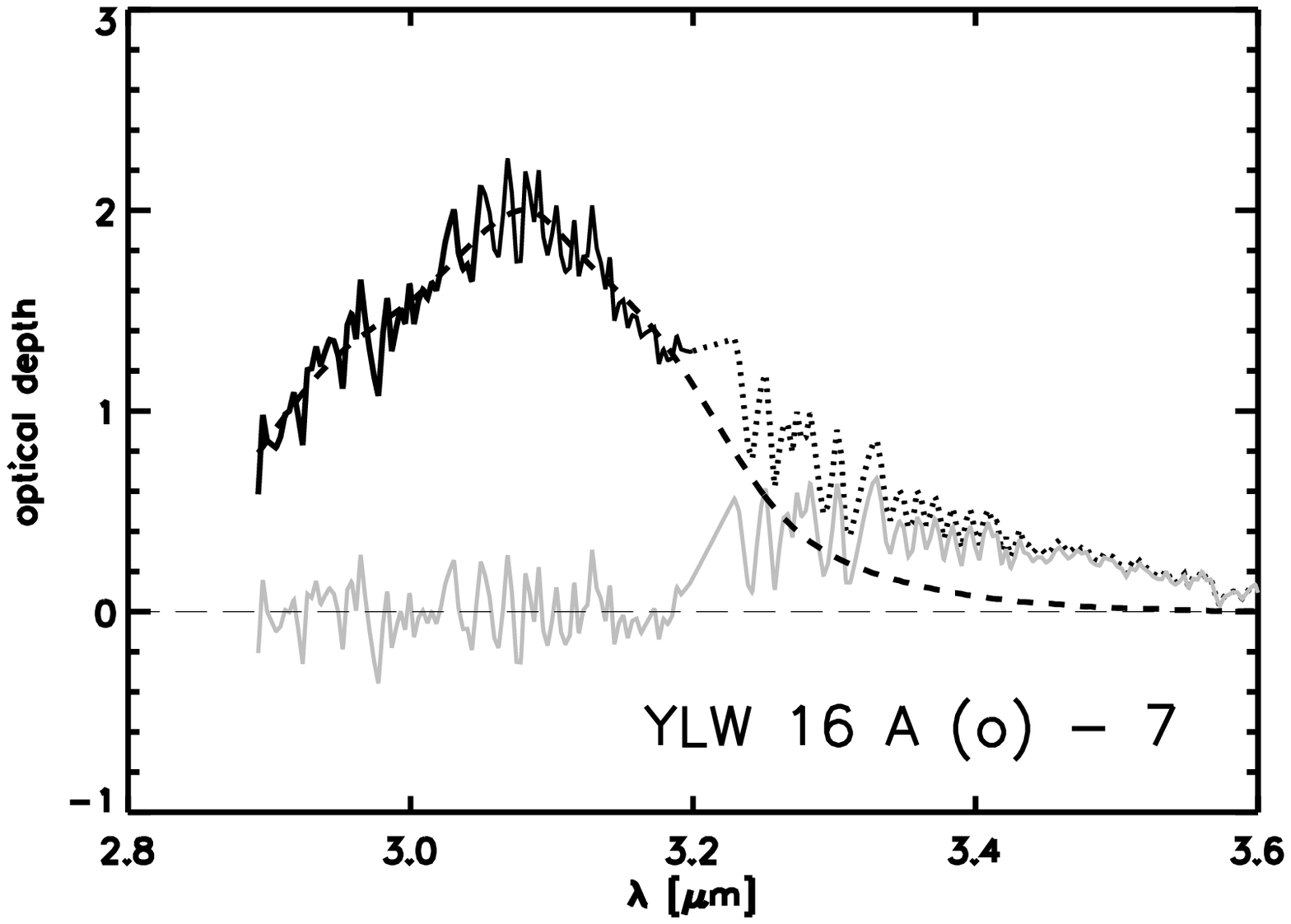}}
    \caption{Continuation of Fig.~\ref{figure:result-water1} but the spectra
      $1$ to $7$ correspond to different regions within the source that could
      be spatially resolved by our observations with NAOS-CONICA
      (Sect.~\ref{section:datareduction}).}
    \label{figure:result-water2}
  \end{figure*}
  
  \subsection{Water ice growth in the disks around T\,Tauri
    stars?}\label{section:graingrowth} 
  The spectra of central source regions derived from a rotational
  center of the continuum's line at $3.8\,\mathrm{\mu m}$ show profiles of
  amorphous water ice grains that exhibit only weak hints of grain growth
  (Table~\ref{table:result-water}). However, when considering the same results for
  a rotational center at $3.6\,\mathrm{\mu m}$, the contribution of the absorption
  profiles of $0.5\,\mathrm{\mu m}$-sized grains is significantly higher
  although the fit worsens by $24\%$ up to $32\%$.
  From Eq.~\ref{eq:scattering}, we find that scattered radiation is
  present only for grain sizes of up $\sim$$0.5\,\mathrm{\mu
    m}$. If we replace the extinction profiles of large
  grains ($a > 0.5\,\mathrm{\mu m}$) with the corresponding
  absorption profiles, the spectroscopic contribution of large grains
  increases by only a few percent. 

\begin{figure}[tb]
    \centering
    \resizebox{0.48\textwidth}{!}{\includegraphics{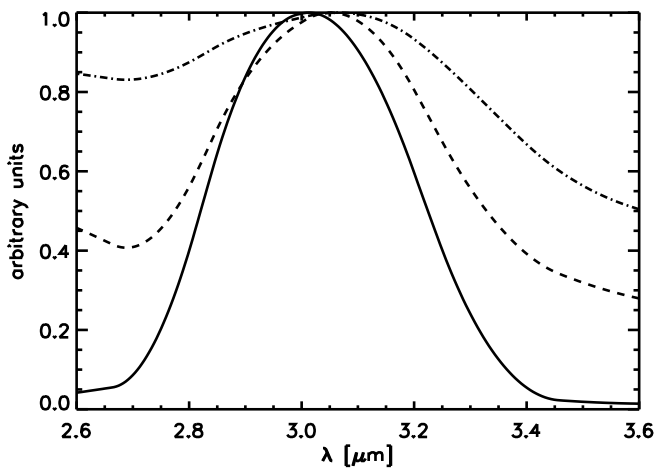}}
    \caption{Relation between extinction profile and the radius of the
      grain core. The core consists of amorphous silicate, while 
      water ice forms a compact shell. The particles shown here have
      a water ice core with a radius of 
      $0.1\,\mathrm{\mu m}$ (solid curve), $0.5\,\mathrm{\mu m}$ (dashed curve), 
      and $0.8\,\mathrm{\mu m}$ (dot-dashed curve). The thickness of the ice
      shell is $0.1\,\mathrm{\mu m}$.\label{figure:ice_on_silicate} The profiles
      were normalized by the maximum value.} 
  \end{figure}

  In previous studies, water ice grains larger than $1\,\mathrm{\mu
    m}$ were not detected in YSOs. Thi et al.~(\cite{thi}) point
  to broadened water ice profiles in the spectra of several YSOs in the molecular
  cloud Vela indicating ice grain growth. There is at least one among
  five objects in their sample where the absorption profile of
  $0.5\,\mathrm{\mu m}$-sized, amorphous ice grains can be used for
  modeling. Whether this result and the other evidence we have found indicate ice
  grain growth remains to be discussed in the context of the 
  following issues. 
  As shown in Schegerer et al.~(\cite{schegerer}), increasing the porosity
  of grains with constant radii broadens the 
  $10\,\mathrm{\mu m}$-silicate feature. We assume that porous water ice
  grains show similar behavior but this assumption can only be confirmed by
  future, theoretical 
  studies. Furthermore, it is often assumed that ice grains have cores,
  such as those of  
  $0.1\,\mathrm{\mu m}$-sized, amorphous silicate (e.g., Jones \&
  Merrill~\cite{jones}). The core serves as a seed where water
  is adsorbed. As shown in Fig.~\ref{figure:ice_on_silicate},
  a growth in the 
  silicate core also results in a broadening of the $3\,\mathrm{\mu m}$-absorption
  band (e.g., Smith et al~\cite{smithII}). A broadening of the
  extinction profile can also be evoked by the shape of the continuum curve
  derived. In our study,
  the continuum is represented by a straight line, while the Planck
  function or spline functions are 
  used in other studies. The Planck function, which is concave down,
  implies that there has been a
  broadening of the water ice band at long wavelengths, which automatically
  indicates ice grain growth. Finally, it is not clear to us why only
  $0.5\,\mathrm{\mu m}$-sized grains are found when large dust grains are
  used in the fit. It has to be clarified whether the ice grain growth to sizes
  $a > 0.5\,\mathrm{\mu m}$ is physically prevented in YSOs or if this finding
  can be ascribed to the fitting approach used.  

  We conclude that an unambiguous identification of
  grain growth is difficult to make. In contrast,
  the discrimination between  
  crystallized and amorphous grains should be easier as argued by Dartois \&
  d'Hendecourt~(\cite{dartois}), as the extinction profile of crystallized ice
  narrows. 

  \subsection{Crystallized water ice in the disks around T\,Tauri
    stars?}\label{section:crystalline?} 
  Because of the low temperatures and weak
  shielding of the interstellar media against hard irradiation, we
  assume that there is a
  negligible contribution of crystallized water ice when YSOs start to form
  from molecular clouds. The finding of crystallized water ice in
  YLW\,16\,A is now discussed. 

  As described in Sect.~\ref{section:datareduction}, information about
  the spatial  
  ice distribution within the circumstellar environment can be extracted
  from the spectra measured with NAOS-CONICA. Instead of summarizing
  the intensities of the aperture, the 
  Gaussian function that is fitted to the intensity distribution in each row
  of the image is symmetrically divided in steps of $20\,\mathrm{AU}$,
  $20\,\mathrm{AU}$, $30\,\mathrm{AU}$, and $40\,\mathrm{AU}$ starting at the
  photocenter of the source. Independently of the rotational center used,
  we mainly find amorphous water ice close to the photocenter
  ($m_\mathrm{a} > 88\% $). However, for increasing distance from the photocenter, the mass contribution of
  crystallized ice strongly increases (Table~\ref{table:result-water},
  Fig.~\ref{figure:result-water2}). 
  The sketch in Fig.~\ref{figure:waterice-sketch} shows a
  circumstellar disk where the 
  location of amorphous and crystallized water ice grains are plotted
  considering our fitting results. We
  assume that the crystallized water ice component has its origin in
  the 
  upper disk layers at large disk radii, where water ice does not evaporate
  and is protected by the surrounding material of the circumstellar envelope
  from irradiation. Because of the higher temperatures, water ice at
  smaller radii close to the snowline is assumed to be crystallized
  inside the 
  more embeded disk layers.  
  The observation of the outer parts of the optically thick disk also
  explains the higher 
  column density $N_\mathrm{A}$ inferred from spectra that
  originates in the outer regions (Table~\ref{table:result-water}). Again,
  this finding
  does not depend on the rotational center used. The
  photocenter of the spectra is assumed to represent the optically
  thin bipolar envelope of the source observed with NICMOS/HST (Allen et
  al.~\cite{allen}). Because of the low effective shielding in this central region,
  crystallized water ice is assumed to be destroyed by irradiation.   

\begin{figure}[tb]
    \centering
    \resizebox{0.48\textwidth}{!}{\includegraphics{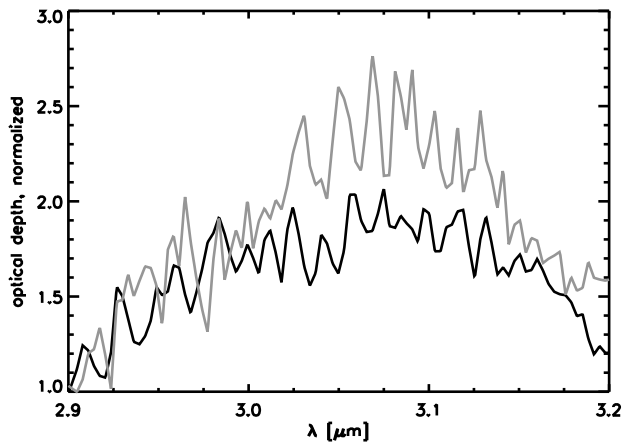}}
    \caption{\label{figure:ueberlagerung} 
      Optical depths derived from the central ({\it black} curve) and peripheral
      ({\it gray} curve) regions of the YSO observed. The optical depths of the region
      3, 4, and 5 were averaged to obtain the black curve while the 
      depth from the region 7 was used to obtain the gray curve. For
      comparison, the curves are normalized at the optical depth at
      $\sim$$2.9\,\mathrm{\mu m}$. A
      larger amount of crystalline ice grains contribute to the narrow gray
      curve with a maximum at $3.075\,\mathrm{\mu m}$. Almost only amorphous
      dust grains contribute to the black curve derived from the more central
      region.}
  \end{figure}

  Figure~\ref{figure:ueberlagerung} shows the normalized optical
  depths derived from the depth of the peripheral
  region 7 ({\it gray} curve) that is overlaid on the mean of the depths of more
  central regions. The optical depths of the more central regions
  are lower than those of the peripheral regions. We note 
  that if only amorphous grains are 
  used in our deconvolution approach, the fitting worsens by up to $14\%$
  according to the parameter $\chi^2$ (Eq.~\ref{eq:chi};
  Table~\ref{table:result-water}) derived. 

  The YSOs that were additionally observed with NAOS-CONICA and whose
  observations are presented in the Appendix of this publication, do not show
  any evidence of crystallization. However, the spectra of these
  objects contain
  the water ice absorption feature. The question of whether these
  features originate in the {\it circumstellar environment} of these objects cannot 
  unequivocally be answered here. 
  
  \section{Summary}\label{section:summary}
  We have presented an L band spectroscopic observation performed with
  NAOS-CONICA of a YSO that is strongly inclined. These
  systems allow the study of optical thick circumstellar regions that shield 
  water ice from hard irradiation and evaporation. The absorption feature of
  water ice at $3.08\,\mathrm{\mu m}$ was identified in the spectra with
  optical depths of between $1.8$ and $2.5$, depending on the region observed. 
  The optical depth, the water-ice column density, and the visual
  extinction are higher in the more peripheral regions at greater
  distance from the photocenter of our observations. The 
  optical depths were derived from the spectra by assuming that a
  straight line represents the continuum flux.
  
\begin{figure*}[!t]
    \centering
    \resizebox{12.cm}{!}{\includegraphics[angle=-90]{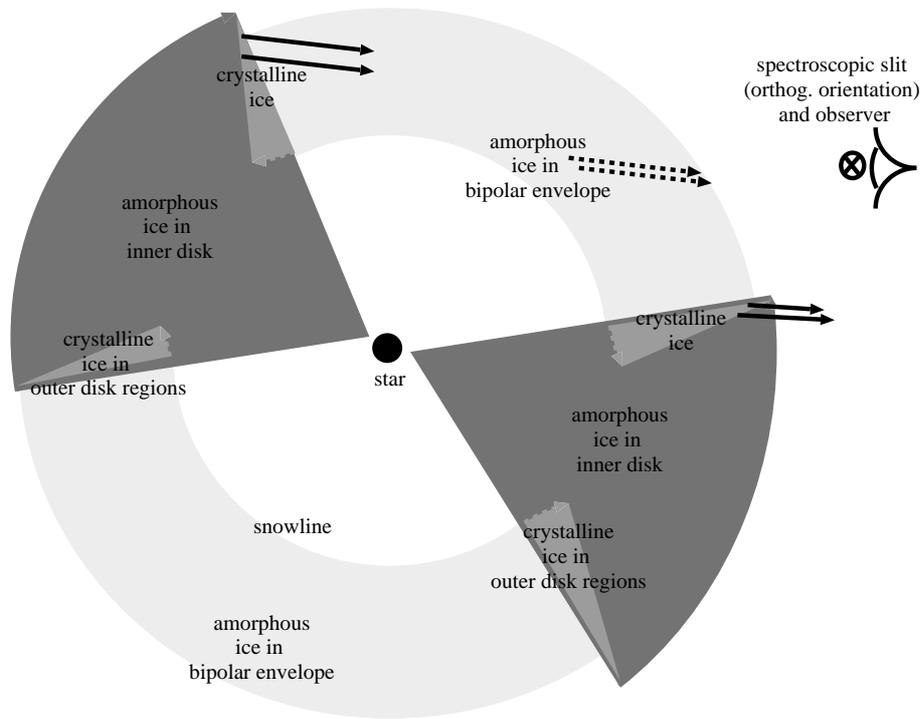}}
    \caption{Sketch for a strongly inclined circumstellar disk such as
      YLW\,16\,A. We assume that 
      amorphous, non-evolved water ice is located 
      in the circumstellar disk and in the bipolar envelope. The assumed
      location of crystallized water ice in the disk is close to upper disk layers, at disk
      radii that are larger than that of the snowline. The
      observer and the spectroscopic slit are also sketched. The
      orthogonal slit orientation is shown. Radiation from amorphous and
      crystallized ice grains are 
      represented by dashed and solid arrows, respectively.}
    \label{figure:waterice-sketch}
  \end{figure*}

  Considering the extinction profiles of amorphous and crystallized water ice
  grains with sizes ranging from $0.1\,\mathrm{\mu m}$ up to $1.5\,\mathrm{\mu m}$,
  the derived optical depths were deconvolved using a fitting routine
  that was already presented in Schegerer et al.~(\cite{schegerer}). 
  The optical depths of the spatially unresolved spectra were found to
  be dominated by the
  extinction profile of small, amorphous (i.\/e., non-evolved) water ice grains. 
  However, in spectra for which the T\,Tauri object YLW\,16\,A could be spatially
  resolved, crystallized water ice could be found in possible outer 
  disk regions, i.\/e., in disk layers at radii $>$$30\,\mathrm{AU}$ in
  projection. The putative 
  bipolar envelope of this source was not found to exhibit any evidence of crystallized
  material (Fig.~\ref{figure:waterice-sketch}). A growth in the
  size of the water ice grains
  cannot unambiguously be determined because the corresponding modification of the
  water ice feature may also be either caused by an increase in the
  porosity and/or a growth of the particle core.
  
  \begin{acknowledgements}
    We thank D.~Semenov and M.~Goto, in particular, for fruitfull discussions
    and their further 
    support. Thanks are also due to J.~Bouwman for his valuable assistance
    during the proposal preparation 
    phase. 
  \end{acknowledgements}

  \onecolumn
  \appendix
  \section{Additional measurements}\label{appendix}
  Table~\ref{table:observation_a} provides details of additional observations with
  NAOS-CONICA. The quantity $T_\mathrm{tot}$ is
  the total exposure time. The standard stars HR\,6070 and
  BS\,7330 were observed directly before or after the
  observation of each target.
    \begin{table*}[!t]
    \centering
    \begin{minipage}{1.0\textwidth}
      \caption{Overview of additional observations with NAOS-CONICA. Coordinates,
        L band magnitude, $FWHM$, date, observing
        time ({\@}$UT${\@}), and the mean airmass ({\@}$AM${\@}) during
        the 
        observations are listed. } 
      \label{table:observation_a}
      \begin{center}
        \begin{tabular}{lrrrrrccr} 
          \hline\hline
          object & $\alpha(\mathrm{J2000})$ & $\delta(\mathrm{J2000})$ & L
          [mag] & $FWHM$ [''] & 
          date & $UT$ & $AM$ & $T_\mathrm{tot}$\\ \hline 
          \object{HR\,6070} & $16\,18\,18$ & $-28\,36\,46$ & $4.807$ & -- & April
          $2^{nd}, 2006$ & $7:02 - 7:04$ & $1.0$ & $2.0\,\mathrm{s}$\\ 
          \object{Elia\,2-23} (p) & $16\,26\,24$ & $-24\,24\,50$ & $5.9$ & $2$ & Apr
          $2^{nd}, 2006$ & $7:45 - 9:34$ & $1.0$ & $72 \times 6 \times
          2.0\,\mathrm{s}$\smallskip\\ 
          HR\,6070 & $16\,18\,18$ & $-28\,36\,46$ & $4.807$ & -- & April
          $2^{nd}, 2006$ & $7:02 - 7:04$ & $1.0$ & $2.0\,\mathrm{s}$ \\ 
          \object{Elia\,2-21} & $16\,26\,24$ & $-24\,24\,50$ & $8.0$ & $1$ & April
          $2^{nd}, 2006$ & $7:45 - 9:34$ & $1.0$ & $72 \times 6 \times
          2.0\,\mathrm{s}$\smallskip\\
          Elia\,2-23 (o) & $16\,26\,24$ & $-24\,24\,50$ & $5.9$ & $2$ & June
          $22^{nd}, 2006$ & $0:00 - 0:30$ & $1.0$ & $76 \times 6 \times
          2.0\,\mathrm{s}$ \\ 
          HR\,6070 & $16\,18\,18$ & $-28\,36\,46$ & $4.807$ & -- & June
          $22^{nd}, 2006$ & $0:45 - 0:47$ & $1.0$ & $2.0\,\mathrm{s}$\bigskip\\ 
          \object{GY92\,244} (p) & $16\,27\,17$ & $-24\,28\,58$ & $8.8$ & $2$ & May
          $27^{th}, 2006$ & $2:48 - 4:00$ & $1.1$ &  
          $54 \times 18 \times 2.0\,\mathrm{s}$ \\
          HR\,6070 & $16\,18\,18$ & $-28\,36\,46$ & $4.807$ & -- & May
          $27^{th}, 2006$ & $4:24 - 4:26$ & $1.0$ & 
          $2.0\,\mathrm{s}$\smallskip\\ 
          GY92\,244 (o) & $16\,27\,17$ & $-24\,28\,58$ & $8.8$ & $2$ & May
          $27^{th}, 2006$  &$5:25 - 6:02$ & $1.0$ &  
          $36 \times 18 \times 2.0\,\mathrm{s}$ \\
          HR\,6070 & $16\,18\,18$ & $-28\,36\,46$ & $4.807$ & -- & May
          $27^{th}, 2006$ &$6:22 - 6:24$ & $1.1$ & $2.0\,\mathrm{s}$\bigskip\\ 
          \object{CRBR\,2422.8-3423} (p) & $16\,27\,05$ & $-24\,36\,31$ & $9.7$ & $1$ &
          August $12^{th}, 2006$ & $2:01 - 2:57$ & $1.3$ & 
          $18 \times 40 \times 4.0\,\mathrm{s}$ \\
          HR\,6070 & $16\,18\,18$ & $-28\,36\,46$ & $4.807$ & -- & Aug
          $12^{th}, 2006$ & $3:15 - 3:17$ & $1.5$ & $2.5\,\mathrm{s}$\smallskip\\ 
          CRBR\,2422.8-3423 (o) & $16\,27\,05$ & $-24\,36\,31$ & $9.7$ & $1$ &
          August $19^{th}, 2006$ & $0:14 - 0:45$ & $1.0$ &  
          $12 \times 60 \times 3.0\,\mathrm{s}$ \\
          HR\,6070 & $16\,18\,18$ & $-28\,36\,46$ & $4.807$ & -- & August
          $19^{th}, 2006$ & $1:11 - 1:13$ & $1.1$ & 
          $2.5\,\mathrm{s}$\bigskip\\ 
          \object{VV\,CrA\,A} (p) & $19\,03\,07$ & $-37\,12\,50$ & $3.7$ & $1$ & July
          $27^{th}, 2006$ & $5:23 - 6:01$ & $1.2$ &  
          $60 \times 8 \times 1.5\,\mathrm{s} + $\hfill{} \\
          & & & & & & & & $+ 16 \times 5 \times 5.0\,\mathrm{s}$ \\
          \object{BS\,7330} & $19\,21\,29$ & $-34\,58\,57$ & $4.9$ & -- & July
          $27^{th}, 2006$ & $6:28 - 6:30$ & $1.3$ 
          &$2.0\,\mathrm{s}$\smallskip\\ 
          VV\,CrA\,A (o) & $19\,03\,07$ & $-37\,12\,50$ & $3.7$ & $1$ & July 
          $4^{th}, 2006$ & $5:17 - 6:03$ & $1.0$ &  
          $60 \times 12 \times 1.0\,\mathrm{s}$ \\
          BS\,7330 & $19\,21\,29$ & $-34\,58\,57$ & $4.9$ & -- & July $4^{th},
          2006$ & $6:20 - 6:22$ & $1.1$ &$2.0\,\mathrm{s}$\bigskip\\ 
          \object{VV\,CrA\,B} & $19\,03\,07$ & $-37\,12\,50$ & $6.2$ & $1$ & July
          $27^{th}, 2006$ & $5:23 - 6:01$ & $1.2$ & 
          $60 \times 8 \times 1.5\,\mathrm{s} + $\hfill{}  \\
          & & & & & & & & $+ 16 \times 5 \times 5.0\,\mathrm{s}$ \\
          BS\,7330 & $19\,21\,29$ & $-34\,58\,57$ & $4.9$ & -- & July
          $27^{th}, 2006$ & $6:28 - 6:30$ & $1.3$ &$2.0\,\mathrm{s}$ \\ 
          \hline
        \end{tabular}
      \end{center}
    \end{minipage}
  \end{table*}
  The brightness of the standard star BS\,7330 is known from the standard star
  catalog of Leggett et al.~(\cite{leggett}). The 
  observation of the standard star BS\,7330 helped to remove 
  the telluric absorption features from the spectrum of VV\,CrA and its subsequent
  photometric calibration, while the object 
  HR\,6070 was used as a standard star for all remaining objects. 

  The L band
  brightness of
  the objects CRBR\,2422.8-3423 and VV\,CrA\,A are found in the
  Gezari-catalog~(\cite{gezari}). The L band brightness of the infrared
  companion VV\,CrA\,B is not known. The $FWHM$
  of the projected diameter of 
  CRBR\,2422.8-3423 could be estimated using a K band image (Pontoppidan et
  al.~\cite{pontoppidan}). The $FWHM$ of the diameter of VV\,CrA\,A and
  VV\,CrA\,B were derived using interferometric
  measurements in N band (Przygodda~\cite{przygodda}). The flux and the diameter
  of the object Elia\,2-23 in the L band were taken from Haisch et
  al.~(\cite{haisch}). The NICMOS/HST-image of GY92\,244 
  in the H~band (Allen et al.~\cite{allen}) allows us to approximate the diameter
  of the corresponding disks in L band. The L band brightness of this
  object was estimated by considering known H band magnitudes and
  (H-K) and (K-N) colors of additional YSOs in the same star-forming
  region (Allen et al.~\cite{allen}).  

  Analogous to YLW\,16\,A, all targets were observed twice, with two orthogonal 
  orientations of the slit: parallel (p) and
  orthogonal (o) to a predefined direction. If the position angle  
  of the disk was not known, the orientation of possible
  companions were considered in the subsequent selection of the slit orientation. 
  A slit length of $28${\arcsec} allowed us to observe the companions
  of VV\,CrA\,A (i.\/e., VV\,CrA\,B at an angular distance of $1.69${\arcsec};
  Przygodda~\cite{przygodda}) and Elia\,2-23 (Elia\,2-21 at an angular
  distance of $10${\arcsec}; Haisch et al.~\cite{haisch}) in the parallel
  orientation of the slit, simultaneously. 
  The data reduction of these observations is described in 
  Sect.~\ref{section:datareduction}. 

  Figures~\ref{figure:result-spek_a} and~\ref{figure:result-spekII_a} show
  the resulting L band spectra of these targets. Considering the derived
  flux, some sources 
  exhibit an intrinsic variability in L band (Leinert et
  al.~\cite{leinert}). For instance, it
  is known that the NIR brightness of VV\,CrA ``conspicuously'' varies
  within a period of few years (Graham~\cite{graham}; Chen \& Graham~\cite{chen};
  Koresko et al.~\cite{koresko}; Przygodda~\cite{przygodda}). 
  \begin{figure*}[!tb]
    \centering
    \resizebox{0.44\textwidth}{!}{\includegraphics{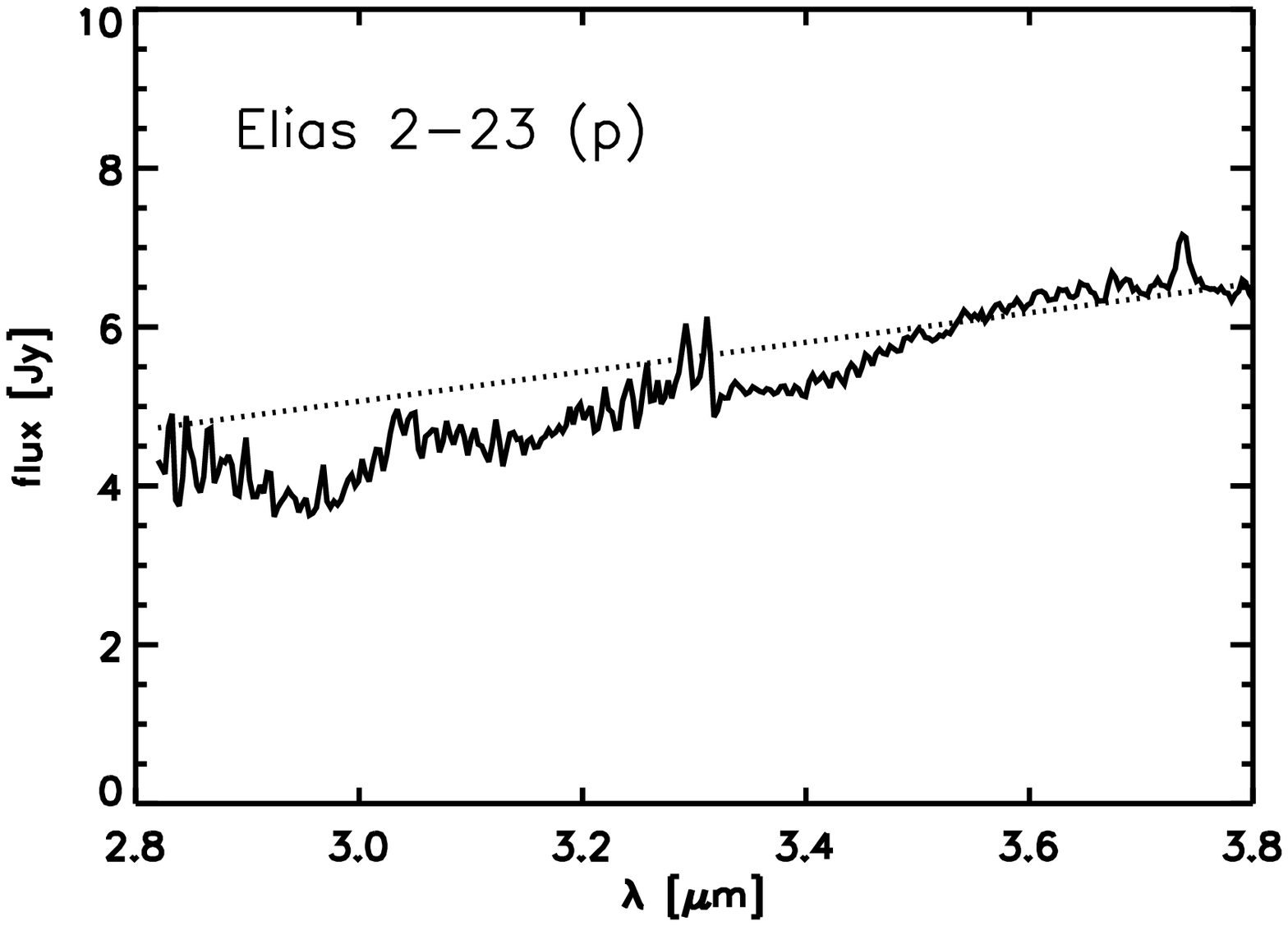}}
    \resizebox{0.44\textwidth}{!}{\includegraphics{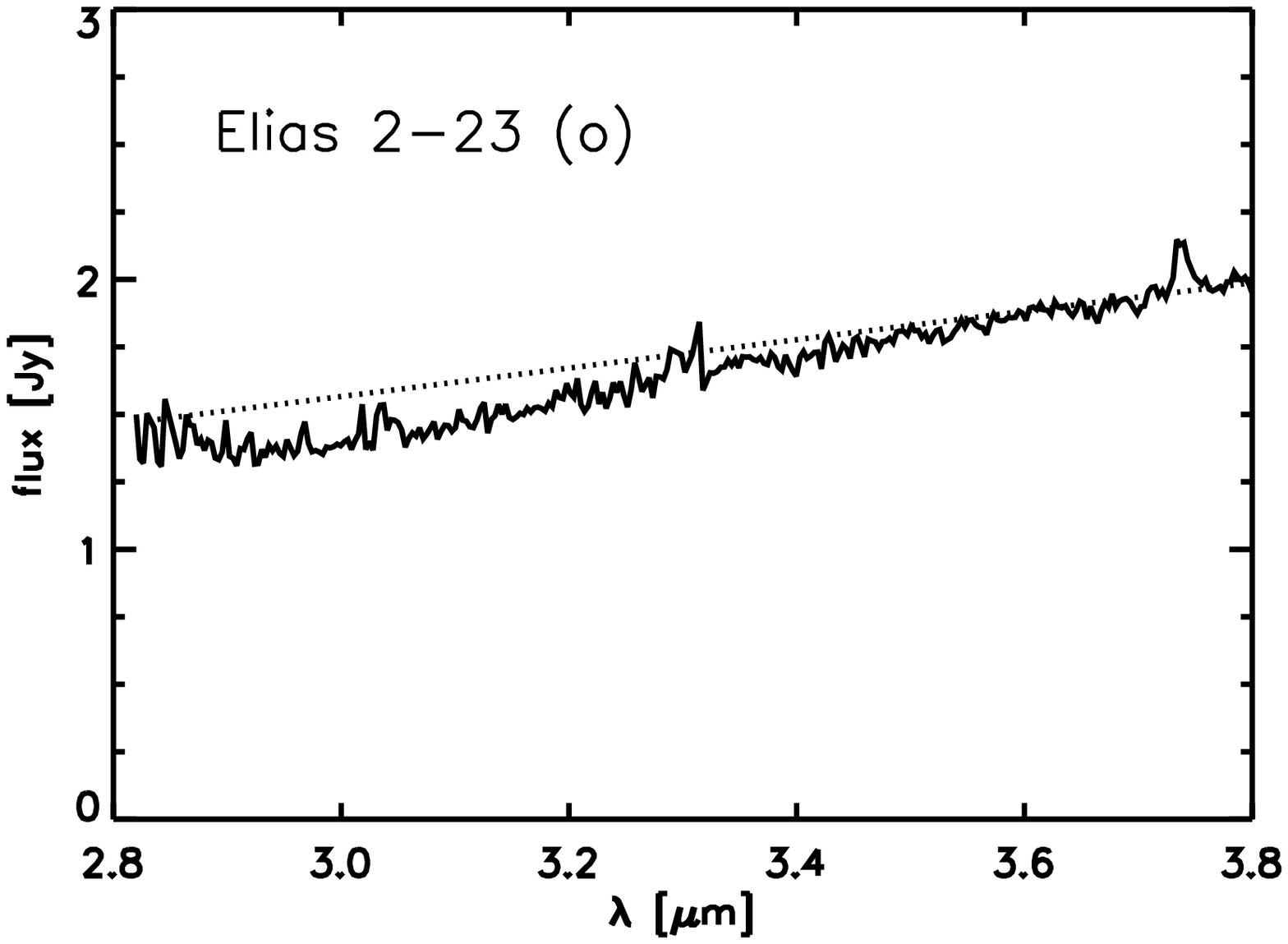}}\newline
    \resizebox{0.44\textwidth}{!}{\includegraphics{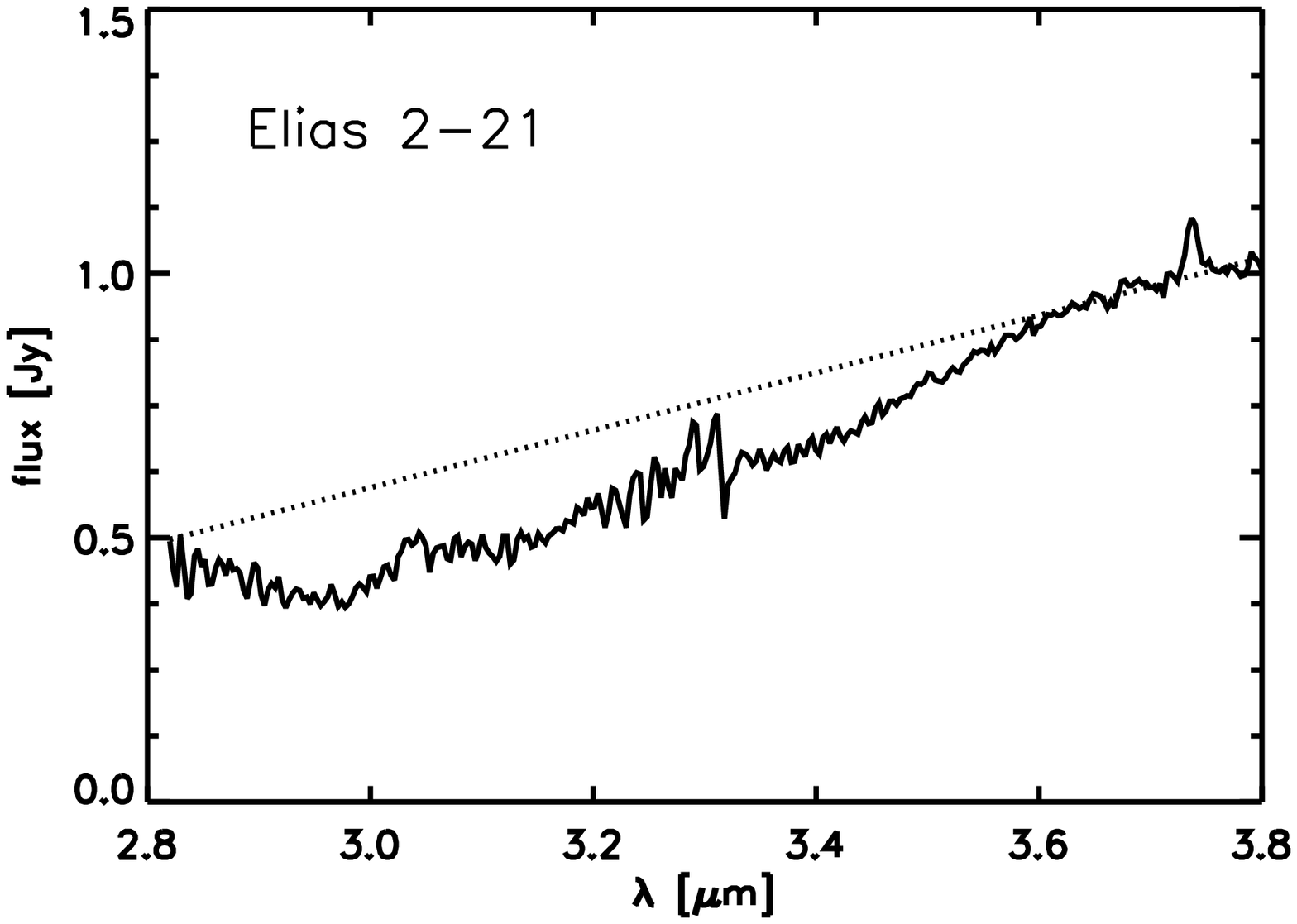}}\newline
    \resizebox{0.44\textwidth}{!}{\includegraphics{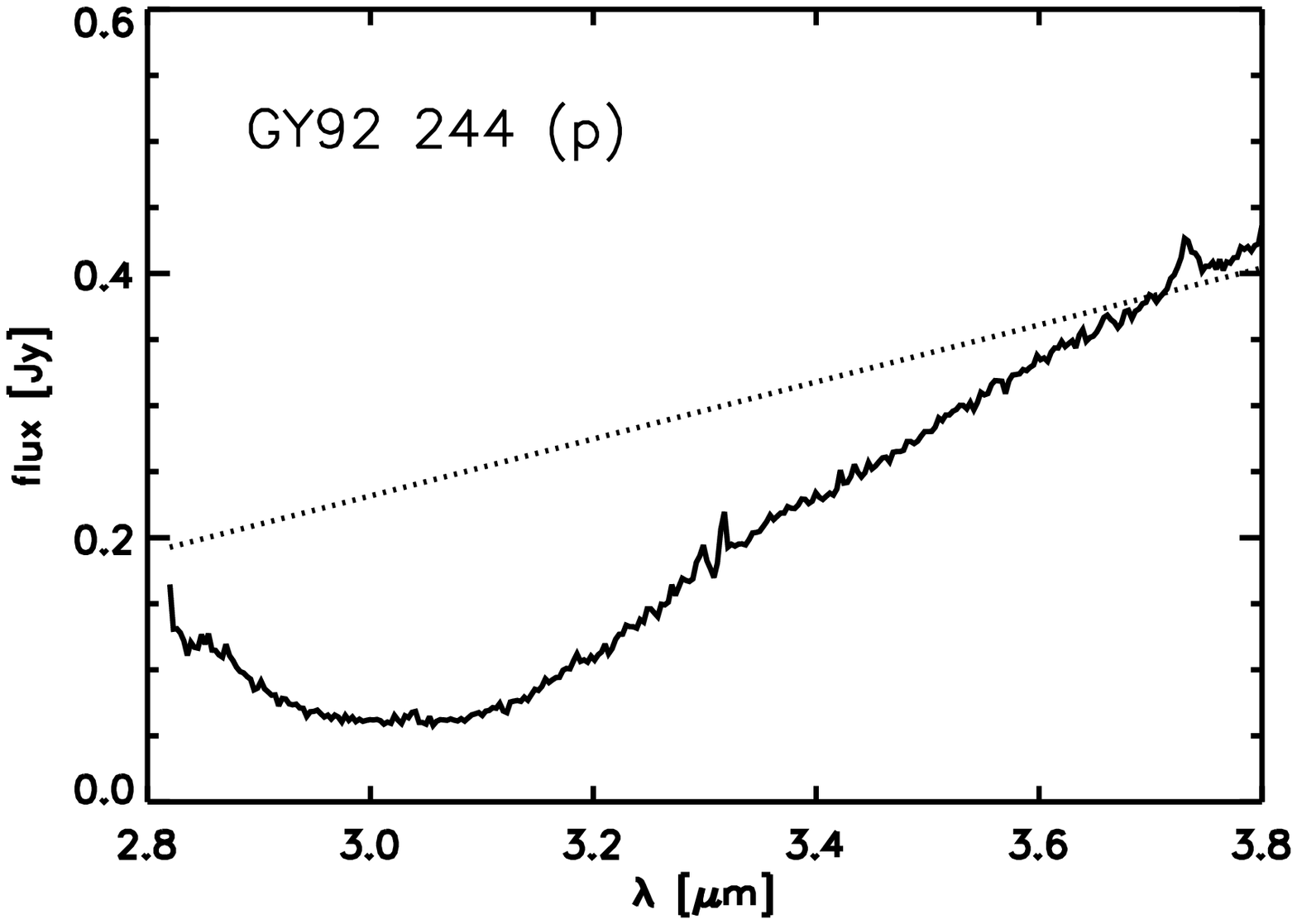}}
    \resizebox{0.44\textwidth}{!}{\includegraphics{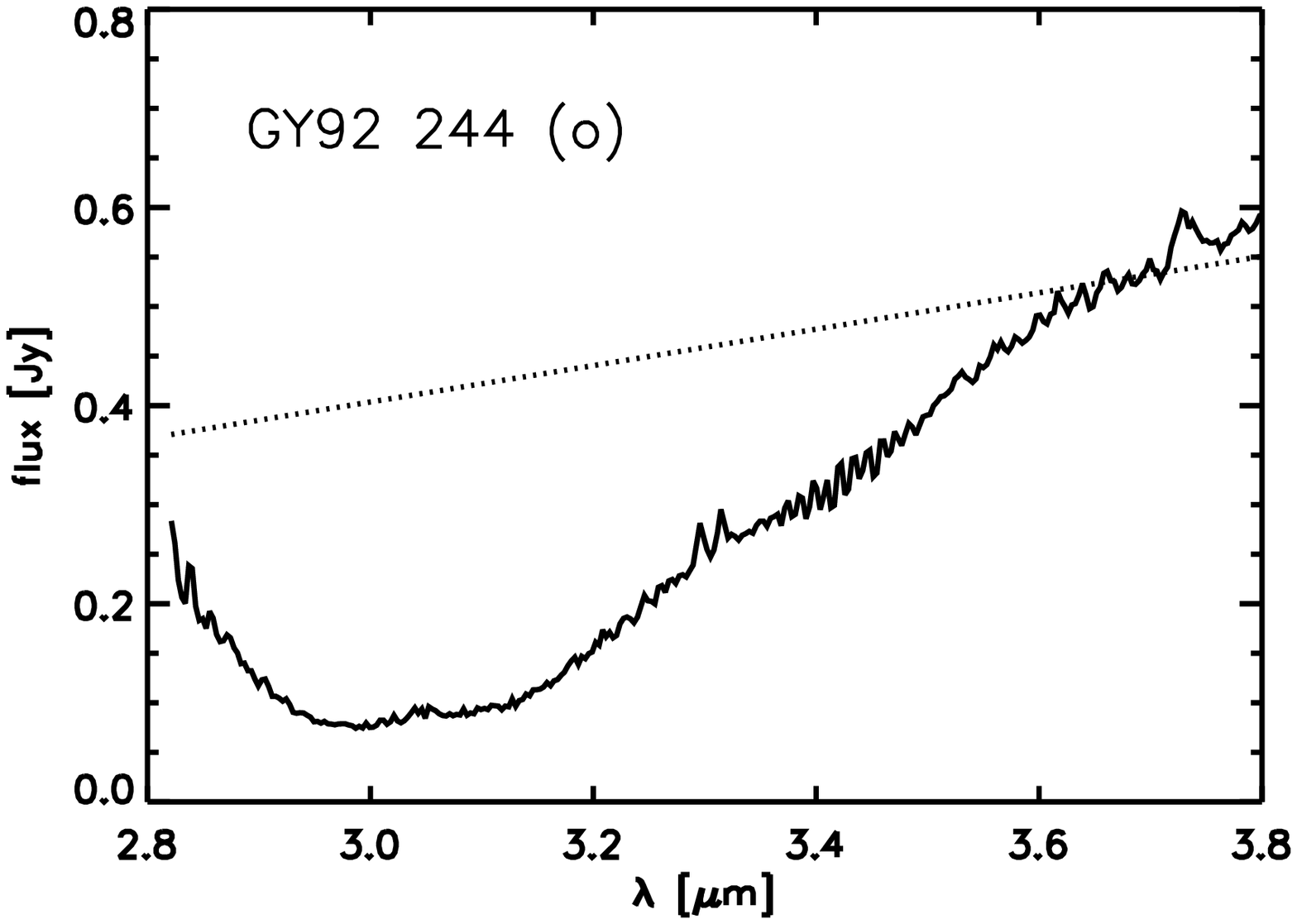}}\newline
    \resizebox{0.44\textwidth}{!}{\includegraphics{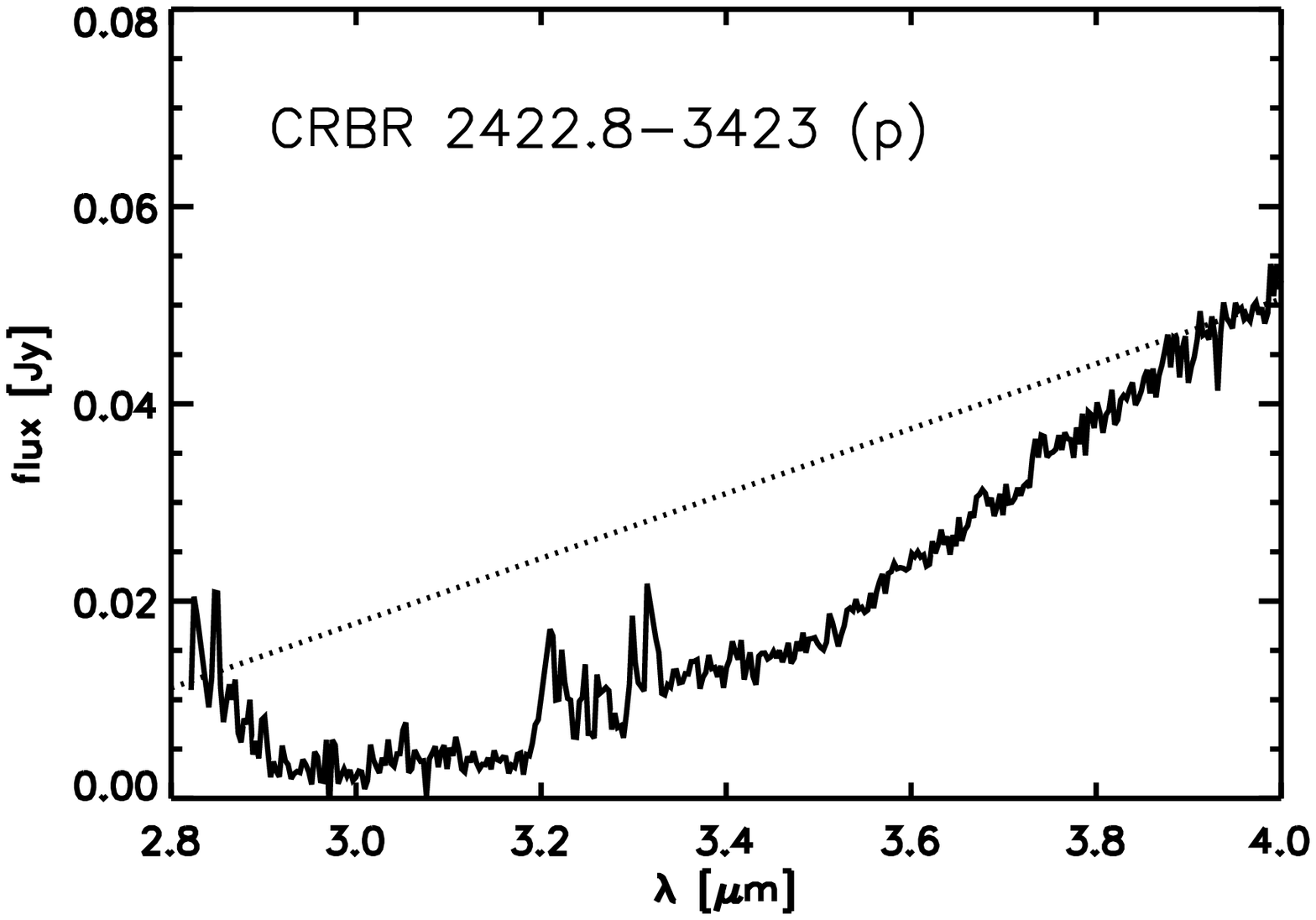}}
    \resizebox{0.44\textwidth}{!}{\includegraphics{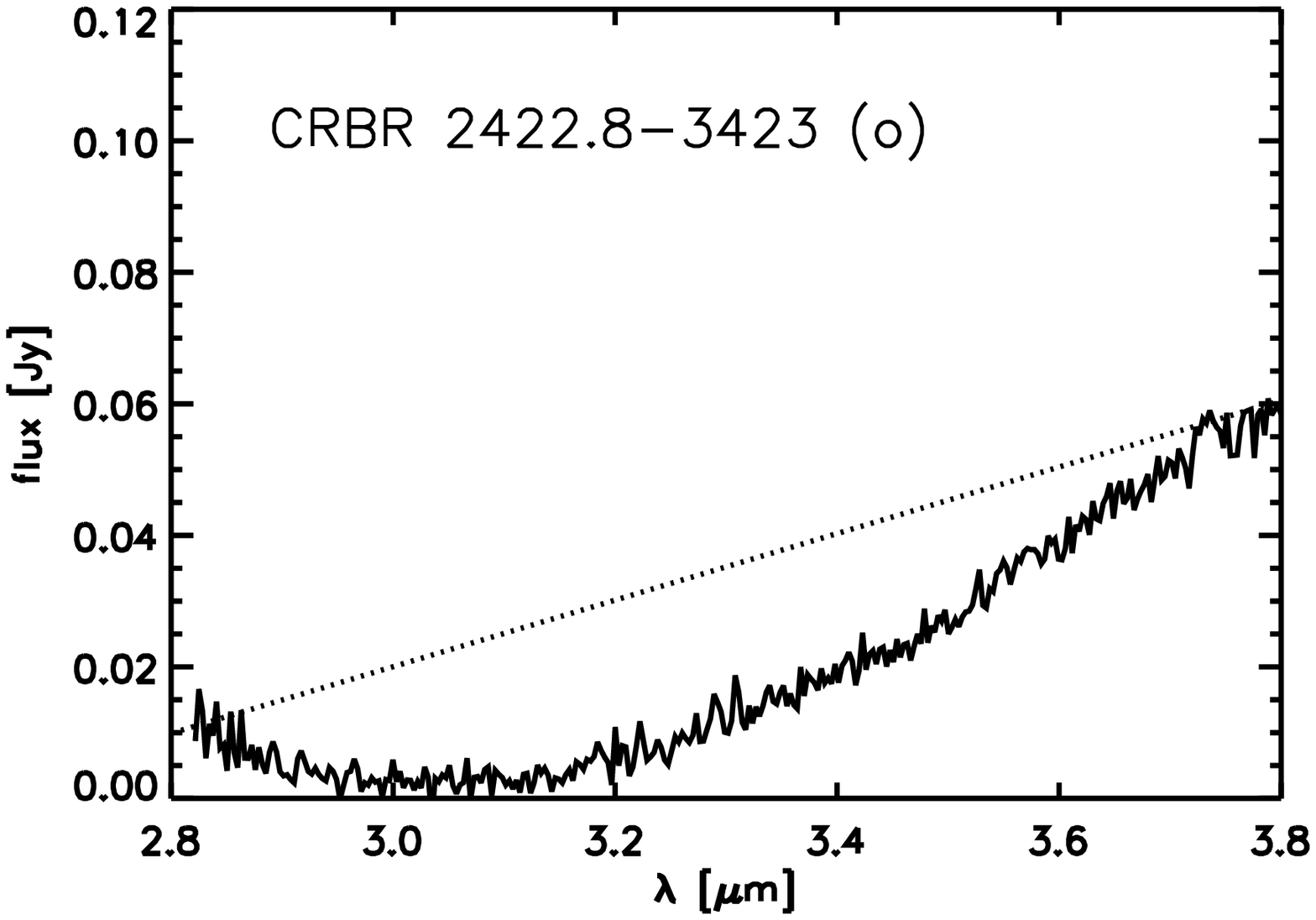}}
    \caption{Extracted and photometrically calibrated L band spectra of further
      targets.}
    \label{figure:result-spek_a} 
  \end{figure*}

  \begin{figure*}[!tb]
    \centering
    \resizebox{0.44\textwidth}{!}{\includegraphics{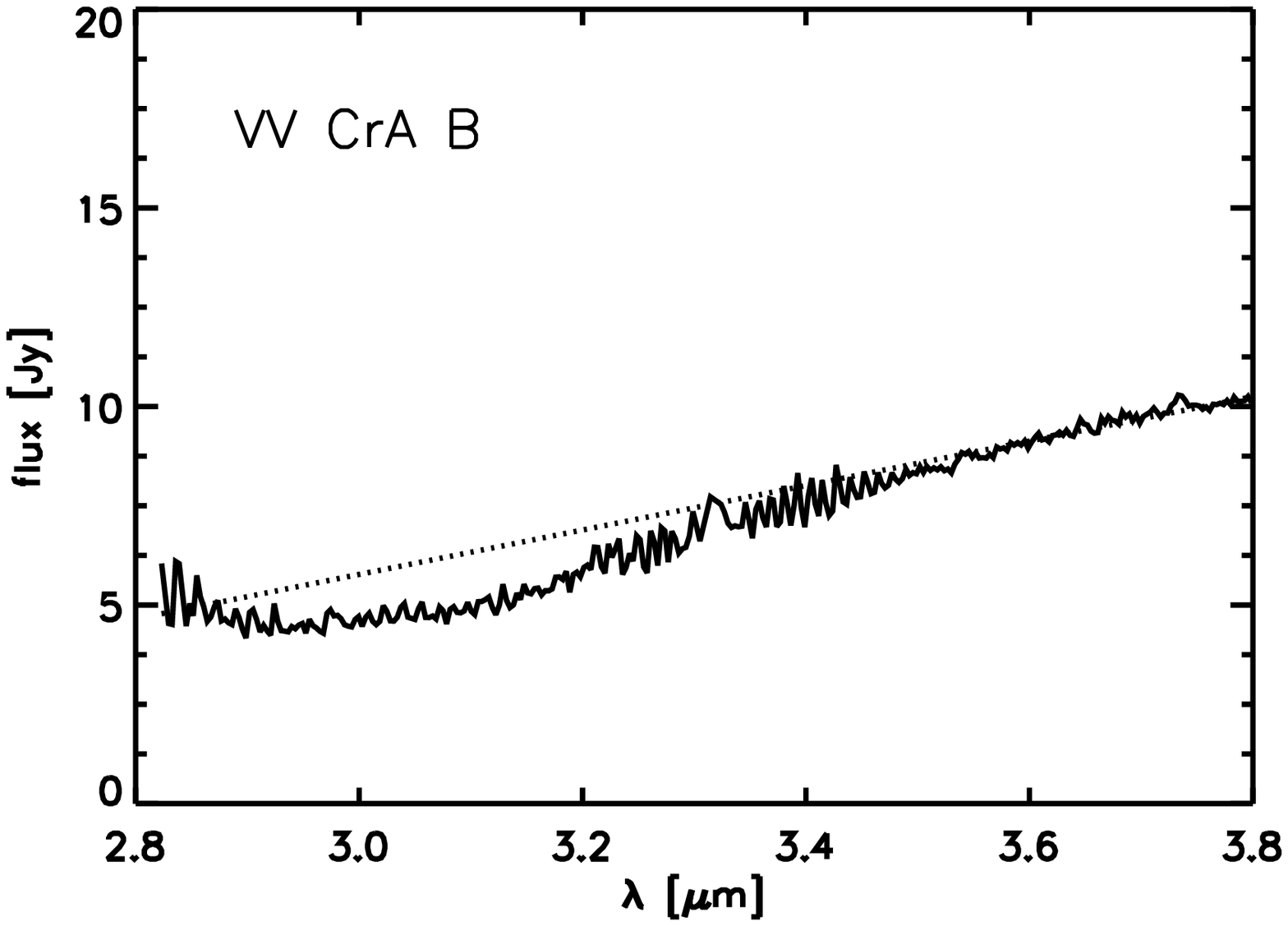}}
    \resizebox{0.44\textwidth}{!}{\includegraphics{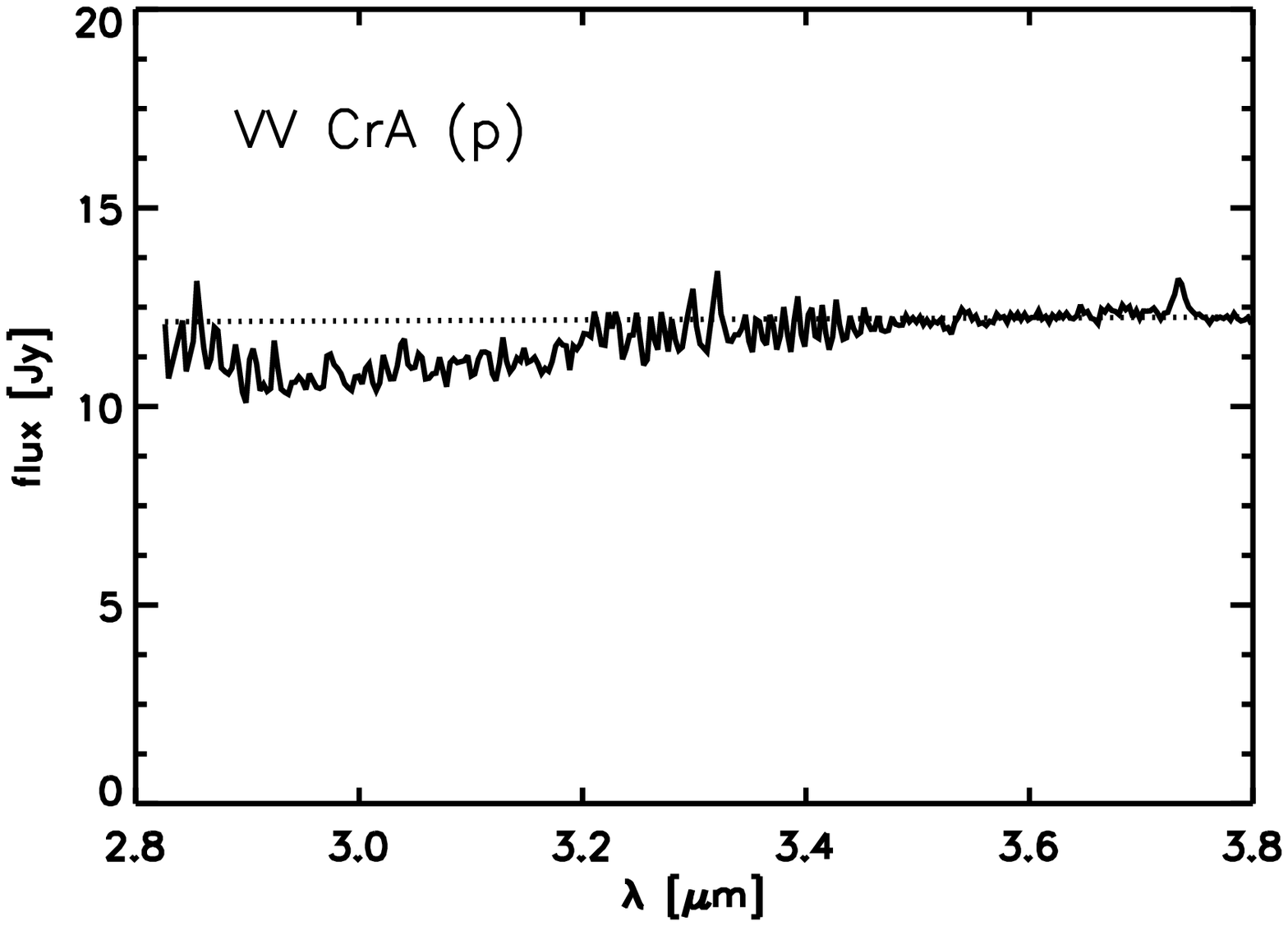}}\newline 
    \resizebox{0.44\textwidth}{!}{\includegraphics{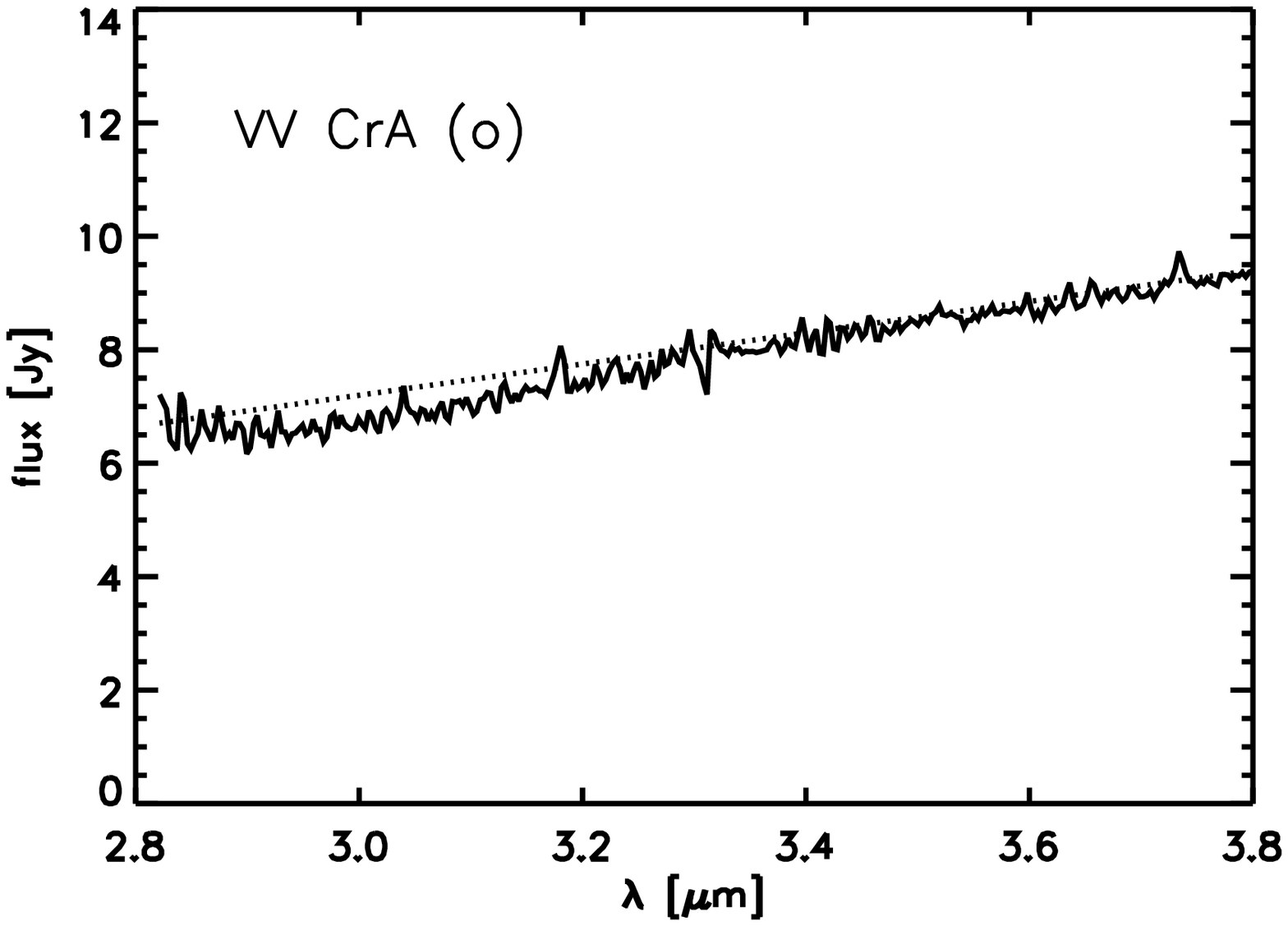}}    
    \caption{Continuation of Fig.~\ref{figure:result-spek_a}.}
    \label{figure:result-spekII_a} 
  \end{figure*} 
  
  The result of our observations of VV\,CrA\,A\footnote{The main component 
    of the binary in N band is the south-west component in the binary
    system. The north-east component is brighter in L band.}
  conflicts with a 
  spectroscopic measurement performed on data acquired using the 4m-telescope of the CTIO (Cerro
  Tololo Inter-American Observatory; Chen \& Graham~\cite{chen}). The
  absorption band of VV\,CrA\,A detected by our observation is weak, while
  the observation of Chen \& Graham~(\cite{chen}) does not show any
  absorption ice band at all. This
  difference may be attributed to the factor-of-two higher spectral re\-solution
  power of NAOS-CONICA and its higher spatial resolution
  power by a factor of $\sim$$11$. Therefore, observations with NAOS-CONICA
  allow us to observe
  more central regions that are strongly influenced by both the star
  and disk. The water ice absorption band of the infrared companion is
  more pronounced. This companion, that is strongly variable in the infrared
  wavelength range (Koresko et al.~\cite{koresko}), exhibits a broad
  absorption band. Our analysis of the observations of this source
  agree with previous measurements with the 4m-telescope at the CTIO. 

\begin{figure*}[!tb]
    \centering
    \resizebox{0.44\textwidth}{!}{\includegraphics{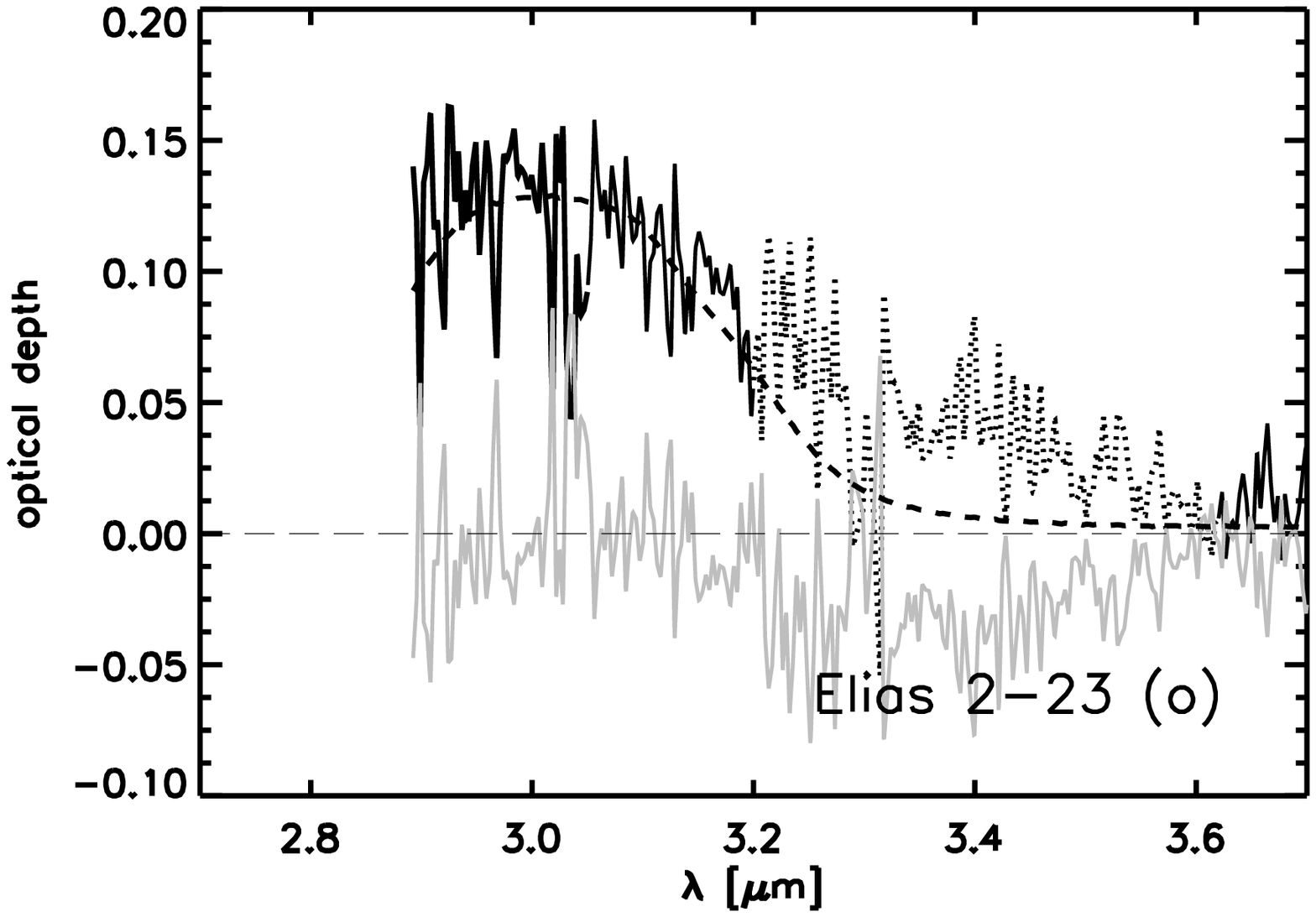}}\newline
    \resizebox{0.44\textwidth}{!}{\includegraphics{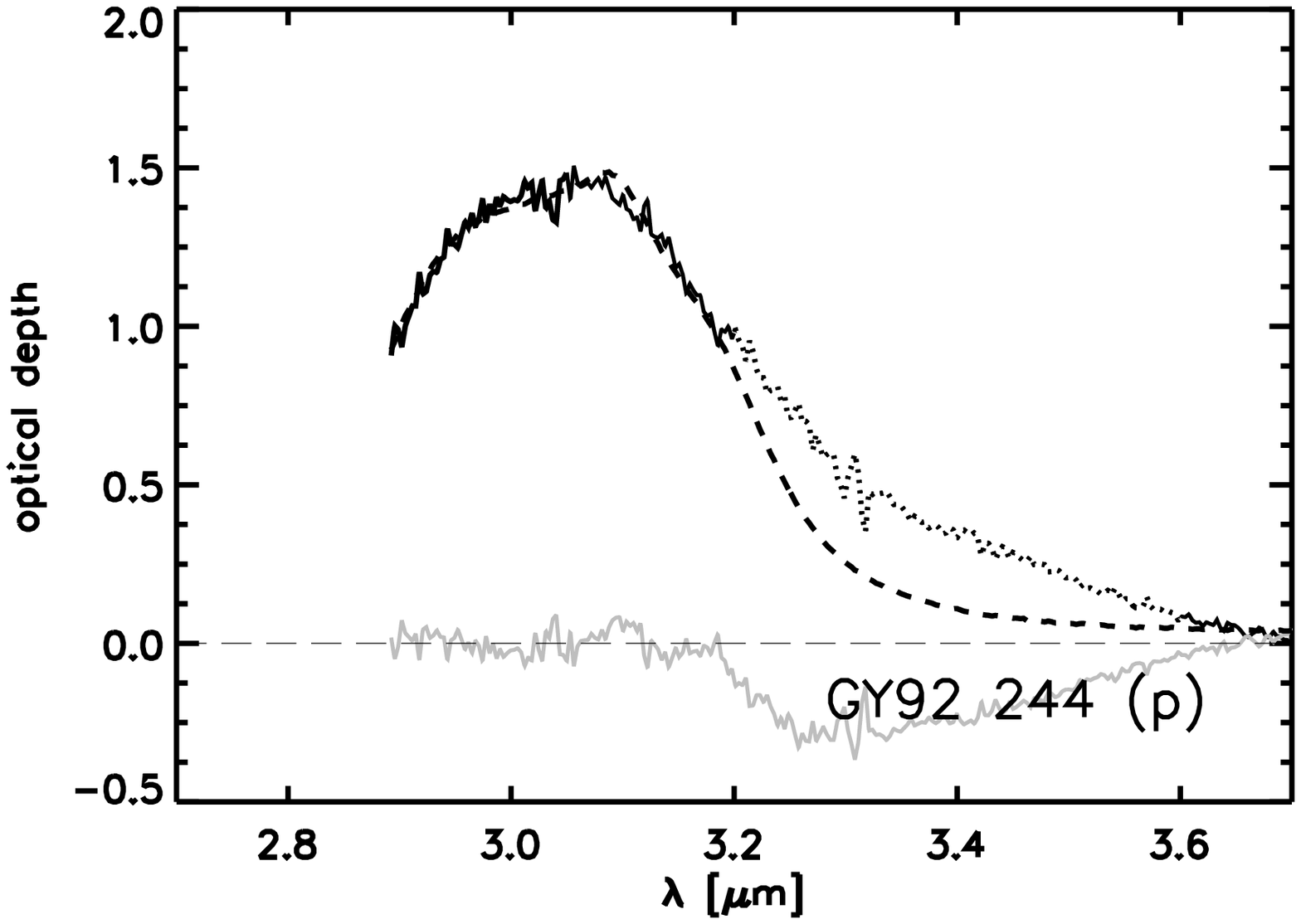}}
    \resizebox{0.44\textwidth}{!}{\includegraphics{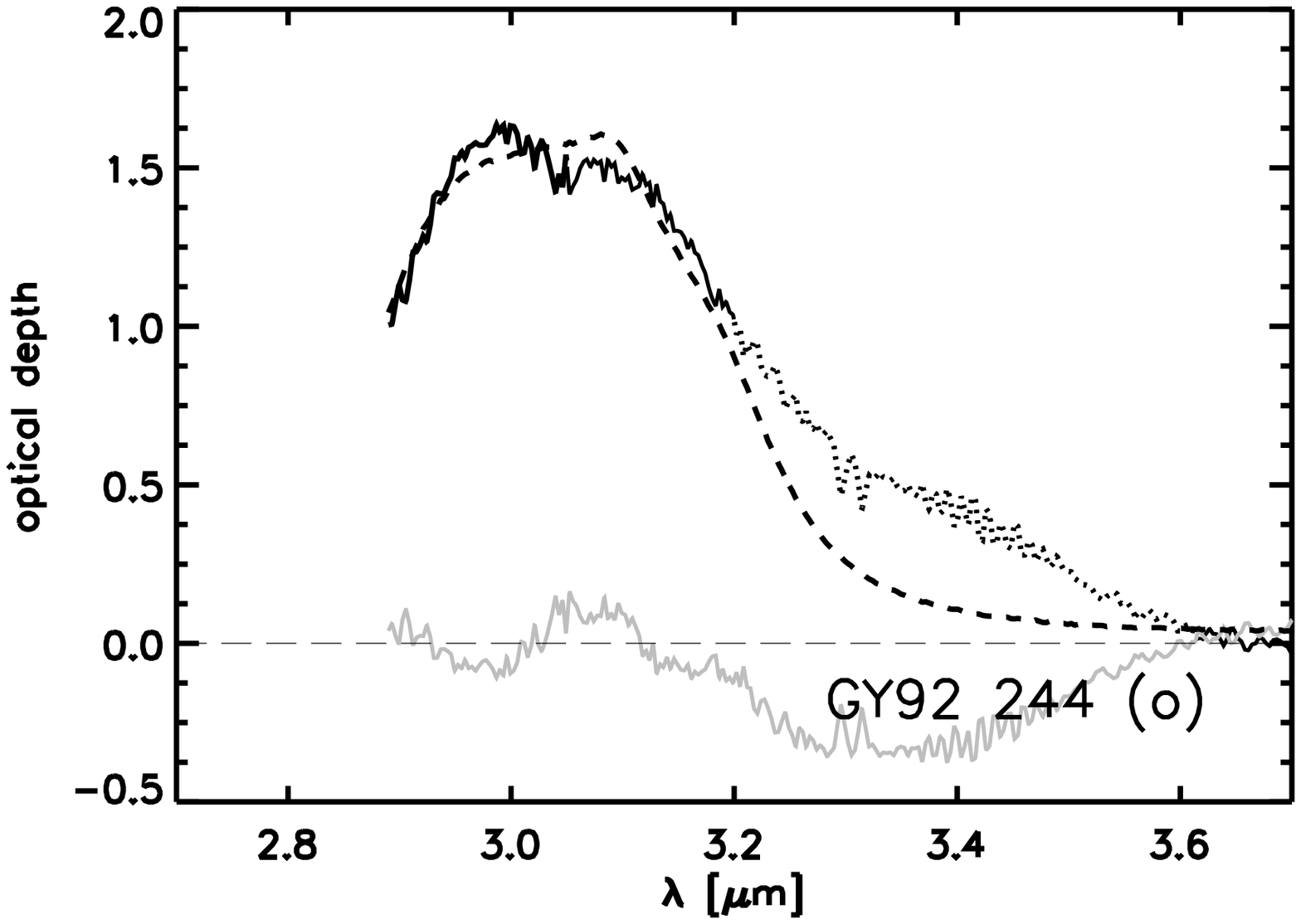}}\newline
    \resizebox{0.44\textwidth}{!}{\includegraphics{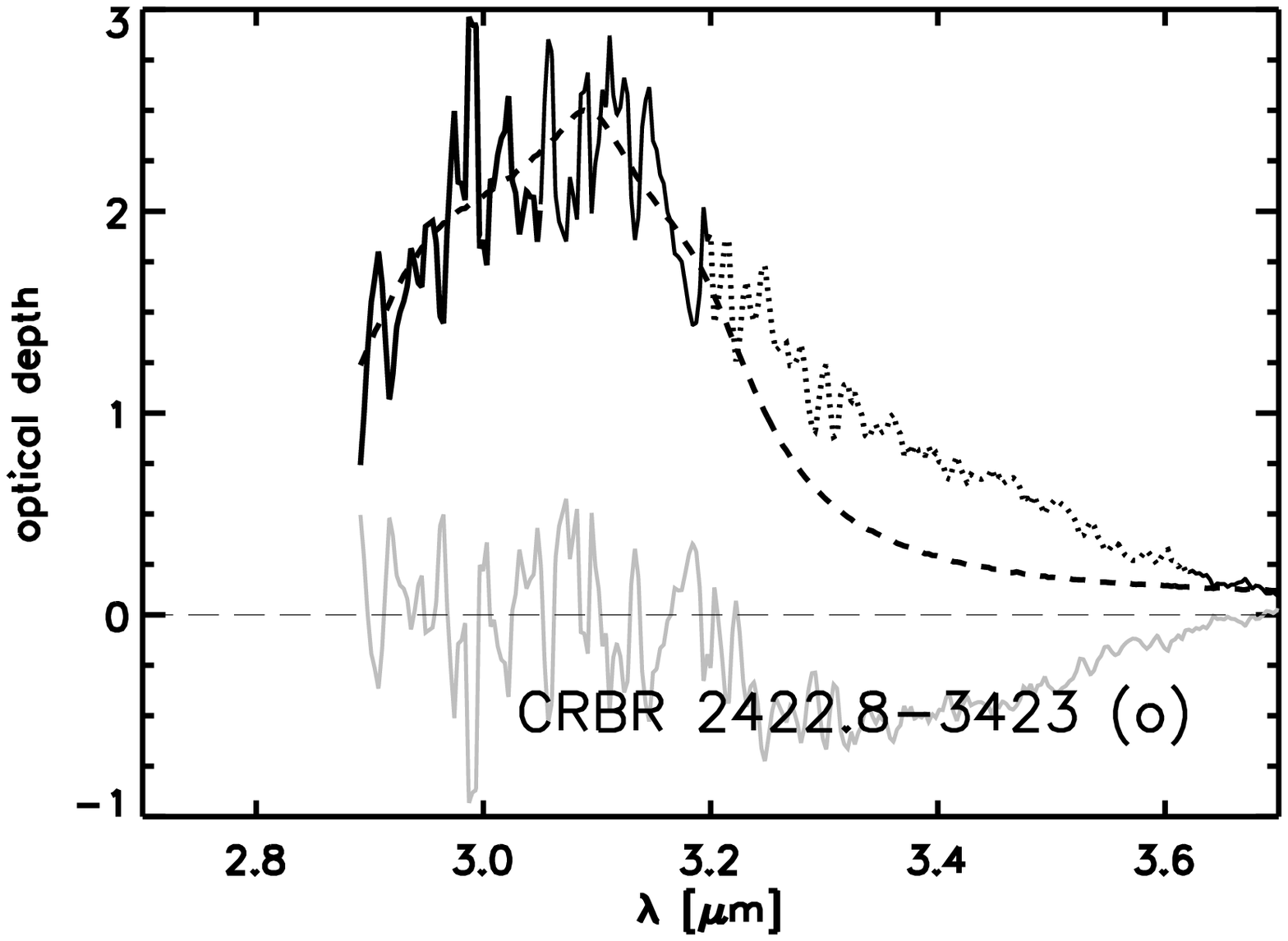}}
    \resizebox{0.44\textwidth}{!}{\includegraphics{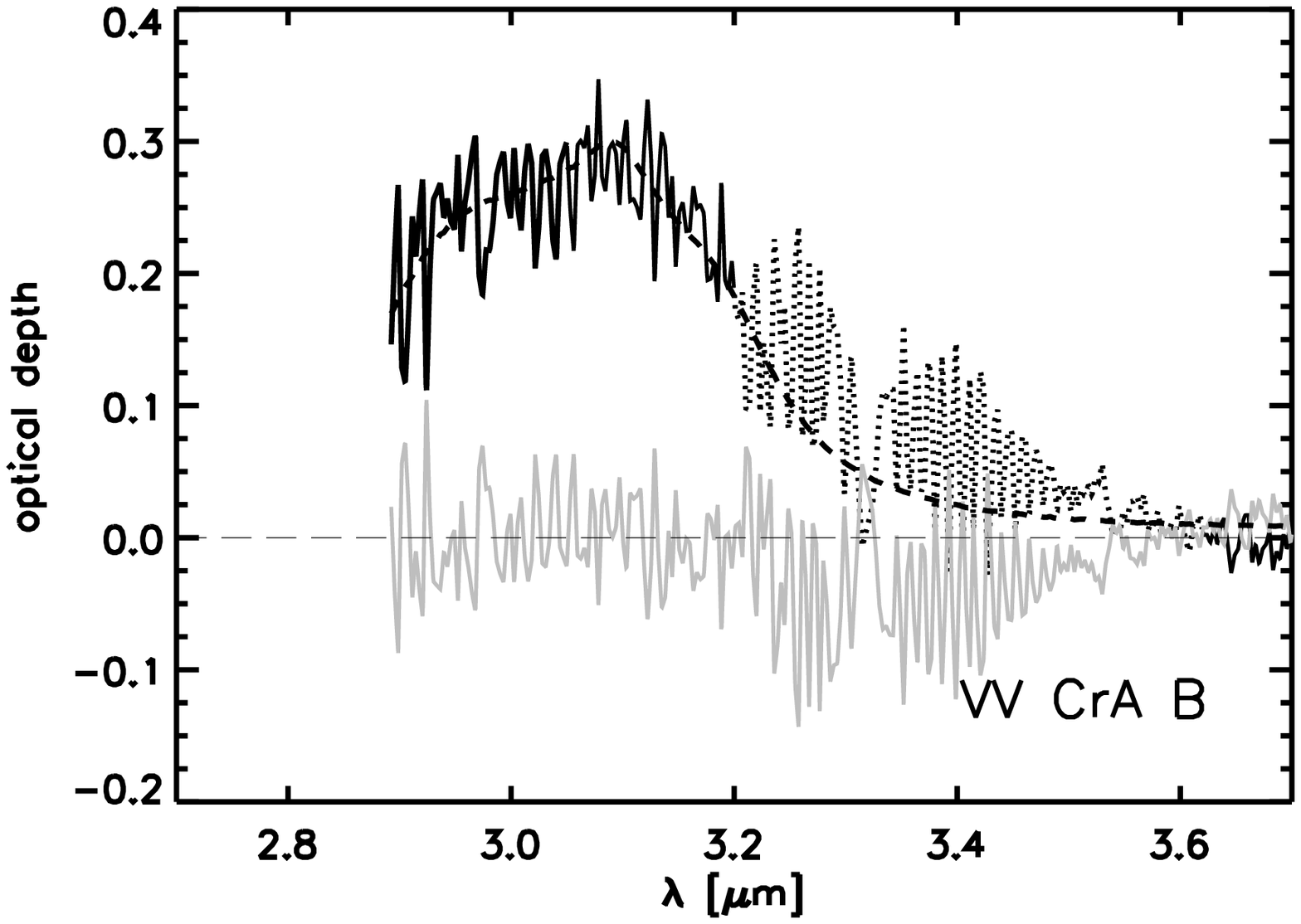}}
    \caption{Optical depth $\tau(\lambda)$ that is derived from the spectra of
      Fig.~\ref{figure:result-spek_a} and Fig.~\ref{figure:result-spekII_a}. The optical depth of the spectra Elias\,2-23~(p), 
      Elias\,2-23~(o), Elias\,2-21, CRBR\,2422.8-2423~(p), VV\,CrA~(p), and
      VV\,CrA~(o) are not modeled because of low signal-to-noise ratios and
      remaining telluric features
      that could not be sufficiently eliminated in data reduction.}
    \label{figure:result-water1_a}
  \end{figure*}

  Owing to the noisy data from which only parts of the telluric features could 
  be removed,
  processed water ice, i.\/e., crystallized and/or grown ice grains cannot
  unambiguously be found. Residual telluric
  features also hamper the determination of the column densities
  $N_\mathrm{A}(\mathrm{H}_{\mathrm{2}}\mathrm{O [ice]})$.
  Figure~\ref{figure:result-water1_a} shows the
  optical depths of the objects for which the determination of the column
  density and a subsequent modelling were possible, given the residual
  telluric features. Table~\ref{table:result-water_a} lists the
  corresponding results.
  \begin{table*}[!t]
    \centering
    \caption{Column density $N_\mathrm{A}$ of water ice along the line of sight to the
      observer (Eq.~\ref{eq:column-density}).}
    \label{table:result-water_a}
    \centering
    {\small \begin{tabular}{lccccccc} 
        \hline\hline
        exposure & $PA$ [$^{\circ}$] & $N_\mathrm{A}$ [$10^{22}$ m$^{-2}$] &
        $A_\mathrm{V}$ [mag] & 
        $m_\mathrm{a;s}$ [\%] & $m_\mathrm{a;l}$ [\%] & $m_\mathrm{c}$ [\%] &
        $\chi$ \\ \hline 
        Elia\,2-23 (p) & $53$ & $0.52$ & $7$ & -- & -- & -- & -- \\
        Elia\,2-23 (o) & $45$ & $0.26$ & $5$ & $100$ & -- & --  & $0.04$ \\ 
        Elia\,2-21 (p) & $53$ & $0.66$ & $9$ & -- & -- & -- & -- \\ 
        GY92\,244 (p) & $178$ &  $1.8$ & $20$ & $> 99$ & $< 1$ & $< 1$  &
        $0.08$ \\ 
        GY92\,244 (o) & $88$ &  $2.1$ & $23$ & $> 99$ & --  &  $< 1$ & $0.1$\\
        CRBR\,2422.8-3423 (p) & $53$ & $3$ & $35.1$ & -- & -- & --  & -- \\
        CRBR\,2422.8-3423 (o) & $45$ & $3$ & $29.1$ & $91$ & $< 1$  & $8$  &
        $0.2$ \\ 
        VV\,CrA\,B (p) & $134$ & $0.66$ & $9$ & $> 99$ & -- & $< 1$ & $0.05$
        \\       
        VV\,CrA\,A (p) & $134$ & $0.13$ & $3$ & -- & -- & -- & -- \\
        VV\,CrA\,A (o) & $43$ & $0.030$ & $2$ & -- & -- & -- & -- \\
        \hline
      \end{tabular}}  
  \end{table*}

 The detection of crystallized water ice in the spectra of CRBR\,2422.8-3423
 cannot definitely be confirmed. 
 Studying the derived profiles in Smith et al.~(\cite{smith}) and Dartois \&
 d'Hendecourt~(\cite{dartois}), a signal-to-noise ratio greater than
 $\sim$$4$ is required to prove crystallized water ice in the
 spectra, i.\/e., to distinguish the extinction profiles of crystallized and
 amorphous ice grains. Our measurement of CRBR\,2422.8-3423 is
 obtained from data with a
 signal-to-noise ratio of $\sim$$6$. However, the profile of
 CRBR\,2422.8-3423 cannot be used because its absorption is saturated
 (Fig.~\ref{figure:result-spek_a}). 

\end{document}